\documentclass[twocolumn,showpacs,aps,prd,amsfonts,
superscriptaddress,nofootinbib]{revtex4-1}

\usepackage{epsfig}
\usepackage{bm} 
\usepackage[usenames]{color}
\usepackage{bbding}

\newcommand{\Tr}{\mbox{Tr}}
\newcommand{\pard}{\partial}
\newcommand{\BB}{\mbox{\boldmath$B$}}
\newcommand{\BW}{\mbox{\boldmath$W$}}
\newcommand{\BV}{\mbox{\boldmath$V$}}
\newcommand{\BX}{\mbox{\boldmath$X$}}
\newcommand{\Bpard}{\mbox{\boldmath$\partial$}}
\newcommand{\dslash}{\!\not{\! \partial}}
\newcommand{\Bdslash}{\!\not{\! \Bpard}}
\newcommand{\Aslash}{\!\not{\!\! A}}
\newcommand{\Wslash}{\!\not{\!\! W}}
\newcommand{\BWslash}{\!\not{\!\! \BW}}
\newcommand{\Bslash}{\!\not{\!\! B}}
\newcommand{\BBslash}{\!\not{\!\! \BB}}
\newcommand{\Zslash}{\!\not{\!\! Z}}
\newcommand{\Vslash}{\!\not{\! V}}
\newcommand{\BVslash}{\!\not{\! \BV}}
\newcommand{\Omslash}{\!\not{\!\! \omega}}
\newcommand{\OmBarslash}{\!\not{\!\! \bar{\omega}}}
\newcommand{\cL}{{\cal L}}
\newcommand{\cM}{{\cal M}}
\newcommand{\Lagr}{{\cal L}}
\newcommand{\vphi}{\varphi}
\newcommand{\Comm}[2]{\mbox{$[$} #1, #2\mbox{$]$}}
\newcommand{\UnitMatrix}{\mbox{\bf I}}
\newcommand{\diag}{\mbox{diag}}
\newcommand{\doublet}[2]
{\left( \begin{array}{c} #1 \\ #2 \end{array}\right)}
\newcommand{\MatrixTwo}[4]
{\left( \begin{array}{cc}
#1 & #2 \\
#3 & #4
\end{array}\right)}
\newcommand{\triplet}[3]
{\left( \begin{array}{c} #1 \\ #2 \\ #3 \end{array}\right)}
\newcommand{\MatrixThree}[9]
{\left( \begin{array}{ccc}
#1 & #2 & #3 \\
#4 & #5 & #6 \\
#7 & #8 & #9
\end{array}\right)}

\begin{document}

\newcommand*{\zilina}{Physics Department, University of \v{Z}ilina, Univerzitn\'{a} 1, 010 26
\v{Z}ilina, Slovakia}
\newcommand*{\praha}{Institute of Experimental and Applied Physics,
Czech Technical University in Prague, \\
Horsk\'{a} 3a/22, 128 00 Prague, Czech Republic}

\title{Top-BESS model and its phenomenology}

\author{Mikul\'{a}\v{s} Gintner}
\email{gintner@fyzika.uniza.sk}
\affiliation{\zilina}
\affiliation{\praha}
\author{Josef Jur\'{a}\v{n}}
\email{josef.juran@utef.cvut.cz}
\affiliation{\praha}
\author{Ivan Melo}
\email{melo@fyzika.uniza.sk}
\affiliation{\zilina}


\begin{abstract}
We introduce the top-BESS model which is the effective description
of the strong electroweak symmetry breaking with a single new $SU(2)_{L+R}$
triplet vector resonance.
The model is a modification of the BESS model in the fermion sector.
The triplet couples to the third generation of quarks only. This
approach reflects a possible extraordinary role of the top quark
in the mechanism of electroweak symmetry breaking.
The low-energy limits on the model parameters found provide hope for
finding sizable signals in the LHC Drell-Yan processes as well as
in the s-channel production processes at the ILC.
However, there are regions of the model parameter space where the interplay
of the direct and indirect fermion couplings
can hide the resonance peak in a scattering process
even though the resonance exists and
couples directly to top and bottom quarks.
\end{abstract}

\maketitle

\section{Introduction}
\label{sec:Intro}

Despite the great success of the Standard model (SM)~\cite{GWS}
one essential component of the theory remains a puzzle: it is the
actual mechanism behind the electroweak symmetry breaking (ESB).
Spontaneous breaking of electroweak symmetry accompanied by the
Higgs mechanism is the way to reconcile the massive gauge bosons
with the principle of gauge invariance. The introduction of the
Higgs complex doublet scalar field of a non-zero vacuum
expectation value to the electroweak theory serves as a benchmark
hypothesis for the mechanism. A direct consequence of this
hypothesis is the presence of the scalar Higgs boson in the
particle spectrum of the SM, not observed as of yet, though.

Nevertheless, there is a host of candidates for alternative
extensions of the SM that offer their own mechanisms of ESB. If
the Large Hadron Collider (LHC) does not discover the SM Higgs
boson, ESB could originate from strongly interacting new physics.
In this scenario the symmetry breaking is triggered by new
non-perturbative forces which form bound states of new elementary
particles. The bound states would appear in the particle spectrum
as new resonances. Typical representatives of this scenario are
the Technicolor model (TC)~\cite{TC} and its
extensions~\cite{ETC,WalkingTC,TopcolorTC}.

More recent extra-dimensional theories~\cite{ExtraDim} predict the
Kaluza-Klein towers of new resonances of which the lowest lying
resonances might be discovered at the LHC. The attractiveness of
this development is strengthened by Maldacena's
conjecture~\cite{Maldacena} on the dual-description relation
between the extra-dimensional weakly interacting theories and the
strongly interacting models in four dimensions.

Obviously, all the alternative extensions must converge to the SM
without Higgs when pre-LHC energies are considered. Facing this
plethora of hypotheses it is desirable to develop unifying
descriptions of their low-energy phenomenologies. For this
purpose, the formalism of effective Lagrangians is very suitable.
The effective Lagrangians can accommodate new particles predicted
by the extensions. The new particles are not only a by-product of a
particular ESB mechanism but they will be needed to tame the
model's unitarity if the Higgs boson below 1~TeV is not
found~\cite{LeeQuiggThacker77}.

In this paper we introduce the {\it top-BESS model} (tBESS)
--- the modified version of the BESS (Breaking Electroweak Symmetry
Strongly) model~\cite{BESS}. The basic ideas of the tBESS model
were formulated already in~\cite{tBESSActaPhysicaSlovaca}.
Both models describe a new $SU(2)$ vector boson triplet that can
represent the spin-1 bound states of hypothetical new strong
interactions. They are effective descriptions of strong Higgsless
ESB based on the $SU(2)_L\times SU(2)_R\times U(1)_{B-L}\times
SU(2)_{HLS}$ global symmetry of which the $SU(2)_L\times
U(1)_Y\times SU(2)_{HLS}$ subgroup is also a local symmetry. ``HLS''
stands for the \textit{hidden local symmetry}~\cite{HLS}, which is
an auxiliary gauge symmetry introduced to accommodate the $SU(2)$
triplet of vector resonances. Beside the triplet, the models
contain only the observed SM particles.

The BESS and tBESS models are gauge equivalent to the nonlinear
sigma model on the $SU(2)_L\times SU(2)_R/SU(2)_{L+R}$ coset space
with the $SU(2)_{L+R}$ vector triplet added in the way introduced
by Weinberg~\cite{WeinbergRho}.

The BESS model also corresponds to the simplest version of the five-dimensional
Higgsless model in the deconstructed picture of three lattice
sites and the $SU(2)_L\times SU(2)\times U(1)_Y$ gauge symmetry,
when direct couplings between SM fermions and new vector bosons
are introduced~\cite{3siteExtraDimBESS}. The four-site Higgsless
model based on the $SU(2)_L\times SU(2)_1\times SU(2)_2\times
U(1)_Y$ gauge symmetry~\cite{4siteExtraDimDBESS} corresponds to
the degenerated BESS model~\cite{DBESS}.

In the tBESS model we modify the \textit{direct} interactions of
the vector triplet with fermions. While in the BESS model there is
a universal direct coupling of the triplet to all fermions of a
given chirality, in our modification we admit direct couplings of
the new triplet-to-top and bottom quarks only. Our modification is
inspired by the speculations about a special role of
the top quark
(or the third quark generation) in the mechanism of ESB~
\cite{snowmass,Han4}. The large top mass is surprisingly close to
the ESB scale: $m_t\approx v/\sqrt{2}$. This suggests that $m_t$
could be generated by the same mechanism as $M_W$ and $M_Z$, i.e.,
by the same strong interactions which are also responsible for
ESB. If this were the case, we would expect the new triplet to
couple significantly to the weak gauge bosons as well as to the
top quark. This happens, for example, for the vector $\rho_T$
resonance of the Extended Technicolor~\cite{ETC}.

On the other hand, the mechanism behind the top mass could differ
from the ESB mechanism. Thus, it can be represented by yet another
sector of new strong interactions introduced just for that sake,
e.g.\ the Topcolor of the Topcolor Assisted
Technicolor~\cite{TopcolorTC}, where $\rho_T$ couples only weakly
to the top quark. The neutral component of the new vector triplet
can also mimic couplings of a $Z'$ spin-1 resonance~\cite{Han3,He}
which has large couplings to the top and bottom, vanishing couplings to
$W$, $Z$, and very small couplings to fermions of the first two
generations. The authors of~\cite{Han1,Han2,Han3} studied the
effective description of the situation when the third quark
generation couples extraordinarily to the new strong resonances,
scalar and vector ones, under some simplifying assumptions. They
also studied various processes as probes of the resonances.

In the tBESS model, we take the possible chirality dependence of
the triplet-to-top/bottom coupling into account multiplying the
$SU(2)_{HLS}$ gauge coupling $g''$ by the $b_L$ and $b_R$
parameters for the left and right fermion doublets, respectively.
In addition, we can disentangle the triplet-to-top-quark right
coupling from the triplet-to-bottom-quark right coupling. This
breaks the $SU(2)_R$ symmetry which is broken by the SM
interactions, anyway. For this sake, we have introduced a free
parameter, $0\leq p\leq 1$. The $p$ parameter can weaken the
strength of the triplet-to-bottom-quark right coupling. However, the $SU(2)_L$
symmetry does not allow us to do the same splitting for the left
quark doublet.

There are two more invariant terms introduced in the tBESS
effective Lagrangian when compared to the BESS model. They are
multiplied by additional free parameters, $\lambda_L$ and
$\lambda_R$. While the $\lambda$ terms do not have a significant
impact on the behavior of the model at energies around the mass of
the vector triplet, they do influence the low-energy limits for
its parameters. Namely, the presence of $\lambda$ terms helps to
relax the low-energy limits for the fermion parameters.

In the tBESS model the vector triplet is introduced as a gauge
field which results in the mixing of the triplet with the
electroweak gauge bosons. Consequently, the \textit{indirect} ---
mixing-induced --- interactions of the vector triplet with
fermions appear on the scene. For the light fermions this is the
only way they can interact with the vector triplet in the
tBESS model. Of course, the interactions are suppressed by the
elements of the mixing matrix.

Since the indirect couplings are suppressed by the mixing factors
it does not have to seem worthwhile to study the tBESS model in
processes where the vector resonances couple to light fermions. We
suggest that despite this naive expectation it is not necessarily
so.

There are regions of the parameter space where the interference of
the direct and indirect couplings suppresses or even zeros a
particular top/bottom decay channel of the vector triplet.
Consequently, the resonance peak might not be visible in a
particular experiment even though the new vector triplet exists.

This paper is organized as follows. In
Section~\ref{sec:tBESSmodel} we introduce the tBESS effective
Lagrangian. While Subsection~\ref{subsec:RHOlagrangian} recalls
the details of the gauge boson and ESB sectors that are shared
with the BESS model, in Subsection~\ref{subsec:Fermions} our
modifications of the fermion sector are explained. In
Section~\ref{sec:Pheno} the basic properties of the tBESS model
are discussed. Beside the vector resonance decay widths, the
unitarity and low-energy limits for the model are derived in
Subsections~\ref{subsec:Unitarity} and \ref{subsec:Limits},
respectively. In Subsection~\ref{subsec:DeathValley} the effect of
suppressing the partial decay widths of the vector triplet for
some values of the parameter space is discussed. In
Subsection~\ref{subsec:ScatteringProcesses} we illustrate the
impact of the suppression on $e^+e^-$ processes. We also suggest
that it might be feasible to use the light fermion enabled
processes at the ILC and the LHC (Drell-Yan) to study the tBESS
model. Section~\ref{sec:End} contains our conclusions followed by
appendices.

\section{The top-BESS model}
\label{sec:tBESSmodel}

The top-BESS effective Lagrangian can be split in three parts
\begin{equation}\label{tBESSLag}
  \cL_{tBESS} = \cL_{GB} + \cL_{ESB} + \cL_{ferm},
\end{equation}
where $\cL_{GB}$ describes the gauge-boson sector including
the $SU(2)_{HLS}$ triplet, $\cL_{ESB}$ is the scalar sector
responsible for spontaneous breaking of the electroweak and
hidden local symmetries, and $\cL_{ferm}$ is the fermion Lagrangian
of the model. The individual terms will be elaborated on
in the subsections below.
Of course, the SM Lagrangian, up to the Higgs doublet,
must be a low-energy approximation of $\cL_{tBESS}$.

\subsection{The \bm{$SU(2)_{HLS}$} vector triplet Lagrangian}
\label{subsec:RHOlagrangian}

We start with reviewing the gauge-field and scalar sectors of the
tBESS model which are --- except for the used notations ---
identical with those of the original BESS model \cite{BESS}. It
contains six unphysical real scalar fields, would-be Goldstone
bosons of the model's spontaneous symmetry breaking. Thus,
naturally, the sector provides the energy scale $v$ of ESB.

Beside the SM gauge fields $W_\mu^a(x)$ and $B_\mu(x)$ there is
the $SU(2)_{HLS}$ gauge triplet
$\vec{V}_\mu=(V_\mu^1,V_\mu^2,V_\mu^3)$ introduced in the model.
Under the $\Gamma^{glob}\times SU(2)_{HLS}^{loc}$ group, where
$\Gamma=SU(2)_L\times SU(2)_R$, it transforms as
\begin{equation}
  \BV_\mu \rightarrow h^\dagger \BV_\mu h +h^\dagger\pard_\mu h,
\end{equation}
where $h(x)\in SU(2)_{HLS}^{loc}$ and $\BV_\mu =
i\frac{g''}{2}V_\mu^a\tau^a$. The $2\times 2$ matrices
$\vec{\tau}=(\tau^1,\tau^2,\tau^3)$ are the $SU(2)$ generators.

The gauge-boson Lagrangian $\cL_{GB}$ is composed of the
Lagrangians for the individual gauge bosons
\begin{eqnarray}
   \cL_{GB} &=& \cL_{W}+\cL_{B}+\cL_{V},
   \\
   \cL_{W} &=& \frac{1}{2g^2}\Tr(\BW_{\mu\nu}\BW^{\mu\nu}),
   \\
   \cL_{B} &=& \frac{1}{2g^{\prime 2}}\Tr(\BB_{\mu\nu}\BB^{\mu\nu}),
   \\
   \cL_{V} &=& \frac{2}{g^{\prime\prime 2}}\Tr(\BV_{\mu\nu}\BV^{\mu\nu}),
\end{eqnarray}
with the field strength tensors
\begin{eqnarray}
  \BW_{\mu\nu} &=& \pard_\mu \BW_\nu - \pard_\nu \BW_\mu + \Comm{\BW_\mu}{\BW_\nu},
  \\
  \BB_{\mu\nu} &=& \pard_\mu \BB_\nu - \pard_\nu \BB_\mu,
  \\
  \BV_{\mu\nu} &=& \pard_\mu \BV_\nu - \pard_\nu \BV_\mu + \Comm{\BV_\mu}{\BV_\nu},
\end{eqnarray}
where $\BW_\mu = i g W_\mu^a\tau^a$, $\BB_\mu = i g' B_\mu Y$ are
$SU(2)_L$ and $U(1)_Y$ gauge fields.

To generate the gauge-boson masses, the six real scalar fields
$\vphi_L^a(x), \vphi_R^a(x),\; a=1,2,3$, are introduced as
parameters of the~$\Gamma$ group elements in the exp form
\begin{equation}\label{XiExpForm}
  \Xi(\vec{\vphi}_L,\vec{\vphi}_R)=\diag(\;\xi(\vec{\vphi}_{L}),\;\xi(\vec{\vphi}_{R})\;)\;\;\in\;\Gamma,
\end{equation}
where $\vec{\vphi}=(\vphi^1,\vphi^2,\vphi^3)$,
$\xi(\vec{\vphi})=\exp(i\vec{\vphi}\vec{\tau}/v)\in SU(2)$ and $v$
is the scale of ESB. The $\Xi$ matrix
transforms\footnote{Transformation properties of the Lagrangian
composing variables are summarized in Appendix \ref{app:Trafos}.}
linearly under $\Gamma^{glob}\times SU(2)_{HLS}^{loc}$:
\begin{equation}
 \Xi(\vec{\vphi}_L,\vec{\vphi}_R)\;\;\rightarrow\;\; G\cdot \Xi\cdot H(x),
\end{equation}
where $G=\diag(g_L,g_R)$, $H(x)=\diag(h(x),h(x))$, $g_{L,R}\in
SU(2)_{L,R}$, $h\in SU(2)_{HLS}$. The scalar fields couple to the
gauge bosons in the form given by the $[SU(2)_L\times U(1)_Y\times
SU(2)_{HLS}]^{loc}$ invariant Lagrangian
\begin{equation}\label{eq:LagrESB}
  \Lagr_{ESB} = -\frac{v^2}{2}\left[\Tr\left(\bar{\Omega}_\mu^\perp\right)^2
                +\alpha\Tr\left(\bar{\Omega}_\mu^\parallel\right)^2\right],
\end{equation}
where $\alpha$ is a free parameter and $\bar{\Omega}_\mu^\perp$
and $\bar{\Omega}_\mu^\parallel$ are $SU(2)_{L-R}$ and
$SU(2)_{L+R}$ projections of the gauged Maurer-Cartan 1-form
$\bar{\Omega}_\mu$
\begin{eqnarray}
  \bar{\Omega}_\mu(\vec{\vphi}_L,\vec{\vphi}_R) &=&
  \Xi^\dagger(\vec{\vphi}_L,\vec{\vphi}_R)\cdot D_\mu\Xi(\vec{\vphi}_L,\vec{\vphi}_R),
  \label{OmegaMu}\\
  \bar{\Omega}^{\parallel,\perp}_\mu(\vec{\vphi}_L,\vec{\vphi}_R) &=&
  \frac{1}{2}\left[\bar{\Omega}_\mu(\vec{\vphi}_L,\vec{\vphi}_R) \pm
                   \bar{\Omega}_\mu(\vec{\vphi}_R,\vec{\vphi}_L)\right].
  \label{OmegaMuParPerp}
\end{eqnarray}
The projections have a block-diagonal form
\begin{equation}
  \bar{\Omega}^{\parallel,\perp}_\mu =
  \diag(\bar{\omega}^{\parallel,\perp}_\mu,\pm\bar{\omega}^{\parallel,\perp}_\mu),
\end{equation}
where the expressions for $\bar{\omega}^{\parallel,\perp}_\mu$ in
terms of $\xi$'s can be inferred from Eqs.~(\ref{XiExpForm}),
(\ref{OmegaMu}), and (\ref{OmegaMuParPerp}). The covariant
derivative $D_\mu\Xi$ reads
\begin{equation}
  D_\mu\Xi(\vec{\vphi}_L,\vec{\vphi}_R)\;=\;\pard_\mu\Xi \;+\; \BX_\mu\cdot\Xi \;-\; \Xi\cdot\BV_\mu,
\end{equation}
where $\BX_\mu=igW_\mu^a T_L^a +ig'B_\mu Y$,
$\BV_\mu=i\frac{g''}{2}V_\mu^a T^a$, $T_L^a=\diag(\tau^a,0)$,
$T_R^3=\diag(0,\tau^3)$, $T^a=\diag(\tau^a,\tau^a)$,
$Y=T_R^3+\frac{1}{2}(B-L)\UnitMatrix^{(4)}$,
$\UnitMatrix^{(4)}=\diag(1,1,1,1)$, and $B$, $L$ denote the baryon
and lepton numbers, respectively.

It can be shown that all the six scalar fields can be transformed
away by an appropriate gauge transformation. Thus they are
unphysical. Namely, the scalar triplet
$\vec{\sigma}=(\vec{\vphi}_L+\vec{\vphi}_R)/2$ can be gauged away
by the $SU(2)_{HLS}^{loc}$ transformation
$h(x)=\xi(\vec{\sigma})$, leaving us with the pseudoscalar
triplet $\vec{\pi}=(\vec{\vphi}_L-\vec{\vphi}_R)/2$. The gauge
transformation turns the Lagrangian (\ref{eq:LagrESB}) into the
gauged nonlinear sigma model on the $SU(2)_L\times
SU(2)_R/SU(2)_{L+R}$ coset space. The triplet $\vec{\pi}$ plays a
role of the Goldstone bosons which supply masses to the
electroweak gauge bosons through the Higgs mechanism. The
$SU(2)_{HLS}$ vector triplet enters the resulting nonlinear sigma
model Lagrangian in the way introduced originally by Weinberg
\cite{WeinbergRho}.

To obtain the masses of the electroweak gauge bosons as well as of
the new vector resonances their mass matrix has to be
diagonalized. The eigenstate transformation matrices of the
neutral and charged gauge-boson sectors, $O^{N}$ and $O^{C}$,
transform the mass eigenstates to the flavor eigenstates
\begin{equation}\label{eq:MassToFlavorBasisTrafo}
  \triplet{W^3}{B}{V^3}_{flavor}=O^{N}\triplet{A}{Z}{V^0}_{mass},\;
\end{equation}
\begin{equation}
\doublet{W^\pm}{V^\pm}_{flavor}=O^{C}\doublet{W^\pm}{V^\pm}_{mass}.
\end{equation}
Note that $X^\pm=(X^1\mp iX^2)/\sqrt{2}$ where $X=W,V$. In the
limit $M_{W^\pm}, M_Z \ll M_{V^0}$ which is equivalent to the
condition $g\ll \sqrt{\alpha}g''$, the mixing matrices
read\footnote{ We are not showing exact formulas for the mass
matrices in this paper. These can be found in the papers on the
original BESS model~\cite{BESS}. Nevertheless, in the process
calculations we have used the exact formulas.}
\begin{equation}\label{eq:ONapprox}
  O^{N}=\MatrixThree{g'/G}{g/G}{-g/g''}
                      {g/G}{-g'/G}{-g'/g''}
                      {2\frac{gg'}{Gg''}}{\frac{g^2-g^{\prime 2}}{Gg''}}{1},
\end{equation}
\begin{equation}\label{eq:OCapprox}
  O^{C}=\MatrixTwo{1}{-g/g''}{g/g''}{1},
\end{equation}
where $G=\sqrt{g^2+g^{\prime 2}}$. In the same limit the gauge
masses can be approximated by the following formulas:
\begin{eqnarray}
  M_{W^\pm} &=& \frac{vg}{2}\left(1-\frac{g^2}{2g^{\prime\prime 2}}\right),
  \\
  M_Z &=& \frac{vG}{2}\left[1-\frac{(g^2-g^{\prime 2})^2}{2g^{\prime\prime 2}G^2}\right],
\end{eqnarray}
\begin{eqnarray}
  M_{V^\pm}&=&\frac{\sqrt{\alpha}vg''}{2}\left(1+\frac{g^2}{2g^{\prime\prime 2}}\right),
  \label{eq:MassVc}\\
  M_{V^0}&=&\frac{\sqrt{\alpha}vg''}{2}\left(1+\frac{G^2}{2g^{\prime\prime 2}}\right).
  \label{eq:MassVn}
\end{eqnarray}
Of course, the mass of the photon $A$ is zero.

\subsection{Fermion Lagrangian}
\label{subsec:Fermions}

In our approach, we modify the interactions of the new vector
triplet with fermions. No new fermions beyond the SM have been
introduced in the model. The modification singles out the new
physics role of the third quark generation, and of the top quark
in particular. Hence, we call the obtained effective Lagrangian
the \textit{top-BESS model}, or \textit{tBESS} in short. It can be
split in two parts
\begin{equation}
  \cL_{ferm} = \cL_{ferm}^{SM} + \cL_{(t,b)}^{tBESS},
\end{equation}
where $\cL_{ferm}^{SM}$ is the SM part of the fermion Lagrangian
and $\cL_{(t,b)}^{tBESS}$ contains the modification concerning the
third quark generation.

The fermions are grouped into six $SU(2)_L$ doublets and six
$SU(2)_R$ doublets $\psi_h^a, a=1,\ldots,6$ where $h=L,R$ denotes
the chirality of the fields. Under $\Gamma^{glob}\times
SU(2)_{HLS}^{loc}$
\begin{equation}
  \psi_{L,R}^a\;\;\rightarrow\;\; g_{L,R}\;\psi_{L,R}^a,\;\;\; g_{L,R}\in SU(2)_{L,R}.
\end{equation}
The leptonic and light quark doublets are indexed by the
$a=1,\ldots,5$ range, $a=6$ is reserved for the top-bottom
doublet. The useful construct for building the fermion Lagrangian
is the matrix
\begin{equation}
  \chi_h^a \equiv \chi(\vec{\vphi}_h,\psi_h^a) = \xi^\dagger(\vec{\vphi}_h)\cdot\psi_h^a.
\end{equation}
Under $\Gamma^{glob}\times SU(2)_{HLS}^{loc}$ it transforms as
\begin{equation}
  \chi_{L,R}^a\;\;\rightarrow\;\; h^\dagger(x)\cdot \chi_{L,R}^a,\;\;\; h(x)\in SU(2)_{HLS}^{loc}.
\end{equation}

The $[\Gamma\times U(1)_{B-L}]^{glob}\times SU(2)_{HLS}^{loc}$
invariant Higgsless effective Lagrangian describing the SM physics
of the fermion doublets reads
\begin{equation}
  \cL_{ferm}^{SM} = \sum_{a=1}^6 \left[ I_c^L(\psi_L^a)+I_c^R(\psi_R^a) - I_{mass}(\psi^a) \right],
\end{equation}
where
\begin{eqnarray}
  I_c^L(\psi_L^a) &=& i\bar{\psi}_L^a(\Bdslash+\BWslash+\BBslash) \psi_L^a,
  \\
  I_c^R(\psi_R^a) &=& i \bar{\psi}_R^a(\Bdslash+\BBslash) \psi_R^a,
\end{eqnarray}
and
\begin{equation}
  I_{mass}(\psi^a) = \bar{\psi}_L^a U M_f \psi_R^a + \mbox{h.c.},
\end{equation}
where $M_f$ is a $2\times 2$ diagonal matrix with the masses of
the upper and bottom fermion doublet components on its diagonal
and
$U=\xi(\vec{\pi})\cdot\xi(\vec{\pi})=\exp(2i\vec{\pi}\vec{\tau}/v)$.
Note that while $I_c^L$ and $I_c^R$ are invariants of
$[\Gamma\times U(1)_{B-L}]^{glob}\times SU(2)_{HLS}^{loc}$ as well
as of $[SU(2)_L\times U(1)_Y]^{loc}$, the $I_{mass}$ terms break
$SU(2)_R\rightarrow U(1)_{R3}$.

The additional $[\Gamma\times U(1)_{B-L}]^{glob}\times
SU(2)_{HLS}^{loc}$ and $[SU(2)_L\times U(1)_Y]^{loc}$ invariants
read ($h=L,R$)
\begin{equation}\label{IbInvariants}
  I_b^h(\psi_h) = i\bar{\chi}_h\left[\Bdslash+\BVslash+ig'\Bslash (B-L)/2\right]\chi_h,
\end{equation}
and
\begin{eqnarray}\label{IlambdaInvariants}
  I_\lambda^h(\psi_h) &=& i\bar{\chi}_h \OmBarslash^\perp \chi_h
  \nonumber\\
  &=& i \bar{\chi}_h\left[\;\Omslash^\perp+
      (\xi_L^\dagger\BWslash \xi_L-\xi_R^\dagger\BBslash^{R3} \xi_R)/2\right]\chi_h,\;\;\;\;\;\;
\end{eqnarray}
where $\BBslash^{R3}=ig'\Bslash\tau^3$, and $\;\Omslash^\perp =
(\xi_L^\dagger\dslash\xi_L-\xi_R^\dagger\dslash\xi_R)/2$. Note
that the $I_b^h$ terms contain the direct interactions of the
vector triplet with fermions as opposed to the $\lambda$
invariants where there is no such interaction. However, the
$\lambda$ terms do modify the couplings of the electroweak gauge
bosons with fermions. The $\lambda$ terms were not present in the
original BESS formulation \cite{BESS}. Even though their values do
not have a significant impact on the observed signals at the
triplet peaks, they do influence the low-energy limits for the
fermion parameters.

We use the invariants (\ref{IbInvariants}) and
(\ref{IlambdaInvariants}) to build the fermion sector of the tBESS
model as follows:
\begin{eqnarray}
  \cL_{(t,b)}^{tBESS} &=& \phantom{+}
  b_L\left[ I_b^L(\psi_L^6)-I_c^L(\psi_L^6) \right]
  \nonumber\\
  && +b_R\left[ I_b^R(P\psi_R^6)-I_c^R(P\psi_R^6) \right]
  \nonumber\\
  && +2\lambda_L I_\lambda^L(\psi_L^6) +2\lambda_R I_\lambda^R(P\psi_R^6).
  \label{eq:LagrFermTBESS}
\end{eqnarray}
The matrix $P=\diag(1,p)$, where $0\leq p\leq 1$, serves to
disentangle the direct interaction of the vector triplet with the
right top quark from the interaction with the right bottom quark.
While $p=1$ leaves the interactions equal, the $p=0$ turns off the
right bottom quark interaction completely and maximally breaks the
$SU(2)_R$ part of the Lagrangian symmetry down to $U(1)_{R3}$.

If $I(\psi)$ is an $SU(2)_R$ invariant then $I(P\psi)$ is a
$U(1)_{R3}$ invariant due to the fact that $U(1)_{R3}$
transformations are generated by the $\tau^3$ matrix which
commutes with the $P$ matrix. Hence, after inserting the $P$
matrix into $\cL_{(t,b)_R}^{tBESS}$, the global symmetry of the
overall theory gets lowered down to $SU(2)_L\times U(1)_{R3}\times
U(1)_{B-L}\times SU(2)_V$. The gauge symmetry $[SU(2)_L\times
U(1)_{Y}\times SU(2)_V]^{loc}$ is maintained, though. As will be
seen in the next section the lower the $p$ value is set the more
relaxed the low-energy limits on the allowed values of $b_R$ and
$\lambda_R$ are.

While the $\cL_{GB}+\cL_{ESB}$ part of the tBESS Lagrangian
(\ref{tBESSLag}) is parity invariant, this is not generally true
for $\cL_{ferm}$. Under the parity transformation,
$I_b^L\leftrightarrow I_b^R$ and $I_\lambda^L\leftrightarrow
-I_\lambda^R$. Therefore, the new physics interactions in the
fermion Lagrangian (\ref{eq:LagrFermTBESS}) break parity, unless
$p=1$, $b_L=b_R$, and $\lambda_L=-\lambda_R$.

For the sake of comparison between the original BESS model
\cite{BESS} and the top-BESS model the following remark should be
made: our parameterization of the Lagrangian
(\ref{eq:LagrFermTBESS}) differs from the BESS model
parameterization of the direct triplet-to-fermions interactions
usually used by the authors of \cite{BESS}. Should we follow the
approach of \cite{BESS} our parameterization would change in the
following way:
\begin{equation}
  b_h\rightarrow \frac{b_h}{1+b_h},\;\;\;
  \lambda_h\rightarrow \frac{1}{2}\frac{\lambda_h}{1+b_h}.
\end{equation}
The parameterization we have used in (\ref{eq:LagrFermTBESS}) is
linear in $b$ and avoids introducing the artificial singularity at
$b_h=-1$.

In addition, the intergeneration universality in the
triplet-to-fermion direct couplings forced the authors of the BESS
model to switch off the direct couplings to the right fermion
fields. In the leptonic sector of the BESS model the direct right
interaction is absent if there are no right-handed neutrinos. In
its hadronic sector the direct right interaction contributes to
$K_L-K_S$ mass difference which results in a strict upper bound on
the interaction \cite{BESS}. The tBESS model avoids these limits
by admitting the direct interactions with the third generation
quarks only.

In the gauge where the six scalar fields $\vec{\sigma}$ and
$\vec{\pi}$ are transformed away the fermion Lagrangian
$\cL_{ferm}^{SM}+\cL_{(t,b)}^{tBESS}$ takes on a more transparent
structure as far as the individual interaction vertices are
concerned. The top-bottom sector reads
\begin{widetext}
\begin{eqnarray}
  \cL_{(t,b)_R} &=&  \cL_{(t,b)_R}^{SM}
  + \bar{t}_R \left[ -\frac{1}{2}\lambda_R g \Wslash^3 + \frac{1}{2}(b_R+\lambda_R) g' \Bslash
                     - \frac{1}{2}b_R\frac{g''}{2}\Vslash^3 \right]t_R
  \nonumber\\
  && \phantom{\cL_{(t,b)_R}^{SM}}
  + \bar{b}_R \left[ \frac{1}{2}p^2 \lambda_R g \Wslash^3 - \frac{1}{2}p^2(b_R+\lambda_R) g' \Bslash
                        + \frac{1}{2}p^2 b_R\frac{g''}{2}\Vslash^3 \right]b_R
  \nonumber\\
  && \phantom{\cL_{(t,b)_R}^{SM}}
  - \left\{\bar{t}_R
    \left[ \frac{1}{\sqrt{2}}p\lambda_R g\Wslash^+ + \frac{1}{\sqrt{2}}pb_R\frac{g''}{2}\Vslash^+ \right]b_R
    + \mbox{h.c.}\right\},
    \label{eq:LagrFermTBR}
\end{eqnarray}
\begin{eqnarray}
  \cL_{(t,b)_L} &=& \cL_{(t,b)_L}^{SM}
                       + \bar{t}_L \left[ \frac{1}{2} (b_L-\lambda_L)g \Wslash^3
                       + \frac{1}{2}\lambda_L g'\Bslash - \frac{1}{2}b_L\frac{g''}{2}\Vslash^3 \right]t_L
  \nonumber\\
  && \phantom{\cL_{(t,b)_L}^{SM}}
     + \bar{b}_L \left[ -\frac{1}{2} (b_L-\lambda_L)g \Wslash^3
                        -\frac{1}{2}\lambda_L g'\Bslash + \frac{1}{2}b_L\frac{g''}{2}\Vslash^3 \right]b_L
  \nonumber\\
  && \phantom{\cL_{(t,b)_L}^{SM}}
   + \left\{\bar{t}_L \left[ \frac{1}{\sqrt{2}}
   (b_L-\lambda_L)g\Wslash^+ - \frac{1}{\sqrt{2}}b_L\frac{g''}{2}\Vslash^+ \right]b_L
        + \mbox{h.c.}\right\},
    \label{eq:LagrFermTBL}
\end{eqnarray}
\end{widetext}
where the gauge fields are considered in the flavor eigenstate
basis. The SM parts of (\ref{eq:LagrFermTBR}) and
(\ref{eq:LagrFermTBL}) read
\begin{eqnarray}
  \cL_{(t,b)_R}^{SM} &=& \phantom{+}i(\bar{t}_R\dslash t_R) + i(\bar{b}_R\dslash b_R)
  \nonumber\\
   && -\frac{2}{3}g'(\bar{t}_R\Bslash t_R) + \frac{1}{3}g'(\bar{b}_R\Bslash b_R),
    \label{eq:LagrFermTBRSM}
\end{eqnarray}
\begin{eqnarray}
  \cL_{(t,b)_L}^{SM} &=& \phantom{+} i(\bar{t}_L\dslash t_L) + i(\bar{b}_L\dslash b_L)
  \nonumber\\
  &&-\frac{1}{6}g'(\bar{t}_L\Bslash t_L) - \frac{1}{6}g'(\bar{b}_L\Bslash b_L)
  \nonumber\\
  && -\frac{1}{2}g(\bar{t}_L\Wslash^3 t_L) +\frac{1}{2}g(\bar{b}_L\Wslash^3 b_L)
  \nonumber\\
  &&     -\frac{1}{\sqrt{2}}g (\bar{t}_L\Wslash^+ b_L + \mbox{h.c.}).
    \label{eq:LagrFermTBLSM}
\end{eqnarray}
Of course, to complete physics of the top and bottom quarks we
have to add the mass terms $m_t(\bar{t}_R t_L + \mbox{h.c.})$ and
$m_b(\bar{b}_R b_L + \mbox{h.c.})$.

To obtain the full fermion Lagrangian terms for the remaining
fermions must be added. The light quark terms
$\cL_{(u,d)_{L,R}}^{SM}$ and $\cL_{(c,s)_{L,R}}^{SM}$ can be
obtained from (\ref{eq:LagrFermTBRSM}) and
(\ref{eq:LagrFermTBLSM}) by simply replacing the $(t,b)$ fields
with $(u,d)$ or $(c,s)$. The lepton terms
$\cL_{(\nu_\ell,\ell^-)_{L,R}}^{SM}$, where $\ell=e,\mu,\tau$,
read
\begin{equation}
  \cL_{(\nu_\ell,\ell^-)_{R}}^{SM} = i\bar{\nu}_R\dslash \nu_R + i\bar{\ell}_R\dslash \ell_R
                             + g'(\bar{\ell}_R\Bslash \ell_R),
\end{equation}
\begin{eqnarray}
  \cL_{(\nu_\ell,\ell^-)_{L}}^{SM} &=&
  \phantom{+} i\bar{\nu}_L\dslash \nu_L + i\bar{\ell}_L\dslash \ell_L
  \nonumber\\
  && +\frac{1}{2}g'(\bar{\nu}_L\Bslash \nu_L) + \frac{1}{2}g'(\bar{\ell}_L\Bslash \ell_L)
  \nonumber\\
  && -\frac{1}{2}g(\bar{\nu}_L\Wslash^3 \nu_L) +\frac{1}{2}g(\bar{\ell}_L\Wslash^3 \ell_L)
  \nonumber\\
  &&     -\frac{1}{\sqrt{2}}g (\bar{\nu}_L\Wslash^+ \ell_L + \mbox{h.c.}).
\end{eqnarray}
The mass terms for the light quarks and leptons possess the same
form as those of the top and bottom quarks.

When the fermion interaction Lagrangians are expressed in terms of
the mass eigenstates of the electroweak gauge bosons the electric
charge $e$ can be defined in the vertex of photon with charged
fermions. It implies the relation of the electric charge to the
gauge couplings $g$, $g'$, and $g''$
\begin{equation}
  \left(\frac{1}{g}\right)^2+\left(\frac{1}{g'}\right)^2+\left(\frac{1}{g''/2}\right)^2=\left(\frac{1}{e}\right)^2.
\end{equation}

Obviously, once the gauge-boson fields are expressed in the mass
eigenstate basis the mixing generated interactions of the vector
triplet with fermions will appear on the scene. Typically, these
indirect interactions will be suppressed by the mixing matrix
factors. Despite of the suppression, the LHC and ILC processes
based on the indirect couplings might provide sizable signals of
the tBESS physics as will be discussed in
Subsection~\ref{subsec:ScatteringProcesses}.

\section{Phenomenology}
\label{sec:Pheno}

\subsection{Properties of the \bm{$SU(2)_{HLS}$} vector triplet}
\label{subsec:Properties}

The masses of the $SU(2)_{HLS}$ vector triplet depend on the three
gauge couplings $g, g', g''$, the free parameter $\alpha$, and the
ESB scale $v$. Of these, $g''$ and $\alpha$ parameterize new
physics beyond the SM. In the limit when $g$ and $g'$ are
negligible compared to $g''$ the masses of the neutral and charged
resonances are degenerate, $M_{V} = \sqrt{\alpha}g''v/2$. If
higher order corrections in $g/g''$ are admitted the mass
splitting occurs such that $M_{V^0}>M_{V^\pm}$. However, the
relative difference is less than one \textit{per mil} if $g''\geq
8$.

The values of $M_V$ below $1$~TeV seem to be disfavored by the CDF
and D0 experiments which have not found a significant excess over
the SM expectations in the measured $M_{t\bar{t}}$ spectrum in
this mass range \cite{TopKeyNotes}.

While the masses of the vector triplet are identical in the both,
tBESS and BESS, models, the total decay widths of the resonances
are different. For corresponding values of the $b$ parameters of
the two models, the tBESS model total widths are smaller than the
BESS model ones. It is caused by the differences in the
triplet-to-fermion couplings. Recall that while in the BESS model
the vector triplet couples directly to all fermions, in the tBESS
model it couples directly to the third quark generation only.

The partial decay widths of the vector resonances to the
electroweak bosons, $V^0\rightarrow W^+W^-$, $V^\pm\rightarrow
W^\pm Z$, in the tBESS model are the same as in the BESS model.
They read
\begin{eqnarray}
  \Gamma_{V^0\rightarrow W^+W^-} &=& \frac{g_{VWW}^2}{192\pi}
  \frac{M_{V^0}}{x_W^4} \left( 1-4x_W^2 \right)^{3/2}
  \nonumber \\
  & & \times  \left( 1+20x_W^2+12x_W^4 \right),
  \\
  \Gamma_{V^\pm\rightarrow W^\pm Z} &=& \frac{g_{VWZ}^2}{192\pi}
  \frac{M_{V^\pm}}{y_W^2 y_Z^2} \left[ 1-(y_W+y_Z)^2 \right]^{3/2}
  \nonumber \\
  & & \times \left[ 1-(y_W-y_Z)^2 \right]^{3/2}
  \nonumber \\
  & & \times \left\{ \left[ 1-(y_W+y_Z)^2 \right] \left[ 1-(y_W-y_Z)^2 \right] \right.
  \nonumber \\
  & & \left. \left. \ \ \ \ \ +12(y_W^2+y_Z^2+y_W^2 y_Z^2) \right]\right\},
\end{eqnarray}
where $x_{W,Z}=M_{W,Z}/M_{V^0}$, $y_{W,Z}=M_{W,Z}/M_{V^\pm}$. The
couplings $g_{VWW}$ and $g_{VWZ}$ are shown in
Table~\ref{tab:couplings}.
\begin{table}
\caption{Some couplings of the top-BESS vector triplet.}
\label{tab:couplings}
\begin{ruledtabular}
\begin{tabular}{cccc}
 $g_{VWW}$ & \multicolumn{3}{c}{$-\left(g {O_{11}^C}^2 O_{13}^N +
                                \frac{g''}{2} {O_{21}^C}^2 O_{33}^N \right)$} \\
 $g_{VWZ}$ & \multicolumn{3}{c}{$-\left(g O_{11}^C O_{12}^C O_{12}^N +
                            \frac{g''}{2} O_{21}^C O_{22}^C O_{32}^N\right)$} \\
 $g_{Vtt}^L$ & \multicolumn{3}{c}{$g_{Vuu}^L + \frac{1}{2} \left[ (b_L-\lambda_L) g O_{13}^N + \lambda_L g' O_{23}^N
                        - \frac{b_L}{2} g'' O_{33}^N \right] $} \\
 $g_{Vtt}^R$ & \multicolumn{3}{c}{$g_{Vuu}^R - \frac{1}{2} \left[ \lambda_R g O_{13}^N - (b_R+\lambda_R) g' O_{23}^N
                        + \frac{b_R}{2} g'' O_{33}^N \right] $} \\
 $g_{Vbb}^L$ & \multicolumn{3}{c}{$g_{Vdd}^L - \frac{1}{2} \left[ (b_L-\lambda_L) g O_{13}^N + \lambda_L g' O_{23}^N
                        - \frac{b_L}{2} g'' O_{33}^N \right] $} \\
 $g_{Vbb}^R$ & \multicolumn{3}{c}{$g_{Vdd}^R + \frac{p^2}{2} \left[ \lambda_R g O_{13}^N - (b_R+\lambda_R) g' O_{23}^N
                        + \frac{b_R}{2} g'' O_{33}^N \right] $} \\
 $g_{Vtb}^L$ & \multicolumn{3}{c}{$g_{Vud}^L + \frac{1}{\sqrt{2}} \left[ (b_L-\lambda_L) g O_{12}^C
                                    - \frac{b_L}{2} g'' O_{22}^C \right]$} \\
 $g_{Vtb}^R$ & \multicolumn{3}{c}{$-\frac{p}{\sqrt{2}} \left[ \lambda_R g O_{12}^C +
                                        \frac{b_R}{2} g'' O_{22}^C \right]$} \\[0.3cm]
\hline
  & $h=L$  & $h=R$ & notes \\
\hline
 $g_{V\nu\nu}^h$ & $-\frac{1}{2} \left(g O_{13}^N - g' O_{23}^N \right)$ & 0 & $\nu=\nu_e,\nu_\mu,\nu_\tau$\\
 $g_{V\ell\ell}^h$ & $ \frac{1}{2} \left(g O_{13}^N + g' O_{23}^N \right)$ & $g' O_{23}^N$ & $\ell=e,\mu,\tau$ \\
 $g_{Vuu}^h$ & $-\frac{1}{2} \left(g O_{13}^N + \frac{1}{3} g' O_{23}^N \right)$ & $-\frac{2}{3} g' O_{23}^N$ & $g_{Vuu}^h=g_{Vcc}^h$\\
 $g_{Vdd}^h$ & $ \frac{1}{2} \left(g O_{13}^N - \frac{1}{3} g' O_{23}^N \right)$ & $ \frac{1}{3} g' O_{23}^N$ & $g_{Vdd}^h=g_{Vss}^h$\\
 $g_{V\nu\ell}^h$ & $-\frac{1}{\sqrt{2}} g O_{12}^C$ & $0$ & $\ell=e,\mu,\tau$ \\
 $g_{Vud}^h$ & $-\frac{1}{\sqrt{2}} g O_{12}^C$ & 0 & $g_{Vud}^h=g_{Vcs}^h$
\end{tabular}
\end{ruledtabular}
\end{table}

The partial decay widths of the vector resonances to the third
quark generation read\footnote{To simplify the analysis the 
Cabibbo-Kobayashi-Maskawa
mixing is ignored throughout the paper.}
\begin{eqnarray}
  \Gamma_{V^0\rightarrow t\bar{t}} &=& \frac{M_{V^0}}{8\pi}\sqrt{1-4x_t^2}
   \left\{ \left[ (g_{V t t}^L)^2+(g_{V t t}^R)^2 \right] (1-x_t^2) \right.
   \nonumber\\
  & & \phantom{\frac{M_{V^0}}{8\pi}\sqrt{1-4x_t^2}\{ }
   \left.  + 6 g_{V t t}^L g_{V t t}^R x_t^2 \right\},
  \\
  \Gamma_{V^0\rightarrow b \bar{b}} &=& \frac{M_{V^0}}{8\pi}
                                    \left[ (g_{Vbb}^L)^2+(g_{Vbb}^R)^2 \right],
  \\
  \Gamma_{V^+\rightarrow t\bar{b}} &=& \frac{M_{V^\pm}}{8\pi}
  \left[ (g_{V t b}^L)^2+(g_{V t b}^R)^2 \right] (1-y_t^2)
  \nonumber \\
  & & \phantom{M_{V^+}\;[ } \times  \left( 2-y_t^2-y_t^4 \right)/2,
\end{eqnarray}
where $x_t=m_t/M_{V^0}$ and $y_t=m_t/M_{V^\pm}$. The mass of the
bottom quark has been neglected.

The mixing of the gauge bosons generates indirect couplings of the
tBESS vector triplet to all fermions. Thus the vector resonances
can also decay to the light fermions, other than top and bottom
quarks. Of course, the indirect couplings are suppressed by the
relevant mixing factors supplied by the mixing matrices
(\ref{eq:ONapprox}) and (\ref{eq:OCapprox}). The light fermion
decay widths can be calculated using the generic massless fermion
formulas
\begin{eqnarray}
  \Gamma_{V^0\rightarrow f\bar{f}} &=& N_C\frac{M_{V^0}}{24\pi}
                                    \left[ (g_{Vf\bar{f}}^L)^2+(g_{Vf\bar{f}}^R)^2 \right],
  \\
  \Gamma_{V^\pm\rightarrow f\bar{f}'} &=& N_C\frac{M_{V^\pm}}{24\pi}(g_{Vf\bar{f}'}^L)^2,
\end{eqnarray}
where $N_C$ is the number of colors the final state is summed
over. The chiral couplings are summarized in
Table~\ref{tab:couplings}.

\begin{figure}
\includegraphics[scale=0.4]{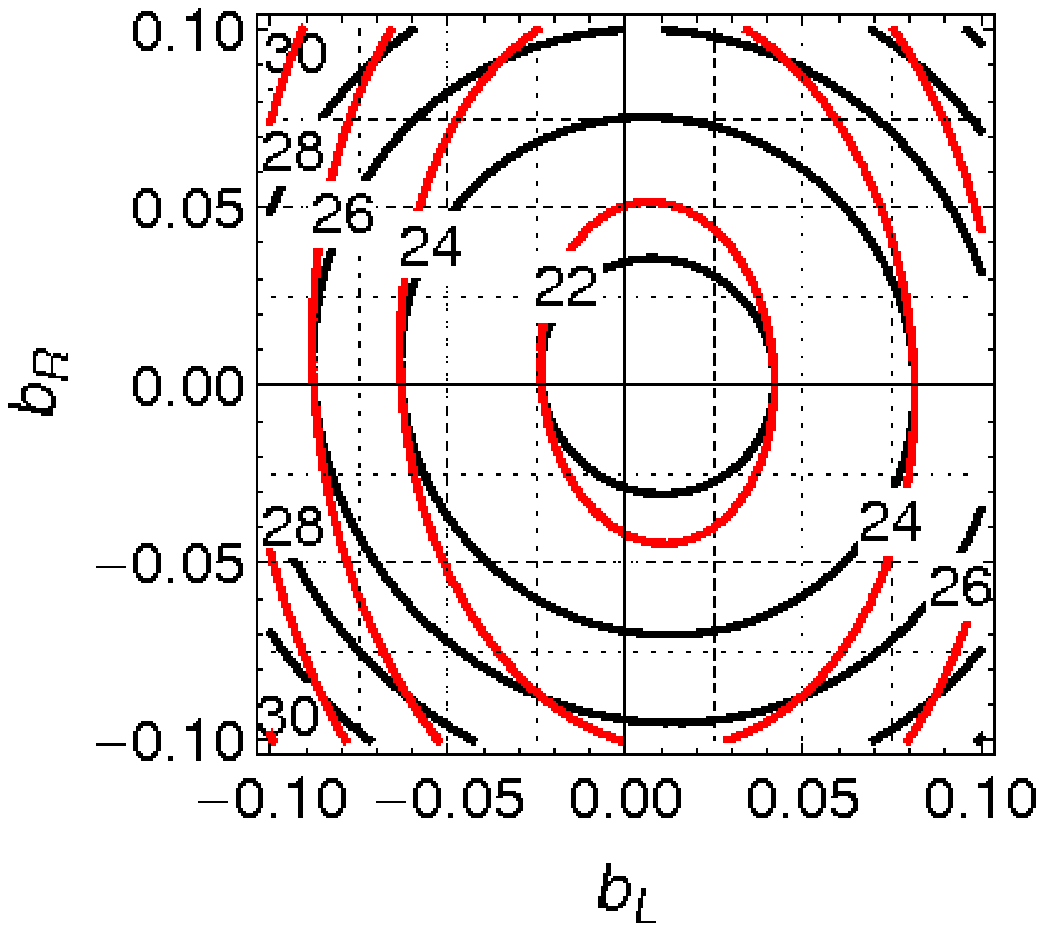}
\includegraphics[scale=0.4]{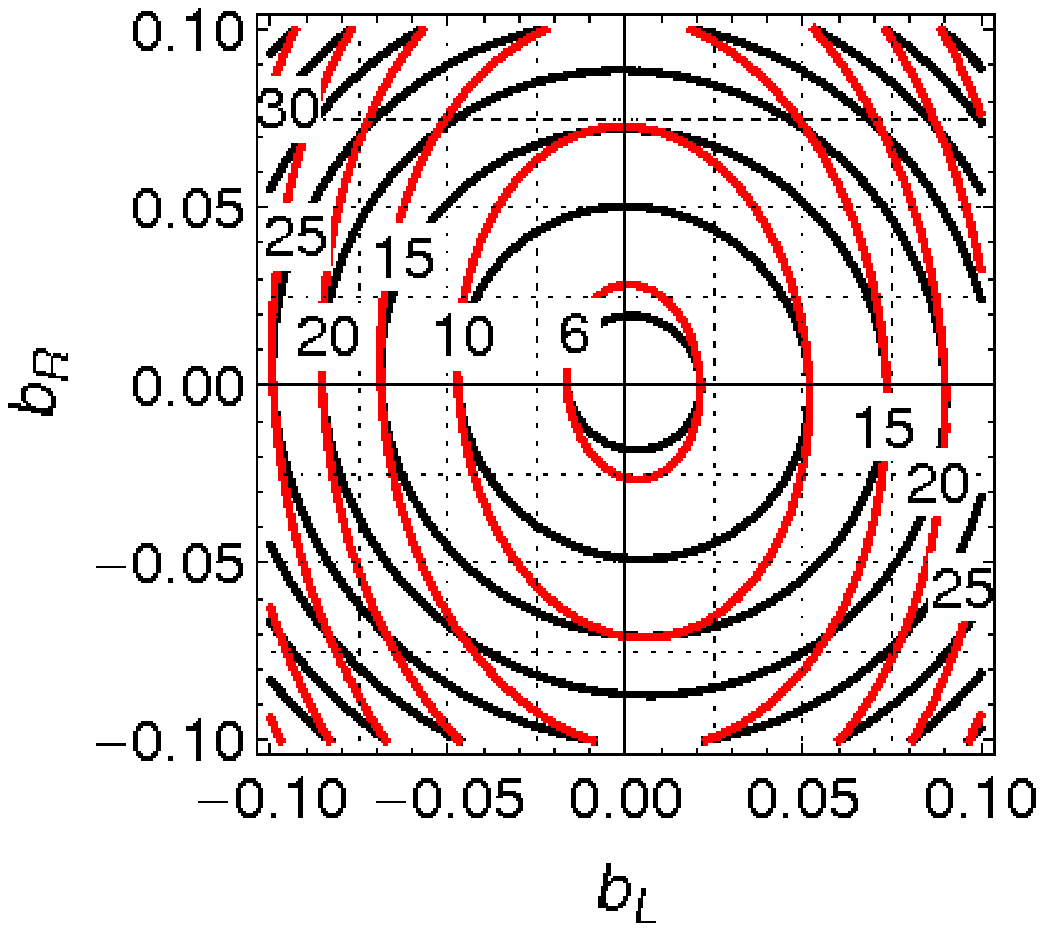}\\[0.3cm]
\includegraphics[scale=0.40]{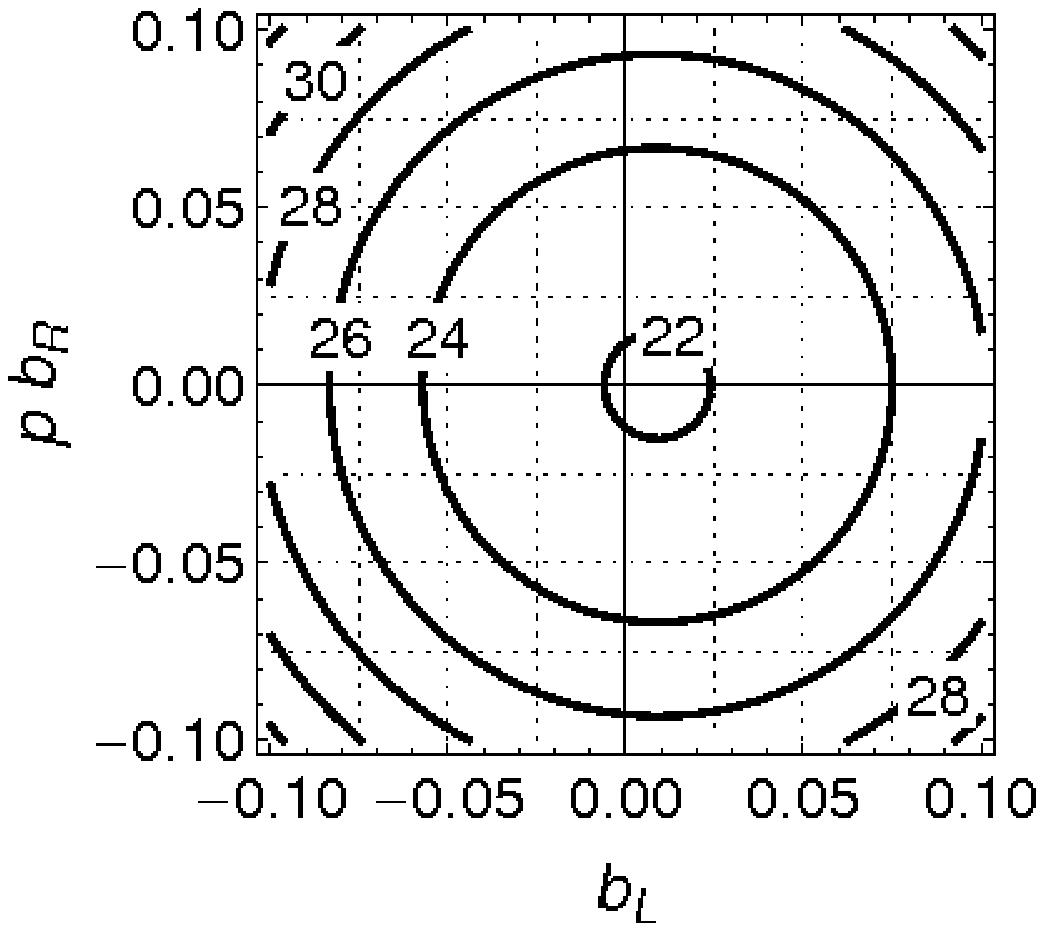}
\includegraphics[scale=0.40]{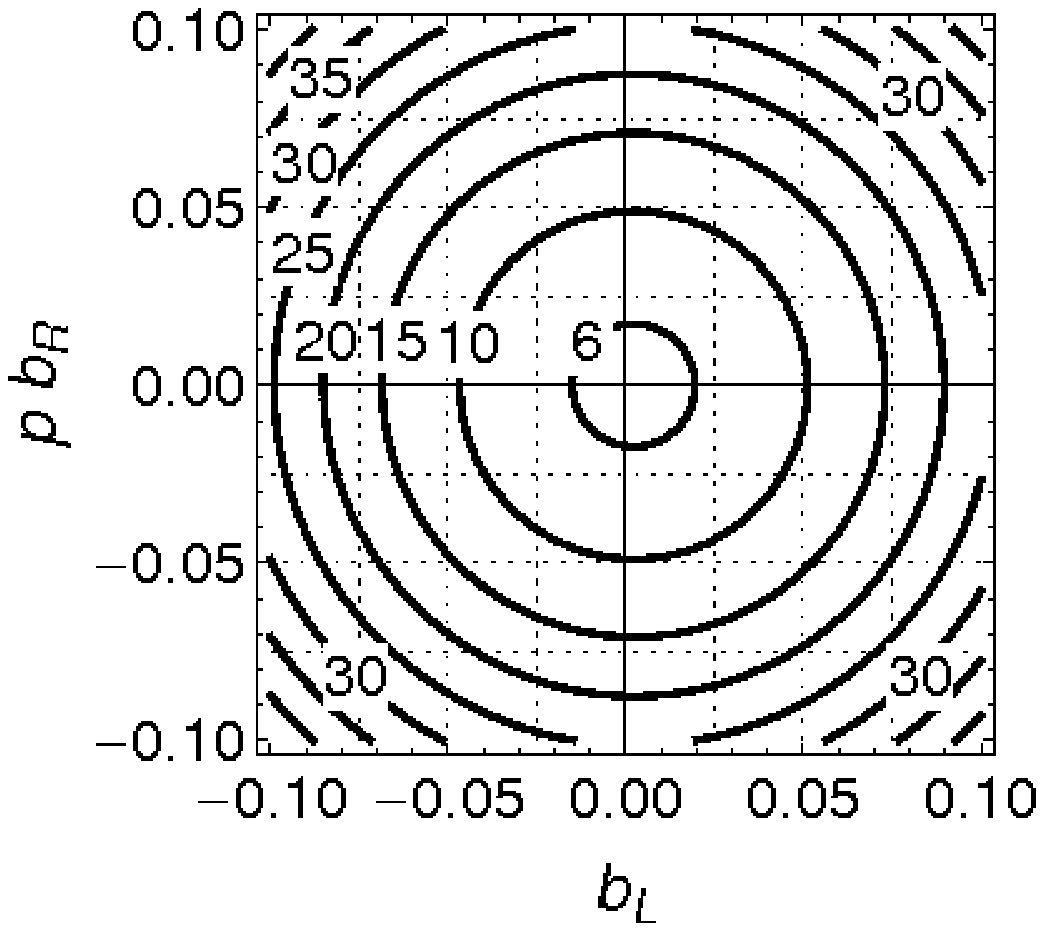}
\caption{\label{fig:DWcontours}(color online)
 The total decay width contours of the tBESS vector triplet.
 The upper row displays the $V^0$ decay widths (numeric labels in GeV)
 in the $(b_L,b_R)$ plane for the cases of $p=1$ (black
 circlelike contours) and $p=0$ (red elliptic contours).
 The bottom row displays the $V^\pm$ decay widths (in GeV)
 in the $(b_L,p b_R)$ plane.
 The graphs in the left and right columns correspond to
 $g''=10$ and $g''=20$, respectively. All graphs have been
 plotted for
 $M_{V^0}=1$~TeV and $\lambda_L=\lambda_R=0$.}
\end{figure}

The total decay widths of the tBESS resonances obtained by summing
up over all decay channels are shown in Fig.~\ref{fig:DWcontours}.
The $\lambda$ parameters were set to zero. As can be seen in
Table~\ref{tab:couplings}, the dominant coupling terms depend
solely on $b$ parameters, while the contributions of the
$\lambda$-dependent terms are always suppressed by the
nondiagonal elements of the mixing matrices $O^N$ and $O^C$.
Hence, the effect of nonzero $\lambda$'s on the decay widths is
negligible if the parameters assume the values dictated by the
low-energy limits (for the limits, see
Subsection~\ref{subsec:Limits}).

Note that the contours of the constant decay widths in the
$b_L$-$b_R$ space form ellipses with the eccentricities depending
on the value of the $p$ parameter. When $p=1$ the ellipses
approach a circular shape. In the case of the charged resonance
the total decay width is not a function of $b_R$ and $p$
separately. It rather depends on the product of the two
parameters. Finally, the ellipses/circles do not have their
centers, which indicate the points of the minimal widths, at the
origin of the parametric space. The centers are shifted from the
origin but the shift decreases with growing $g''$. The centers
reach the origin when $g''\rightarrow\infty$.

\begin{figure}
\includegraphics[scale=0.45]{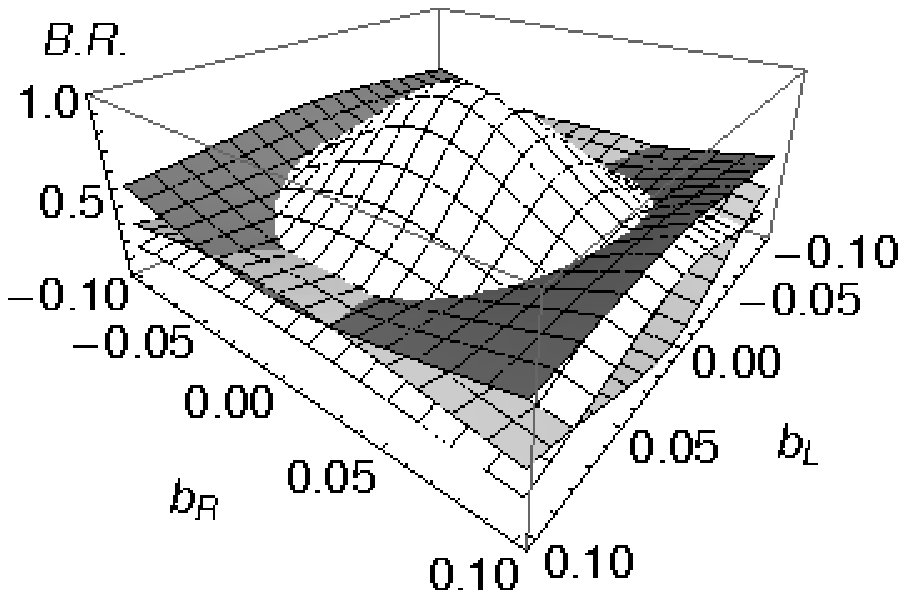}
\includegraphics[scale=0.45]{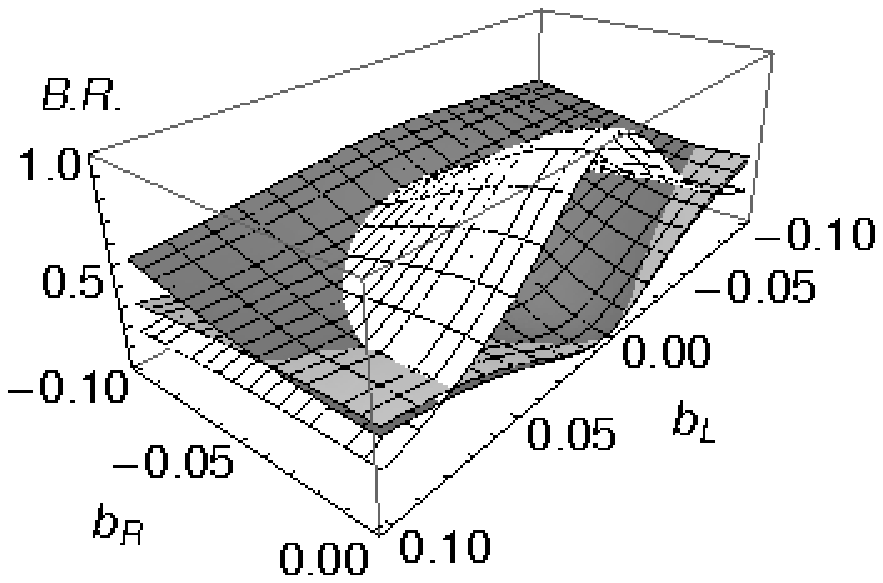}\\
\includegraphics[scale=0.45]{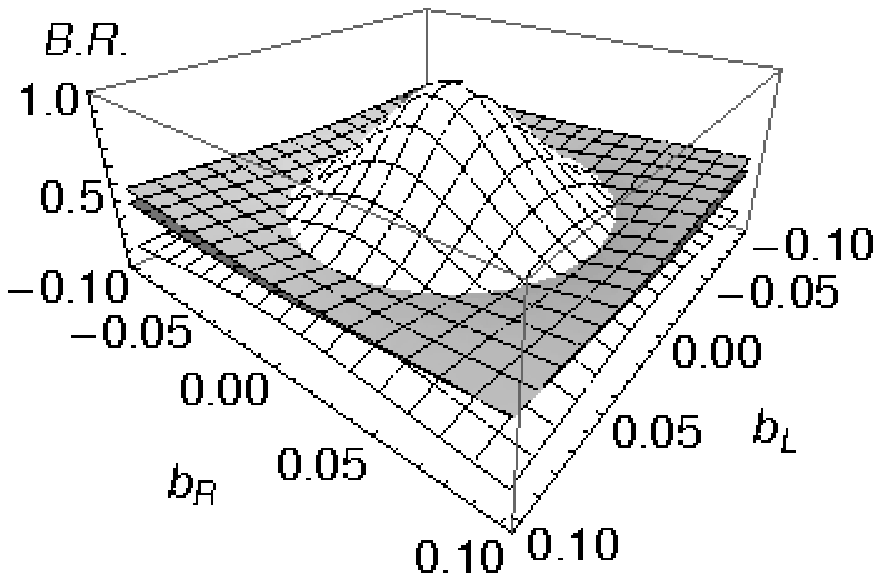}
\includegraphics[scale=0.45]{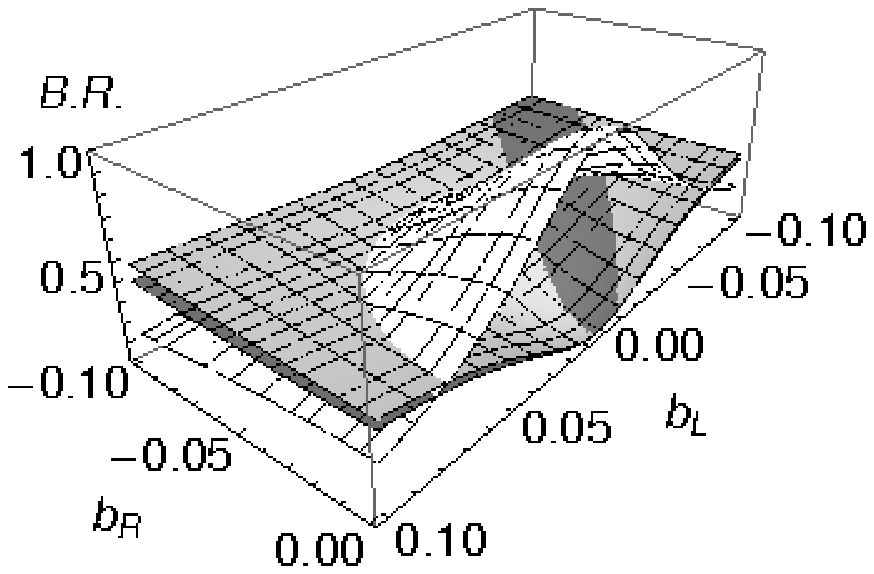}\\
\caption{\label{fig:BRN}
 The branching ratio of  $V^0$ to
 the $W^+W^-$ (white), $t\bar{t}$ (dark gray),
 and $b\bar{b}$ (light gray) decay channels
 in the $(b_L,b_R)$ plane for the cases
 of $p=0$ (upper row) and $p=1$ (bottom row).
 In all graphs $g''=20$, $M_{V^0}=1$~TeV, and $\lambda_L=\lambda_R=0$.
 The $b_R=0$ dissections of the graphs
 are shown in the right column.}
\end{figure}

\begin{figure}
\includegraphics[scale=0.45]{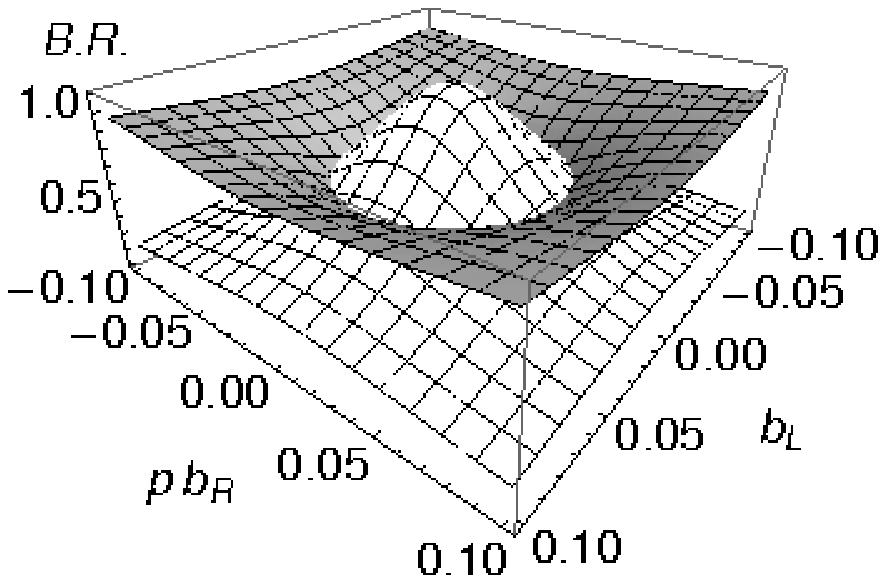}
\includegraphics[scale=0.45]{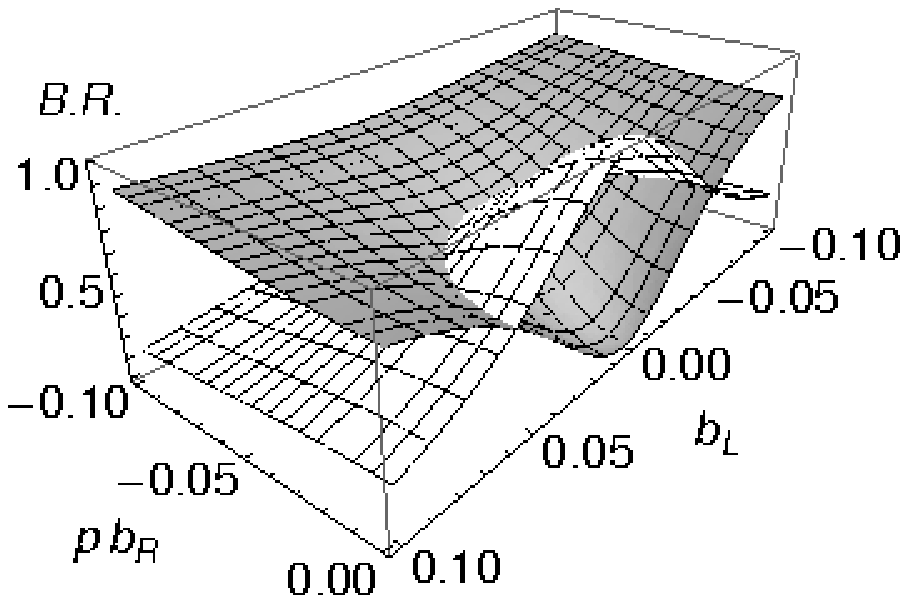}\\
\caption{\label{fig:BRC}
 The branching ratio of $V^\pm$ to
 the $WZ$ (white) and $tb$ (gray) decay channels
 in the $(b_L,p b_R)$ plane.
 Other parameters:
 $g''=20$, $M_{V^0}=1$~TeV, and $\lambda_L=\lambda_R=0$.
 The $pb_R=0$ dissection of the left graph
 is shown on the right-hand side.}
\end{figure}

Except for the special regions of the parametric space which will
be discussed in Sec.~\ref{subsec:DeathValley} the vector
triplet predominantly decays to the electroweak gauge bosons,
$W^\pm$ and $Z$, and/or to the third generation of quarks.
Figures~\ref{fig:BRN} and \ref{fig:BRC} depict the branching ratios
of the neutral and charged resonances, respectively, for the gauge
boson and top/bottom quark channels.
As expected, the quark decay channels prevail when the moduli of
$b$ parameters assume sufficiently large values. Details depend on
other parameters of the model, $g''$, $M_{V^0}$, and $p$. While
the decay widths to the third generation of quarks grow with $g''$
--- namely, they are proportional to $g''^{2}$ --- the decay
widths to $W^\pm$ and $Z$ are proportional to $1/g^{\prime\prime
2}$.

\subsection{Tree-level unitarity constraints}
\label{subsec:Unitarity}

The SM without the Higgs is not renormalizable and its amplitudes
violate unitarity at some energy. In particular, when the
longitudinal electroweak gauge-boson scattering is considered the
partial wave tree-level unitarity is violated at
$\sqrt{s}=1.7$~TeV \cite{DominiciNuovoCim}. The result has been
obtained using the Equivalence Theorem
\cite{EquivalenceTheoremRenorm,EquivalenceTheoremNonRenorm}
approximation of the $W_L^+W_L^-$, $Z_L Z_L$, $W_L^\pm Z_L$, and
$W_L^\pm W_L^\pm$ scattering by the pionic scattering amplitudes
of the $SU(2)_L\times SU(2)_R/SU(2)_{L+R}$ nonlinear sigma model.
The matrix of the $a_0$ partial waves of all the scattering
amplitudes was formed, where the zero index at $a_0$ indicates the
$J=0$ angular momentum. The $S$-matrix unitarity implies that the
maximum of the moduli of the $a_0$ matrix eigenvalues should be
less than 1 \cite{LeeQuiggThacker77}. This condition leads to the
energy restriction cited above.

To obtain the unitarity constraints for the tBESS model an
analogical procedure has been applied. The only difference is that
the $W_L^+W_L^-$, $Z_L Z_L$, $W_L^\pm Z_L$, and $W_L^\pm W_L^\pm$
scattering amplitudes can also proceed through the exchange of the
new resonances. It modifies the amplitude expressions so that they
read
\begin{eqnarray}
  \cM(W_L^+W_L^-\rightarrow W_L^+W_L^-) &=& A(s,t,u)+A(t,s,u),
  \nonumber\\
  \cM(Z_L Z_L\rightarrow Z_L Z_L) &=& 0,
  \nonumber\\
  \sqrt{2}\cM(W_L^+W_L^-\rightarrow Z_L Z_L) &=& A(s,t,u),
  \nonumber\\
  \cM(W_L^\pm Z_L\rightarrow W_L^\pm Z_L) &=& A(t,s,u),
  \nonumber\\
  \sqrt{2}\cM(W_L^\pm W_L^\pm\rightarrow W_L^\pm W_L^\pm) &=& A(t,s,u)+A(u,t,s),
  \nonumber
\end{eqnarray}
\begin{equation}\label{eq:Astu}
  A(s,t,u) = \frac{s}{4v^2}(4-3\alpha)+\frac{\alpha M_V^2}{4v^2}[f(t,u)+f(u,t)],
 \end{equation}
where
\begin{equation}
  f(t,u) =  \frac{u-s}{t-M_V^2+iM_V\Gamma_V}.
\end{equation}

\begin{figure}
\includegraphics[scale=0.75]{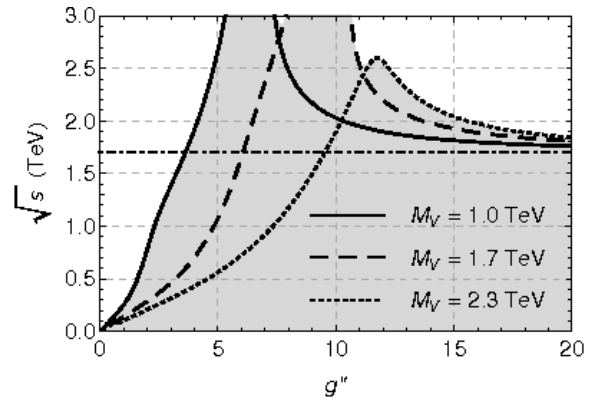}
\caption{\label{fig:Ulimit} The tree-level unitarity constraints
from the gauge-boson scattering obtained for various masses of the
vector triplet: $M_V=1$~TeV (solid line), $1.7$~TeV (dashed),
$2.3$~TeV (dotted). The horizontal dashed-dotted line is the
Higgsless SM unitarity limit of 1.7~TeV. The shaded area indicates
the region where the unitarity holds. No couplings to fermions are
assumed.}
\end{figure}

The eigenvalues of the $a_0$ matrix based on the electroweak gauge-boson 
scattering amplitudes are functions of $\sqrt{s}$, $g''$,
and $M_V$, when the $\alpha$ parameter has been replaced by $M_V$
using the leading order of the mass relation (\ref{eq:MassVn}),
$M_V=\sqrt{\alpha}vg''/2$. No couplings to fermions are assumed,
therefore
\begin{equation}\label{eq:DecayWidthGBonly}
   \Gamma_V=\frac{M_V g''^2}{768\pi}\alpha^2
           =\frac{M_V^5}{48\pi v^4}\frac{1}{g''^2}.
\end{equation}
Then, constraining the maximal eigenvalue modulus by unity results
in the unitarity constraints depicted in Fig.~\ref{fig:Ulimit}.
If we require that the tBESS model amplitudes unitarity holds up
to the same energy as for the Higgsless SM --- $1.7$~TeV --- the
$g''$ parameter is restricted only from below. This bottom limit
depends on $M_V$: $g''\stackrel{>}{\sim}3,6$, and $9$, when
$M_V=1.0, 1.7$, and $2.3$~TeV, respectively. The tBESS model
amplitudes can satisfy the unitarity also at higher energies, if
$g''$ is properly restricted from above. One has to remember that
nonrenormalizability of the model implies the upper limit on the
applicability of the Equivalence Theorem
\cite{EquivalenceTheoremNonRenorm}. It holds for $E\leq 4\pi
v\approx 3$~TeV. This sets the upper energy limit on any
conclusions inferred from the use of the theorem.

\begin{figure}
\includegraphics[scale=0.73]{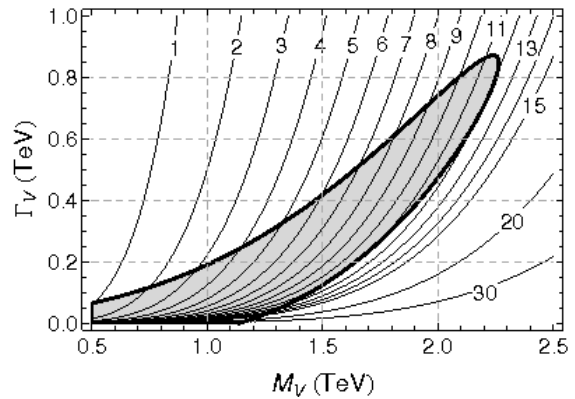}
\caption{\label{fig:MVGammaVunitarity} The allowed values (the
shaded region) of the width and mass of the vector resonance
assuming the unitarity saturation at or above $E=1.5M_V$. The
lines crossing the plane indicate the points of the fixed values
of $g''$. No couplings to fermions are assumed.}
\end{figure}

It seems reasonable to demand that the unitarity constraint
exceeds the mass of the $SU(2)_{HLC}$ resonance. If we require
that the unitarity of the model holds up to the energy of $E=1.5
M_V$ we obtain the unitarity allowed region in the
$\Gamma_V$-$M_V$ plane shown in Fig.~\ref{fig:MVGammaVunitarity}.
The lines of constant $g''$ values are superimposed over the
unitarity allowed region. The graph suggests that the $M_V$ values
which can be accommodated by the tBESS effective model cannot
exceed $2.26$~TeV. If we wish to avoid wide resonances, we should
stay at somewhat lower masses, say, up to about $1.5$~TeV. Recall
that no decays to fermions have been involved when obtaining these
conclusions.

The expression (\ref{eq:Astu}) is identical with that of the BESS
model except for the decay width $\Gamma_V$ of the vector
resonance. To reflect the impact of the fermion sector on the
tBESS vector boson decay widths the third quark generation decay
channels assuming no gauge-boson mixing and the massless quarks
will be added. In this approximation, the neutral tBESS resonance
decays to $W_L^+W_L^- + b\bar{b} + t\bar{t}$ and the charged one
to $W_L^\pm Z_L + t\bar{b}/\bar{t}b$. Thus, the total decay width
(\ref{eq:DecayWidthGBonly}) will be modified as follows
\begin{equation}\label{eq:DecayWidth}
   \Gamma_V=\frac{M_V g''^2}{768\pi} \left[ \alpha^2 + 12\beta^2(b_L,b_R,p) \right],
\end{equation}
where $\beta=[b_L^2+b_R^2(1+p^4)/2]^{1/2}$ for the neutral
resonance and $\beta=[b_L^2+b_R^2 p^2]^{1/2}$ for the charged one.
In this approximation the decay width (\ref{eq:DecayWidth}) is not
a function of $\lambda$'s. As argued in
Sec.~\ref{subsec:Properties}, dropping the $\lambda$
dependence has negligible consequences.

The vector resonance decay width makes the unitarity constraint
sensitive to the parameters of the fermionic sector which are
neatly packed into the $\beta$ parameter. The unitarity
constraints based on the electroweak gauge-boson scattering
amplitudes should be supplemented by the unitarity constraints
derived from the scattering amplitudes with the participation of
the top and bottom quarks. We have not performed the analysis in
this paper. Thus, at this moment, we cannot tell whether and how
the inclusion of the quark scattering processes influences the
conclusions about the unitarity constraints. Nevertheless, the
question of unitarity of top/bottom quark scattering amplitudes in
similar situation to ours was treated in the
literature~\cite{Han1,Han2}. Their conclusions seem to suggest
that the fermion amplitudes do not place stricter constraints than
those based on the $ww\rightarrow ww$ scattering.

Considering the decay width (\ref{eq:DecayWidth}) we have obtained
the tBESS unitarity constraints which depend also on the $\beta$
parameter. There is an ambiguity which of the $\beta$ parameters
should be used in the calculations of the unitarity limits. Both,
neutral and charged, resonances contribute to the processes under
consideration. This problem is a side effect of merging the
contributions of two different Lagrangians, the nonlinear sigma
model and the gauged tBESS model, into one decay width and as such
it has no rigorous solution. It is the price to be paid for the
shortcut we took in order to estimate the tBESS model behavior.

\begin{figure*}[t]
\includegraphics[scale=0.5]{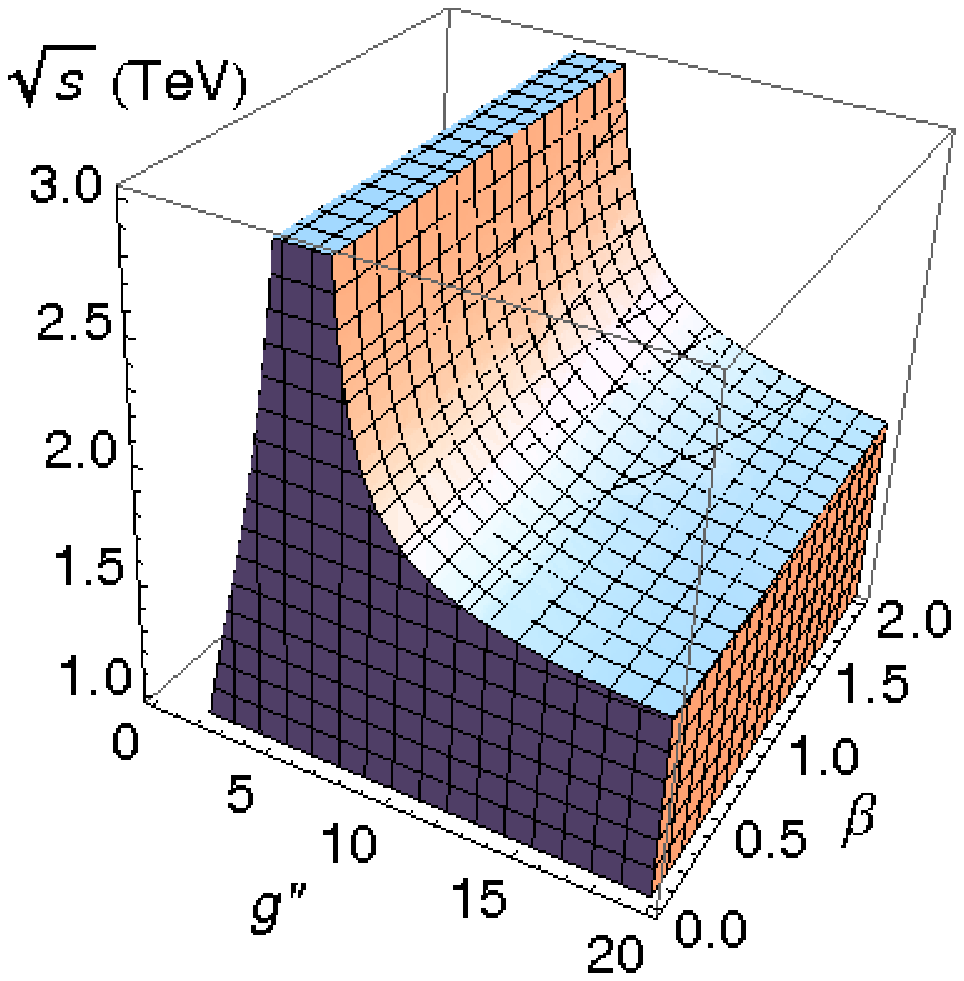}\hspace{0.5cm}
\includegraphics[scale=0.5]{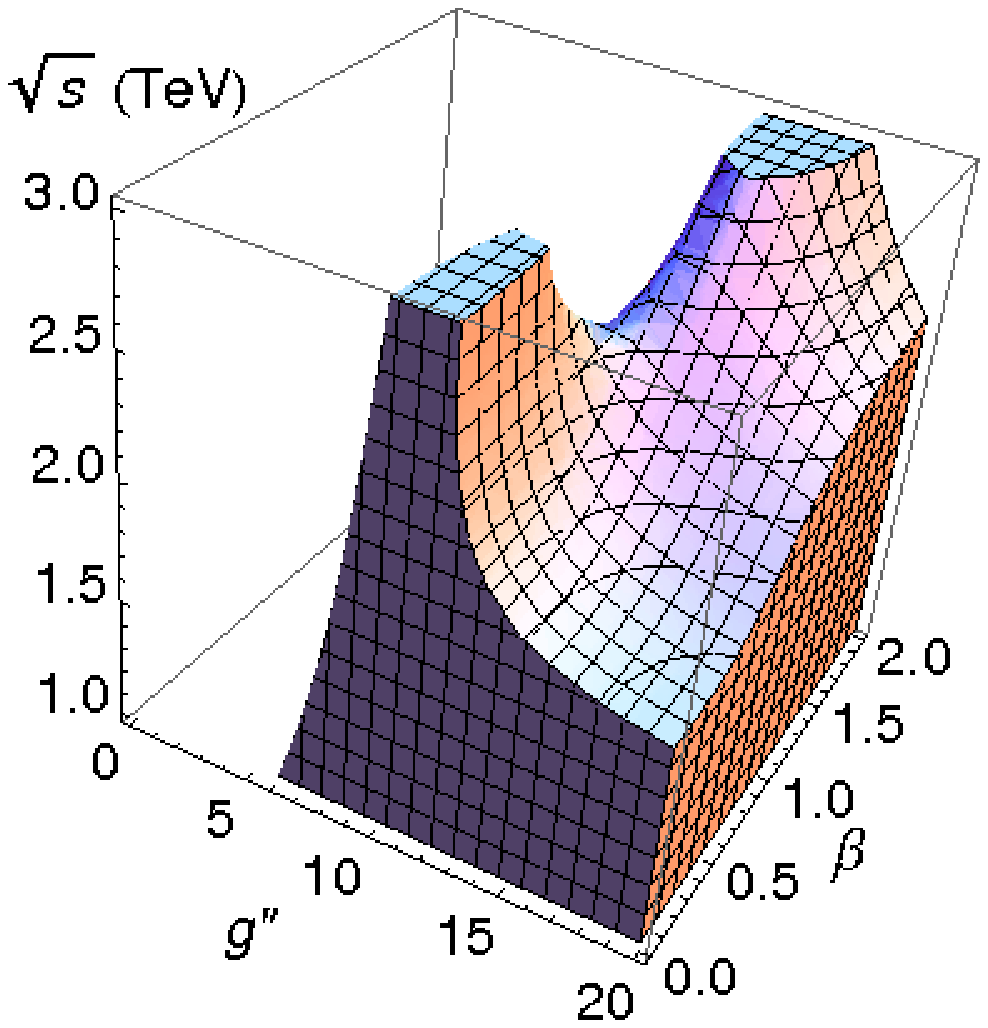}\hspace{0.5cm}
\includegraphics[scale=0.5]{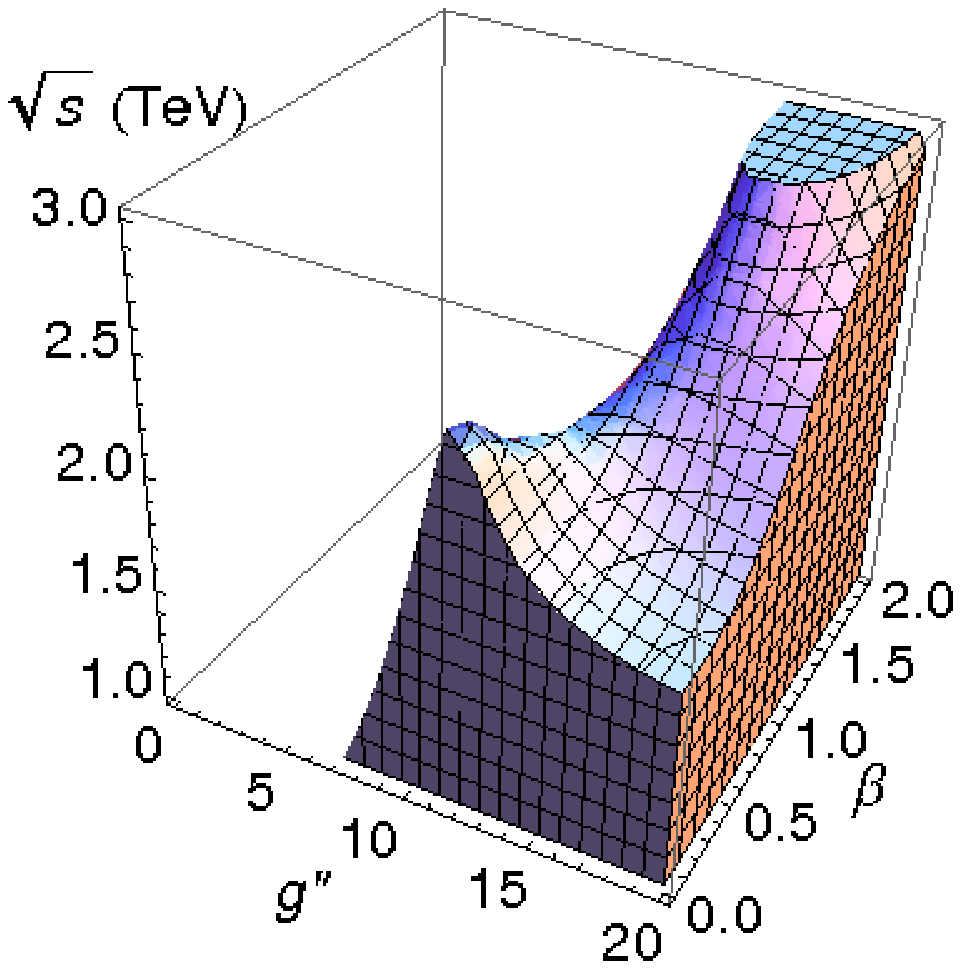}
\caption{\label{fig:Ulimit3D}(color online) The tree-level
unitarity constraints from the gauge-boson scattering obtained for
various masses of the tBESS vector triplet. The decays to the
third quark generation are included in evaluation of the vector
resonance width; $\beta=[b_L^2+b_R^2(1+p^4)/2]^{1/2}$. The graphs
correspond to $M_V=1$~TeV, $2$~TeV, and $2.3$~TeV (from left to
right).}
\end{figure*}

In our calculations, the neutral resonance $\beta$ parameter has
been used. The $(g'',\beta)$-dependent constraints for various
values of $M_V$ can be seen in Fig.~\ref{fig:Ulimit3D}.
It appears that for $M_V$ below $2$~TeV the $\beta$-dependence of
the unitarity limit is negligible certainly when $\beta$ is below
about $0.2$ and the limits of Fig.~\ref{fig:Ulimit} remain valid.
On the other hand, the higher the mass of the vector resonance,
the stronger the effect of $\beta$. However, the higher values of
$\beta$ are disfavored by the low-energy limits which will be
discussed in Sec.~\ref{subsec:Limits}.

\begin{figure}[t]
\includegraphics[scale=0.7]{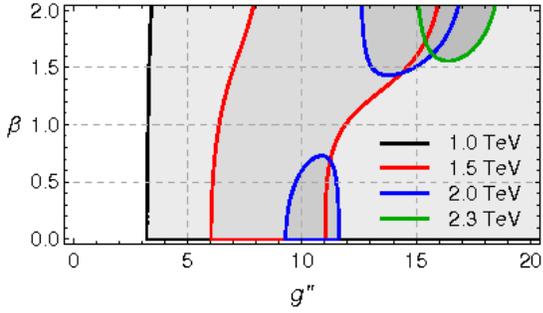}
\caption{\label{fig:BetaGppUnitarity}(color online) The allowed
values (the shaded regions) of $\beta$ and $g''$ parameters
assuming the unitarity saturation at or above $E=1.5M_V$ when the
vector resonance width includes the decays to the third quark
generation. The regions for different values of $M_V$ are
depicted: $M_V=1$~TeV (black contour), $M_V=1.5$~TeV (red),
$M_V=2$~TeV (blue), $M_V=2.3$~TeV (green). The gray color of the
allowed regions darkens as the value of $M_V$ grows.}
\end{figure}

When we ask that the tBESS model unitarity is not violated below
$1.5M_V$ we obtain the allowed regions in the $(\beta,g'')$ plain.
They are depicted in Fig.~\ref{fig:BetaGppUnitarity} for different
values of $M_V$.
For $M_V=1$~TeV there is the rectangular quarter-plane, not bound
from above, neither on the right-hand side, of the allowed values
of the $\beta$ and $g''$ parameters. When the mass grows the
allowed area shrinks and splits into discontinued regions. The
upper bound on $g''$ appears once the required unitarity
constraint crosses $1.7$~TeV which corresponds to the Higgsless SM
unitarity limit plotted in Fig.~\ref{fig:Ulimit}. This occurs at
$M_{V^0}=1.13$~TeV. Raising further the mass value results in the
splitting of the allowed area into two separate regions. Even
higher mass value causes the lower region to disappear. Of course,
the critical mass values as well as all the constraints displayed
in Fig.~\ref{fig:BetaGppUnitarity} are subject to the condition
that the unitarity saturation takes place at least $50\%$ above
the mass of the resonance. Changing the condition would alter the
presented results.

When there were no fermion interactions, the unitarity saturation
at $1.5 M_V$, or higher, restricted the maximal vector resonance
mass, for which the effective description works, to amount to
$M_V=2.26$~TeV. On the other hand, Fig.~\ref{fig:BetaGppUnitarity}
suggests that fermion interactions with $\beta\geq 1.6$ can bring
masses higher than $2.26$~TeV back into the game. Because of the
circumstances mentioned above this result should be supplemented
by the analysis which would include the top/bottom scattering
before reaching final conclusions on this matter.

\subsection{Low-energy limits}
\label{subsec:Limits}

The tBESS model is an effective description of a high-energy
extension of the Higgsless SM. The existing \textit{electroweak
precision data} (EWPD) restrict tBESS induced deviations from the
SM at the relevant energies. This experimental input results in
the \textit{low-energy limits} on the parameters of tBESS.

To obtain these limits we have to derive the low-energy Lagrangian
by integrating out the vector triplet of the tBESS Lagrangian. It
proceeds by taking the limit $M_{triplet}\rightarrow\infty$, while
$g''$ is finite and fixed, and by substituting the equation of
motion for the triplet fields obtained under these conditions.

The low-energy tBESS Lagrangian has been related to
several independent measurements. First of all,
to restrict $g''$ as well as the $b$ and $\lambda$ parameters,
we have used
the standard \textit{epsilon method} for the
EWPD~\cite{EpsilonMethod1,EpsilonMethod2}.
Another independent limit on $g''$ has resulted from
the D0 measurement of $p\bar{p}\rightarrow WZX$~\cite{D0experimentTGV}.
Independently, the $b$ and $\lambda$ parameters have been
restricted by the measurement of the $B\rightarrow X_s\gamma$
decay~\cite{btosgData}.

Let us briefly review the epsilon analysis.
There are four epsilon parameters
$\epsilon_1,\epsilon_2,\epsilon_3,\epsilon_b$ which summarize the
input of the SM weak physics at loop level and non-SM weak physics
at tree and loop levels. The epsilons are extracted from data
independently of $m_t$, $m_H$, and \textit{new physics
parameters}. Both $\epsilon_1$ and $\epsilon_3$ are obtained from
the measurements of $A_{FB}^\ell$ and
$\Gamma(Z\rightarrow\ell\ell)$. To obtain $\epsilon_2$ the
measurement of $M_W/M_Z$ has to be supplemented. To obtain
$\epsilon_b$ the measurement of $\Gamma(Z\rightarrow b\bar{b})$
has to be added. More details on deriving the low-energy limits
from the epsilon analysis can be found in
Appendix~\ref{app:LElimits}.

\begin{figure}
\includegraphics[scale=0.9]{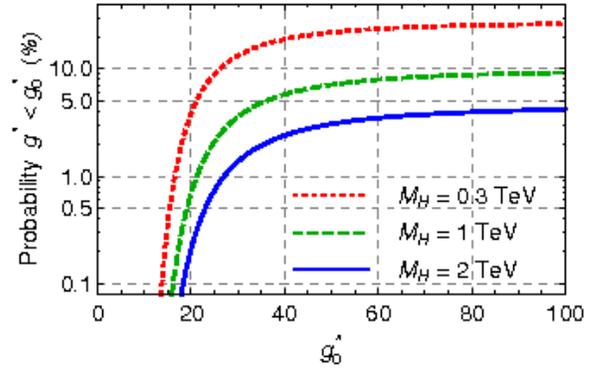}
\caption{\label{fig:gppLimit}(color online)
The probability that $g''$ lies anywhere below a given value of $g_0''$.
It is based on $\epsilon_3$ parameter and depends on $M_H$ used for
calculation of $\delta\epsilon_3^{SM}$. Plots for $M_H=0.3$~TeV (red dotted),
$1$~TeV (green dashed), and $2$~TeV (blue solid) are shown.}
\end{figure}

The EWPD limit on $g''$ can be obtained from the $\epsilon_3$
parameter, using the relation\footnote{For details, see Appendix~\ref{app:LElimits},
the Eq.~(\ref{eps3}), and the related text.}
\begin{equation}\label{epsilon3gpp}
 \epsilon_3=\left(\frac{g}{g''}\right)^2+\delta\epsilon_3^{SM},
\end{equation}
where $\epsilon_3= 0.005\,34 \pm 0.000\,94$
is obtained from experiment~\cite{EpsilonData}, and the value of
$\delta\epsilon_3^{SM}$ is the theoretical prediction which depends on
$M_H$. Namely,
$\delta\epsilon_3^{SM}=0.005\,89$, $0.006\,54$, and $0.006\,92$,
for $M_H=0.3$, $1$, and $2$~TeV, respectively.
Thus, the mean value of $\epsilon_3-\delta\epsilon_3^{SM}$
is negative and the Eq.~(\ref{epsilon3gpp}) has no solution for $g''$.
Nevertheless, the positive values of the difference are statistically
admissible if we assume its normal distribution with the standard
deviation taken from $\epsilon_3$. Then, the probability that
the difference is positive amounts to $28\%$, $10\%$, and $5\%$,
when $M_H=0.3$, $1$, and $2$~TeV, respectively. At the same time these
numbers indicate the confidence level of $g''$ taking on any value.
The likelihood that the $g''$
value lies anywhere below a given value $g_0''$ is depicted in
Fig.~\ref{fig:gppLimit}.

While these numbers may seem low, there are some points to be made
in order to see the situation in proper perspective. First of all,
in the epsilon analysis, the approximation in which the tBESS
loop-level contributions to $\epsilon$'s are replaced with the SM
$M_H$-dependent loop-level contributions plus the net tBESS
loop-level contributions is used. In addition, in the
Eq.~(\ref{epsilon3gpp}) the net tBESS loop terms, which are not
necessarily negligible against $\delta\epsilon_3^{SM}$, have not
been considered (see Table~\ref{tab:epsilons}). Thus, it might be
possible that using more precise formulae would significantly
change the probability numbers shown above.

Secondly, in the original BESS model, $\epsilon_3$ depends, beside
$g''$, also on the universal fermion parameter $b$. This
dependency can compensate for the negativity of
$\epsilon_3-\delta\epsilon_3^{SM}$. By turning off the direct
coupling of the vector triplet to the light fermions, as we have
done in the tBESS model, the dependency disappears and we face the
tension between the negativity of
$\epsilon_3-\delta\epsilon_3^{SM}$ and the positivity of
$(g/g'')^2$. Nevertheless, adding a new independent direct
interaction of the light fermions with the vector triplet would be
straightforward. It would be a natural extension of the tBESS
model which is in line with the original motivation of the
extraordinary role of the top quark. A new parameter, thus
introduced to the Eq.~(\ref{epsilon3gpp}), could compensate for
the negativity of $\epsilon_3-\delta\epsilon_3^{SM}$ in the same
way as the $b$ parameter in the BESS model does.

The gauge coupling $g''$ can also be restricted by the measurement
of the gauge-boson self-interactions. In particular, the D0
measurement of $p\bar{p}\rightarrow WZX$ puts limits on the
anomalous couplings of the effective $WWZ$ vertex
\cite{D0experimentTGV}. If the CP-invariant operators up to
dimension four are considered, there are two free parameters,
$g_1^Z$ and $\kappa_Z$, in the effective $WWZ$ vertex
\cite{HagiwaraTGV}. In the tBESS model, the two parameters
coincide, $g_1^Z=\kappa_Z$, and depend on a single non-SM
parameter, namely $g''$. Note that this is different from the BESS
model where $g_1^Z$ (which, again, equals to $\kappa_Z$) depends
also on the universal couplings of the vector triplet with
fermions. The $D0$ measurements provide separate limits on $g_1^Z$
and $\kappa_Z$. Since $g_1^Z=\kappa_Z$ in the tBESS model, we
consider the stronger of these limits to derive the restriction
for $g''$. The obtained lower bound reads $g''\geq 3.4\;
(95\%\mbox{C.L.})$.

\begin{figure}[t]
\includegraphics[scale=0.7]{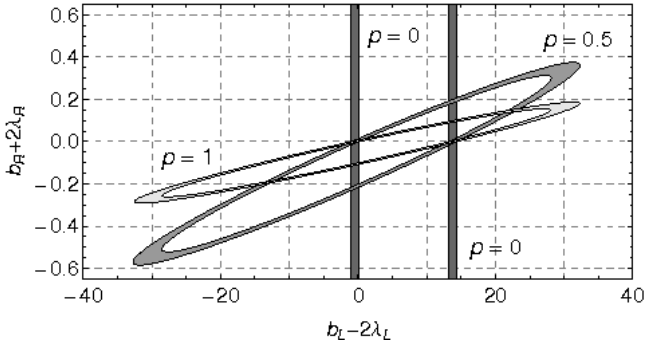}
\caption{\label{fig:blambdaconts3} The $90\%$ C.L. $B\rightarrow
X_s\gamma$ allowed regions of the $b$ and $\lambda$ parameters
when $g''=10$. The allowed regions form closed elliptical bands;
the darkest gray correspond to $p=0$, the lightest to $p=1$, with
$p=0.5$ in between. When $p=0$, only parts of two parallel bands
of an ``infinite'' ellipse can be seen.}
\end{figure}

The $b$ and $\lambda$ parameters are
restricted by the measurement of $B\rightarrow X_s\gamma$ which
puts limits on the anomalous $\kappa_{R,L}^{Wtb}$ parameters of
the $W^\pm t_{R,L} b_{R,L}$
vertices~\cite{btosgData,LariosKappaAnalysis}. In tBESS, these
anomalous couplings are functions of the model's parameters. It
implies the low-energy limits on $b_L-2\lambda_L$ and
$b_R+2\lambda_R$ for given values of $g''$ and $p$. The limits for
various values of $p$ and $g''=10$ are shown in
Fig.~\ref{fig:blambdaconts3}.
The case of $g''\rightarrow\infty$ introduces a change not
distinguishable in the graph.

Both $\epsilon_1$ and $\epsilon_b$ parameters can provide
independent EWPD limits on the same combinations of $b$'s and
$\lambda$'s as in the previous case. However, to the limits based
on the $\epsilon_b$ parameter a qualification applies. The tBESS interactions are more general than the
restriction imposed on the anomalous vector and axial-vector
couplings of the bottom quark in the definition of 
$\epsilon_b$~\cite{EpsilonMethod2}. The definition assumes that these
couplings are not independent of each other. Thus, $\epsilon_b$
can be used to derive the low-energy limits on the tBESS fermion
parameters under this additional assumption only. In particular,
the following condition must hold: either $p=0$, or
$b_R=-2\lambda_R$. As far as the limits derived from $\epsilon_1$
are concerned, no such restrictions apply.

\begin{figure}[b]
\includegraphics[scale=0.48]{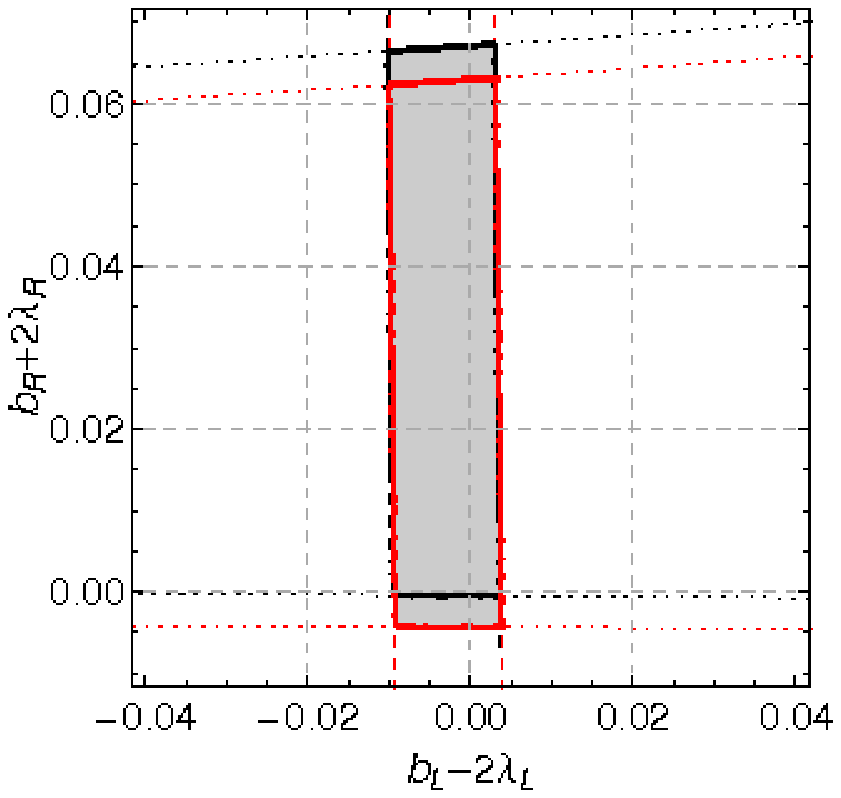}\hfill
\includegraphics[scale=0.48]{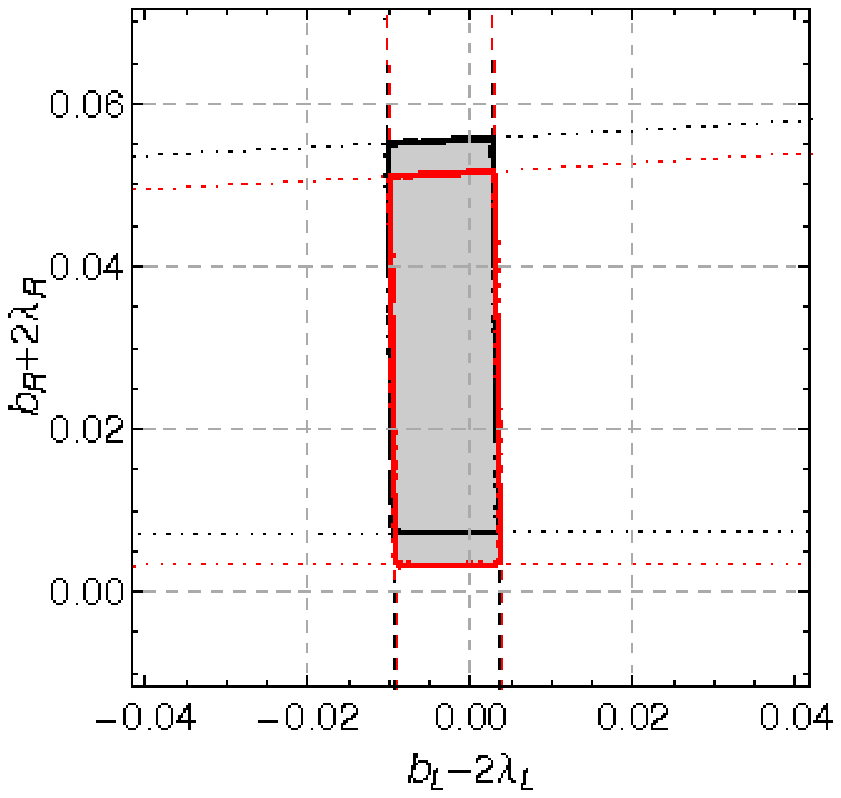}
\caption{\label{fig:blambdaconts1}(color online)
The intersecting
parts of the $90\%$ C.L. allowed regions derived from $\epsilon_1$
(horizontal strip) and $\epsilon_b$ (vertical strip).
The $\epsilon_b$ region assumes that $p=0$.
The $\epsilon_1$ region is not sensitive to $p$. The
black contours correspond to $g''=10$ and the red ones to
$g''\rightarrow\infty$. The cut-off scales considered are
$\Lambda=1$~TeV (left) and $\Lambda=2$~TeV (right).}
\end{figure}

The intersections of the $\epsilon_1$ and $\epsilon_b$ based
regions for $g''=10$ and $\infty$, and $p=0$ are
depicted\footnote{Actually, there are four distinct intersections
of the allowed regions. Only one of them is depicted in
Fig.~\ref{fig:blambdaconts1}. The other three
regions are excluded by the $B\rightarrow X_s\gamma$ decay and/or
allow too large values of the fermion parameters to consider them
reliable. For more detailed discussion, see Appendix
\ref{app:LElimits}.} in Fig.~\ref{fig:blambdaconts1}. The cut-off scale
$\Lambda$ of the low-energy effective theory is reasonable to be
put equal to the mass of the vector resonance. In the figure, the
graphs for $\Lambda=1$~TeV and $\Lambda=2$~TeV are displayed. The
shaded areas of Fig.~\ref{fig:blambdaconts1} lie completely inside
the region allowed by the $B\rightarrow X_s\gamma$ decay. Thus,
they can also be considered as the combined allowed region of the
epsilon and $B\rightarrow X_s\gamma$ methods when $p=0$.

If $b_R=-2\lambda_R$ and $p$ is arbitrary,
the intersection of the $90\%$~C.L. regions
of $\epsilon_1$, $\epsilon_b$, and $B\rightarrow X_s\gamma$,
when $\Lambda=1$~TeV, reads
\begin{equation}\label{eq:LELimitblL1}
   -0.009\leq  \;b_L-2\lambda_L\;  \leq 0.004.
\end{equation}
This interval is virtually independent of $g''$.
When $\Lambda=2$~TeV the regions have no common intersection
at the given confidence level and for any value of $g''$.

If neither $p=0$, nor $b_R=-2\lambda_R$, the low-energy
restrictions are provided by $\epsilon_1$ only
as far as the epsilon parameters are considered.
The restrictions are represented by the horizontal strips in Fig.~\ref{fig:blambdaconts1}.

In this case, the $\epsilon_b$-based restriction can be
substituted for by the low-energy limit obtained directly from the
measurement of the $\Gamma(Z\rightarrow b\bar{b})$ decay employing
Eq.~(11) of~\cite{EpsilonMethod2}. Details of the calculation
can be found in Appendix~\ref{app:LElimitsGammab}.

In Fig.~\ref{fig:blambdaconts4}, the $90\%$~C.L.
regions based on $\epsilon_1$, $\Gamma(Z\rightarrow b\bar{b})$,
and $B\rightarrow X_s\gamma$, and their intersections are shown.
Various combinations of the $p$, $g''$, and $\Lambda$ values are
considered. It can be seen that for some combinations the
intersections are restricted by the $B\rightarrow X_s\gamma$
measurement. In some cases some combinations are excluded
completely; e.g.\ when $p=1, g''=10, \Lambda=2$~TeV.

In Fig.~\ref{fig:blambdaconts4}, for the sake of
comparison, the intersections based on $\epsilon_b$ are also
shown. Even though they are not identical with the
$\Gamma(Z\rightarrow b\bar{b})$ based contours for $p=0$, they are
reasonably close to each other.

\begin{figure}[t]
\includegraphics[scale=0.42]{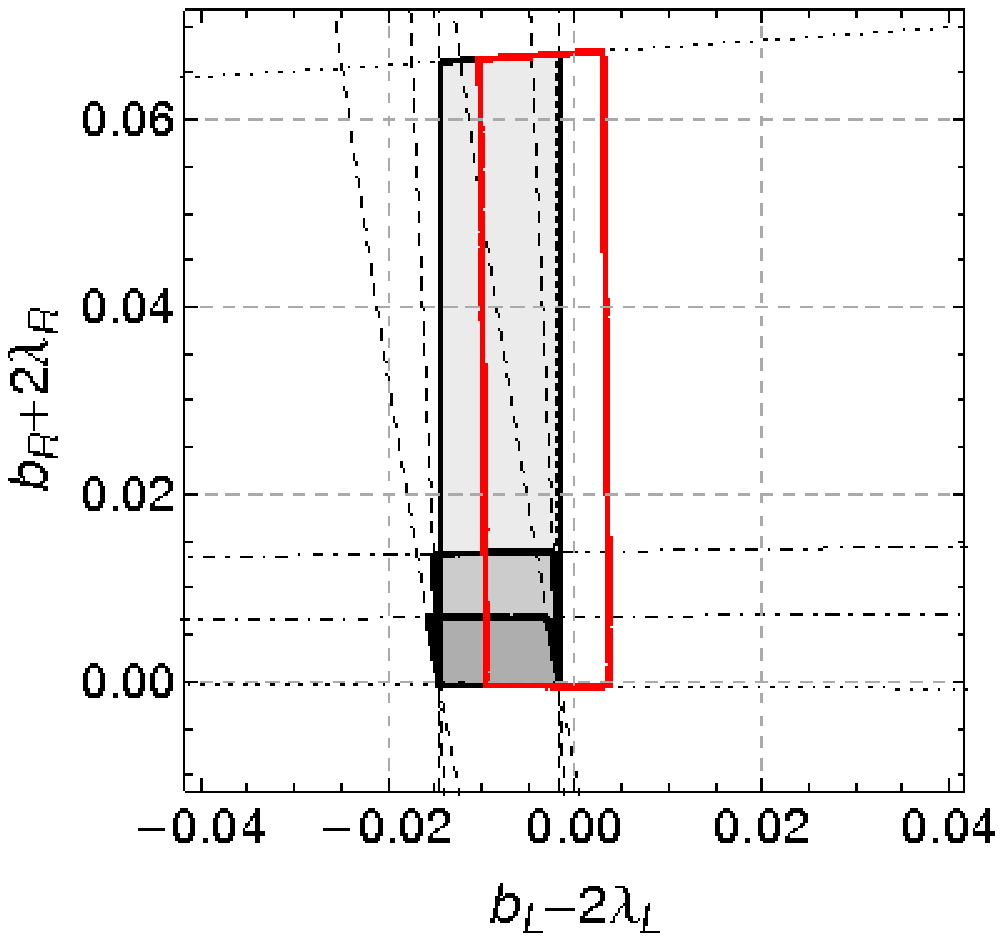}\hfill
\includegraphics[scale=0.42]{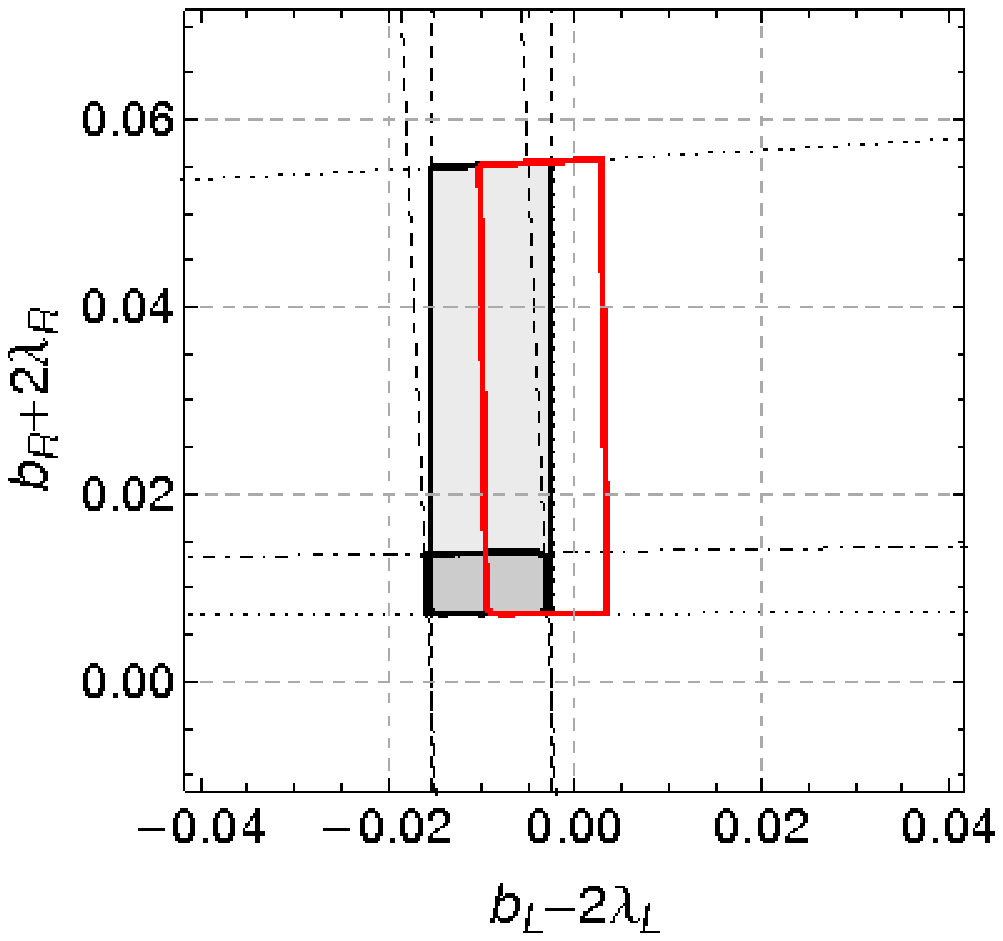}
\\[0.2cm]
\includegraphics[scale=0.42]{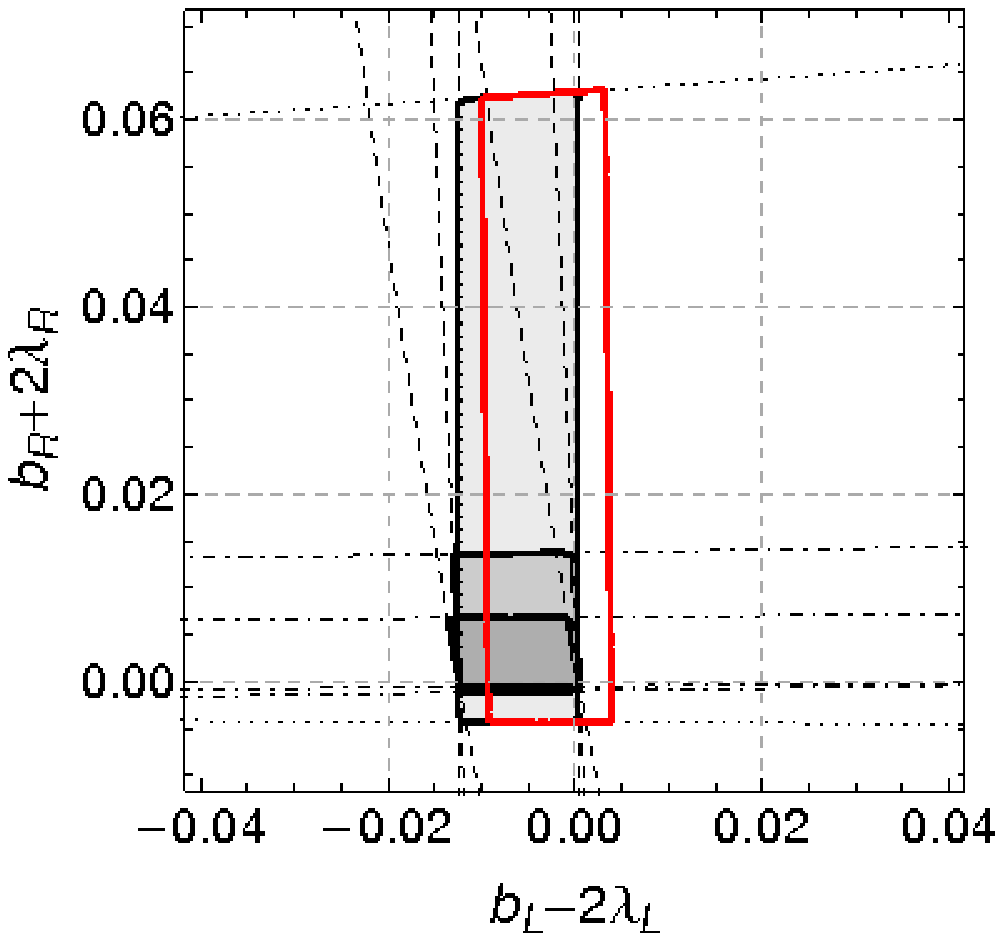}\hfill
\includegraphics[scale=0.42]{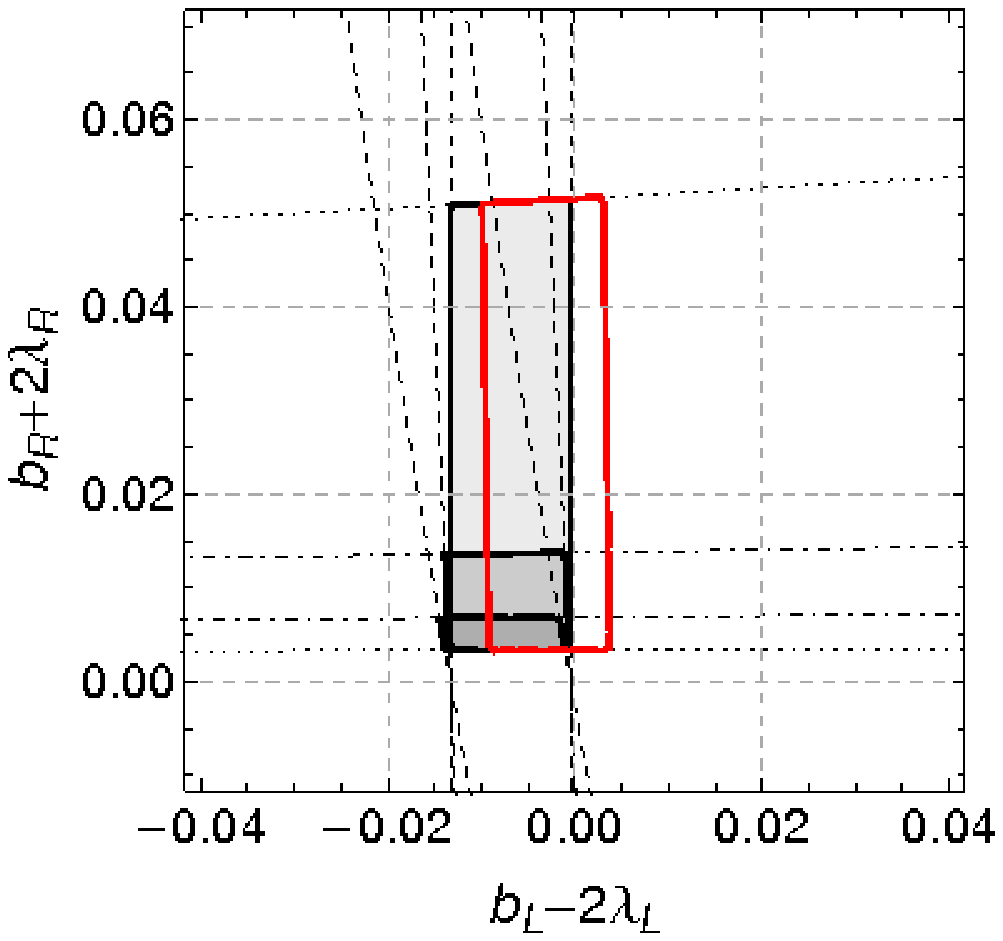}
\caption{\label{fig:blambdaconts4}(color online)
The intersections
of the $\epsilon_1$ allowed regions
(dotted lines) with the $\Gamma(Z\rightarrow b\bar{b})$
allowed regions (dashed lines) and the
$B\rightarrow X_s\gamma$ allowed regions (dash-dotted lines).
All regions are $90\%$~C.L.
The $\Gamma(Z\rightarrow b\bar{b})$ regions are shown
for $p=0$ (vertical strip),
$p=0.5$ (middle tilted), and $p=1$ (the most tilted).
The first row of the figures corresponds to $g''=10$,
the second row to $g''=\infty$.
The left column of figures corresponds to $\Lambda=1$~TeV,
the right column to $\Lambda=2$~TeV.
The shaded areas represent the intersections of all the regions
for $p=0$ (the lightest gray), $p=0.5$ (middle gray), and $p=1$ (the darkest gray).
The empty region with the red solid boundary corresponds to the $\epsilon_b$
based intersection taken from Fig.~\ref{fig:blambdaconts1}.
}
\end{figure}

There are no low-energy limits on the values of the $b$ and
$\lambda$ parameters individually. Thus, in principle, $b$'s and
$\lambda$'s can be tuned to \textit{any} values if their
sum/difference falls within the allowed interval. However, if one
does not wish to admit a fine-tuning of the parameters, it is in
place to add some \textit{ad hoc} restriction; say, the absolute
values of the $b_{L,R}$ or $\lambda_{L,R}$ parameters should not
be greater than 10 times the size of the allowed interval for
$b_{L,R}\mp 2\lambda_{L,R}$. This way, the fine-tuning would not
go below $10\%$.
For example, if we apply this restriction to the
limit for $p=0$, we obtain $|b_L| \leq
0.13$. Of course, at the same time the $\lambda$ parameters must fall
in the strip $b_L-0.003\leq 2\lambda_L\leq b_L+0.010$.

In the BESS model, as well as in many other models of strong ESB
and in the most common extra-dimensional Higgsless theories, the
new vector resonances must be rather fermiophobic in order to
satisfy the EWPD limits. There are ways how to remove this
restriction found in the literature: e.g.\ the degenerated BESS
model~\cite{DBESS} and the four-site Higgsless
model~\cite{4siteExtraDimDBESS}. The top-BESS model provides
another alternative which does not suffer from this restriction.

The parameters $b_{L}$ and $b_R$ correspond to the BESS parameters
$b$ and $b'$ through the relations $b_L=b/(1+b)$ and
$b_R=b'/(1+b')$. The authors of the BESS model~\cite{BESS} used
$\epsilon_3$ to derive the low-energy limits for
$b$~\cite{LowEBESS}. We have updated the limits for the BESS model
using the same epsilon values~\cite{EpsilonData} as for deriving
the limits of the tBESS fermion parameters. When $g''=10$, we have
obtained
\begin{equation}
  0.008\leq b\leq 0.015\;\;\; (90\%\; \mbox{C.L.}).
\end{equation}
Thus, the limit on $b_{L}$  obtained by the combination of the
low-energy bounds and the no-fine-tuning requirement is
significantly less restrictive, than the low-energy limit for $b$.

In the BESS model the universal right fermion coupling $b'$ is
usually set to zero due to the reasons mentioned before. In the
tBESS model, the low-energy limits on $b_R$ can be even less
restrictive than those on $b_L$ when $p$ approaches zero.

\subsection{The Death Valley effect}
\label{subsec:DeathValley}

The interplay of the direct and indirect couplings of the vector
triplet with fermions can diminish or even zero a particular
top/bottom quark channel decay width of the vector resonance for
some nonzero values of the $b$ parameters. Thus, it might happen
that even though the direct couplings of the vector resonance to
the top and/or bottom quark are nontrivial the resonance will not
decay through the given quark channel. Or, the particular decay
will be suppressed below the value that would be implied by the
indirect couplings alone.

\begin{figure}
\includegraphics[scale=0.42]{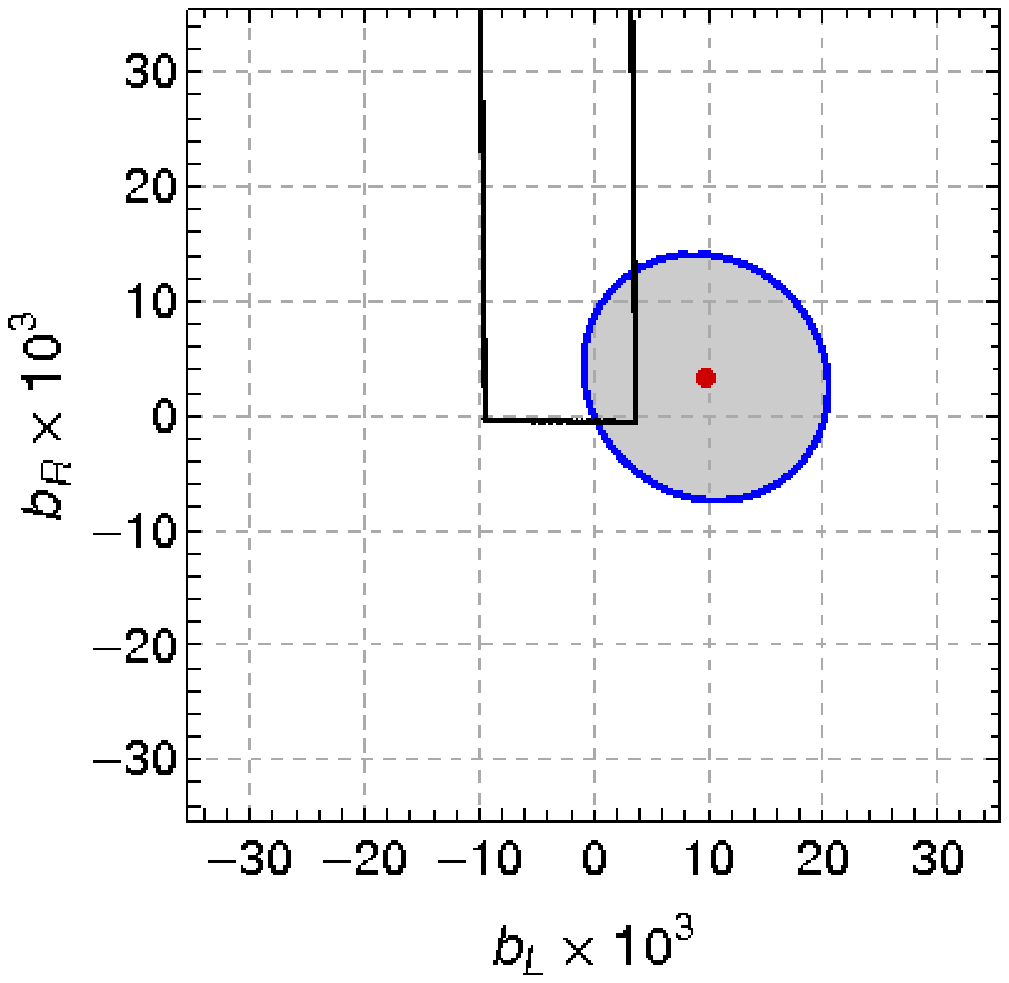}\hfill
\includegraphics[scale=0.41]{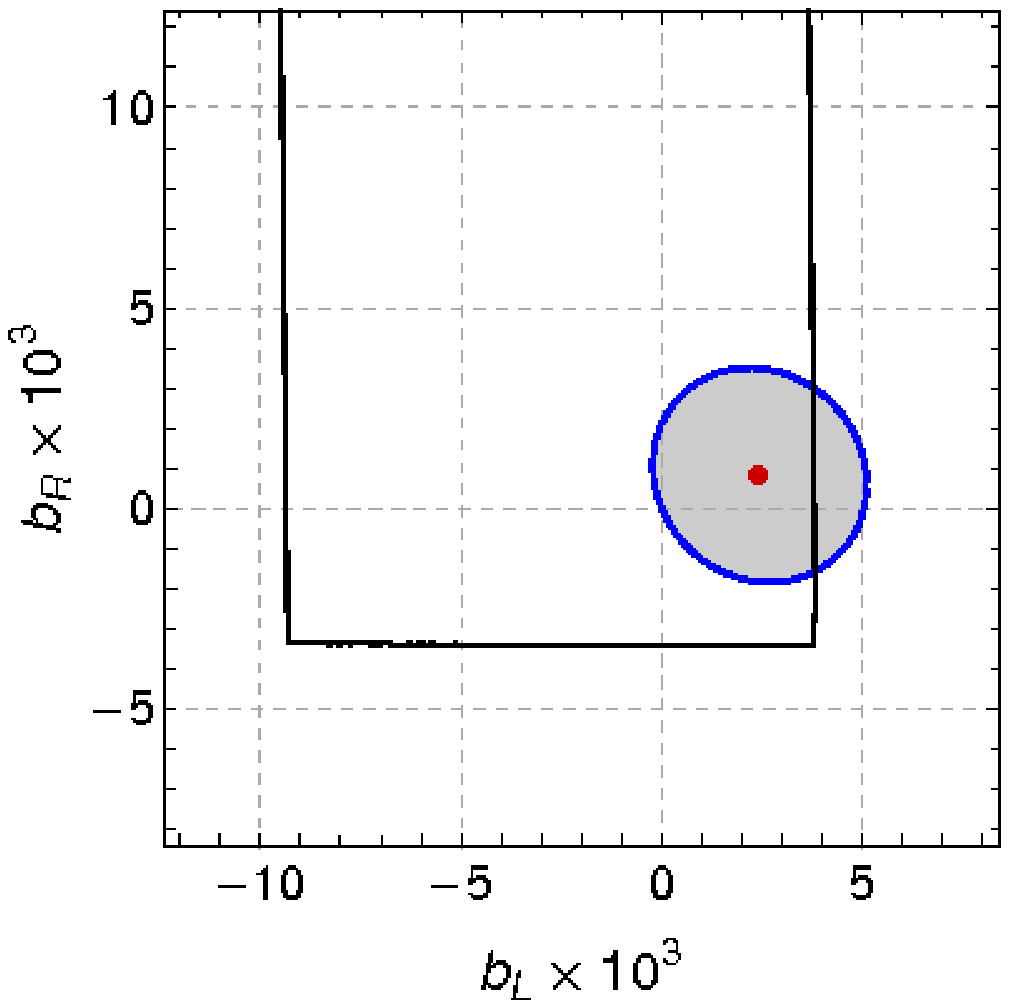}
\caption{\label{fig:DVtt}(color online)
The Death Valley regions
(shaded areas) of the $V^0\rightarrow t\bar{t}$ decay for
$M_{V^0}=1$~TeV and $g''=10$ (left) and $g''=20$ (right). The red
dot indicates values for which the corresponding partial decay
width is equal to zero. The low-energy allowed region for
$\lambda_L=\lambda_R=0$, $p=0$, and $\Lambda=1$~TeV (solid line) is
superimposed on the graphs.}
\end{figure}

Figure~\ref{fig:DVtt} shows the area of the $b_L$-$b_R$ parametric
space where the decay width of $V^0\rightarrow t\bar{t}$ is equal
to or lower than the corresponding value generated by the indirect
couplings alone. We call this region the \textit{Death Valley}
(DV) because that is where the resonance decay through the
particular decay channel deteriorates or even dies out.
The dot in the middle of the area indicates the parameter values
for which the partial decay width is equal to zero. The DV shrinks
and the zero width point moves to the origin of the parametric
space as $g''$ grows. The DV region for $t\bar{t}$ channel does
not depend on $p$. Its dependence on $\lambda$'s can be neglected.

There are the EWPD contours superimposed over the DV graphs in the
figure to show which part of the allowed parameter values overlaps
with the DV. Recall that the low-energy limits apply to the
combination of $b$'s and $\lambda$'s rather than to the parameters
alone. The low-energy limits depicted in Fig.~\ref{fig:DVtt}
correspond to $\lambda_L=\lambda_R=0$. By choosing nonzero values
for $\lambda_{L,R}$ the low-energy contours get shifted around the
parameter space. There are acceptable values\footnote{Acceptable
in the sense of no more than $10\%$ of the fine-tuning of $b$'s
and $\lambda$'s.} of $\lambda$'s leading to both extrema
--- (a) no overlap, and (b) the maximal overlap --- of
the DV and the low-energy allowed regions.

\begin{figure}
\includegraphics[scale=0.42]{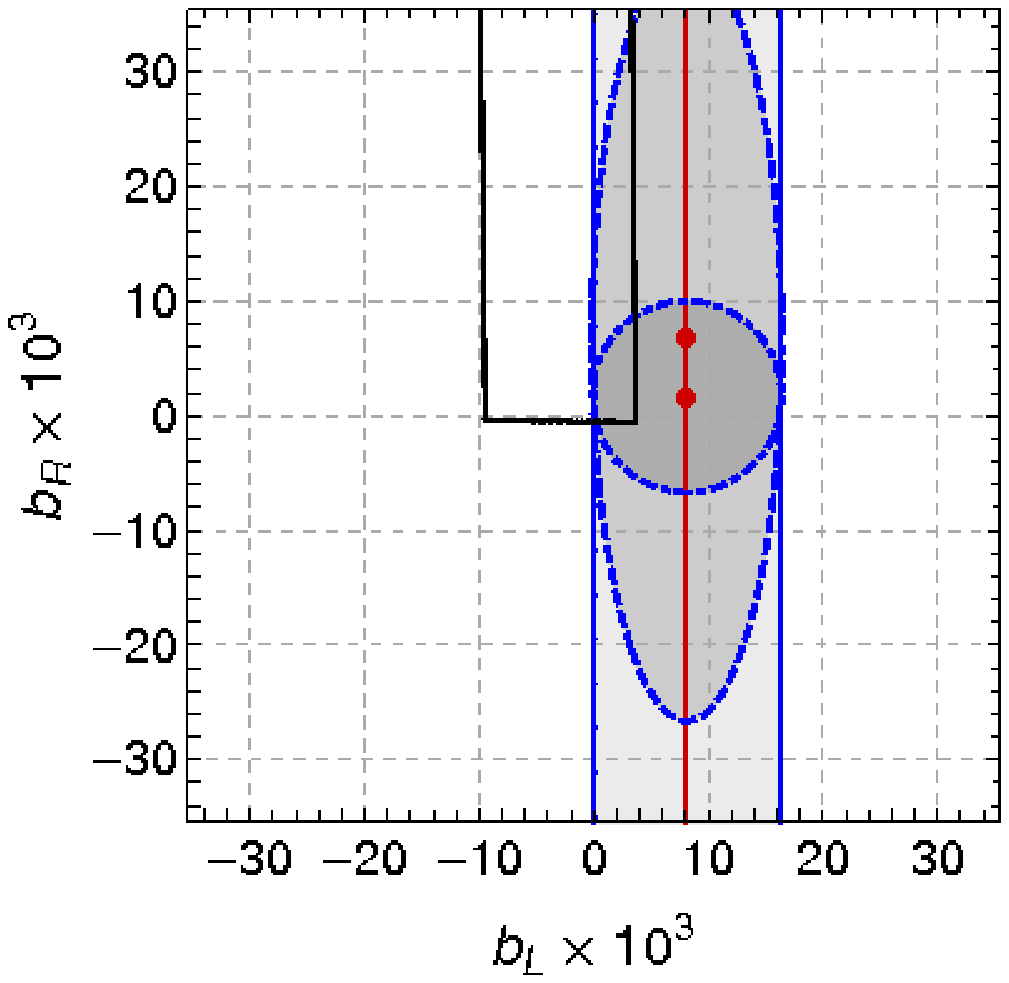}\hfill
\includegraphics[scale=0.41]{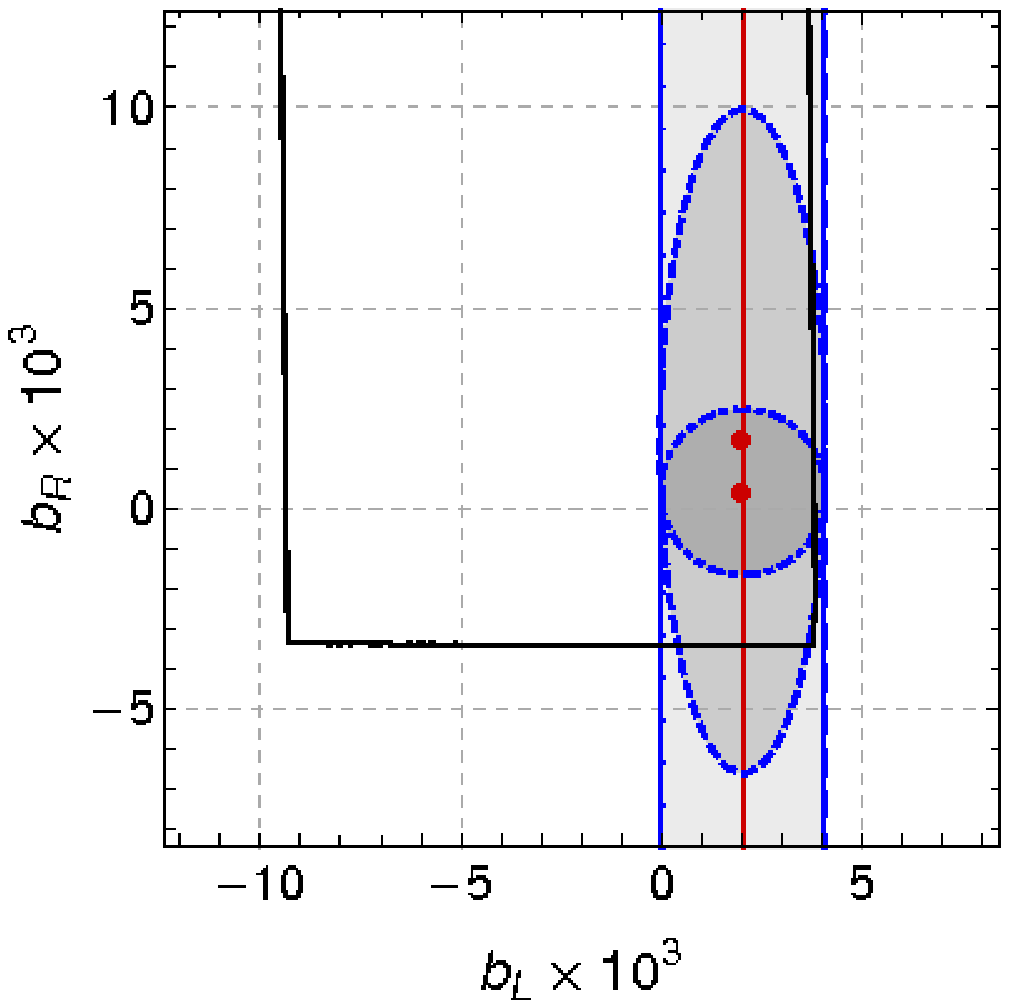}
\caption{\label{fig:DVbb}(color online)
The Death Valley regions
(shaded areas) of the $V^0\rightarrow b\bar{b}$ decay for
$M_{V^0}=1$~TeV and $g''=10$ (left) and $g''=20$ (right). The dark
gray area corresponds to the DV of $p=1$, the medium gray area to
$p=0.5$, and the light gray region to $p=0$. The lower ($p=1$) and
higher ($p=0.5$) red dots indicate the $(b_L,b_R)$ values for
which the partial decay width is equal to zero. The middle red
line corresponds to the $(b_L,b_R)$ values of the minimal
$b\bar{b}$ decay width when $p=0$. The low-energy allowed region
for $p=0$, $\lambda_L=\lambda_R=0$, and $\Lambda=1$~TeV is
superimposed on the graphs.}
\end{figure}

The DV regions of the $V^0\rightarrow b\bar{b}$ decay for
$M_{V^0}=1$~TeV are shown in Fig.~\ref{fig:DVbb}.
In this case the DV region is of elliptical shape and depends on
$p$. When $p=0$ the DV is an unbound strip in the $b_R$ direction.
As $p$ decreases from 1 to 0 the $b_R$ coordinate of the zero
width dot grows, reaching infinite value for $p=0$. Since $p=0$
turns off the $b_R$ coupling for any value of $b_R$, the indirect
interaction of the vector triplet with the right bottom quark
cannot be compensated by its direct analogue. Therefore, the
$V^0\rightarrow b\bar{b}$ decay width cannot be equal to zero for
any finite values of the $b$ parameters when $p=0$. Nevertheless,
there will be the minimal value of the width at a fixed value of
$b_L$ and any value of $b_R$.

Note that if $g''=10$ the DV area is equal or larger than the EWPD
region. If we change the $\lambda_L$ value of the left-hand side
graph of Fig.~\ref{fig:DVbb} to $\lambda_L=0.006$ the low-energy
contours get shifted to the right and find themselves inside the
DV's. Thus, in this case the EWPD admit only $b_{L,R}$ values
which lie inside the DV.

\begin{figure}
\includegraphics[scale=0.42]{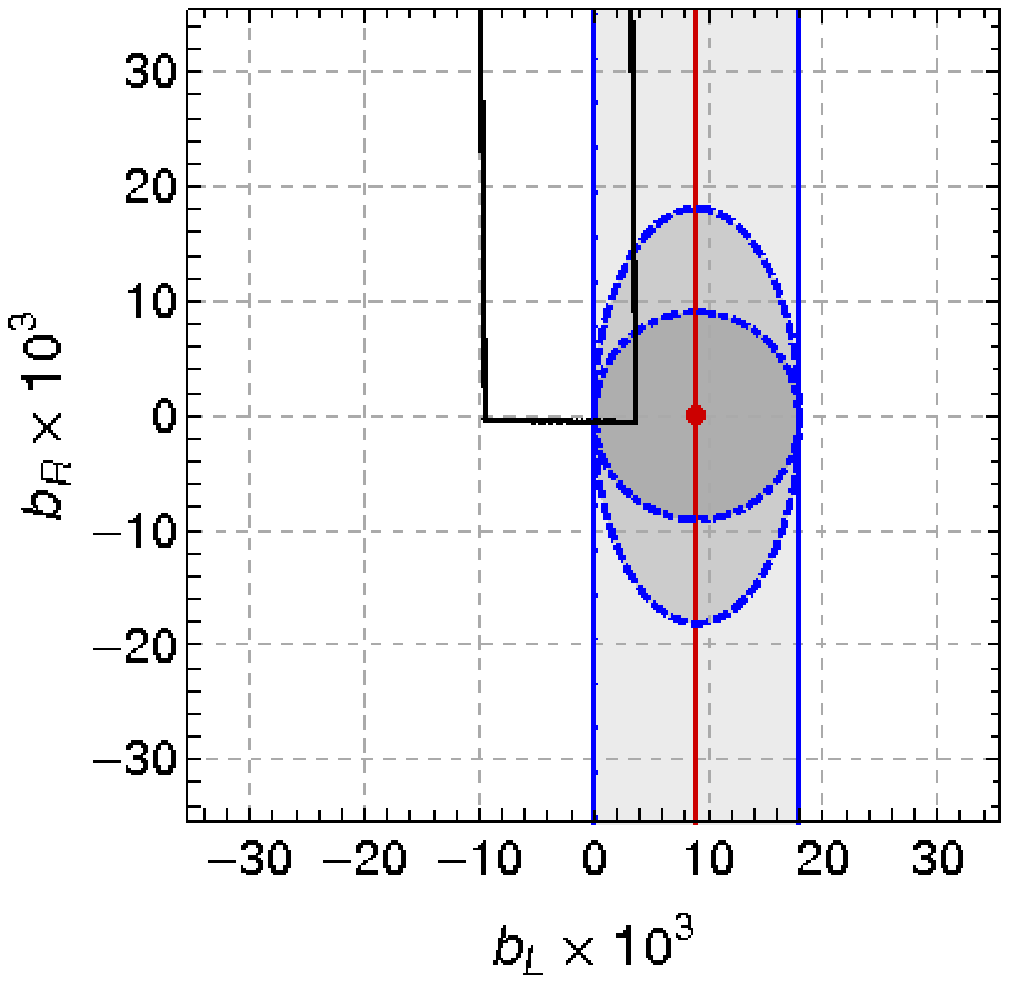}\hfill
\includegraphics[scale=0.41]{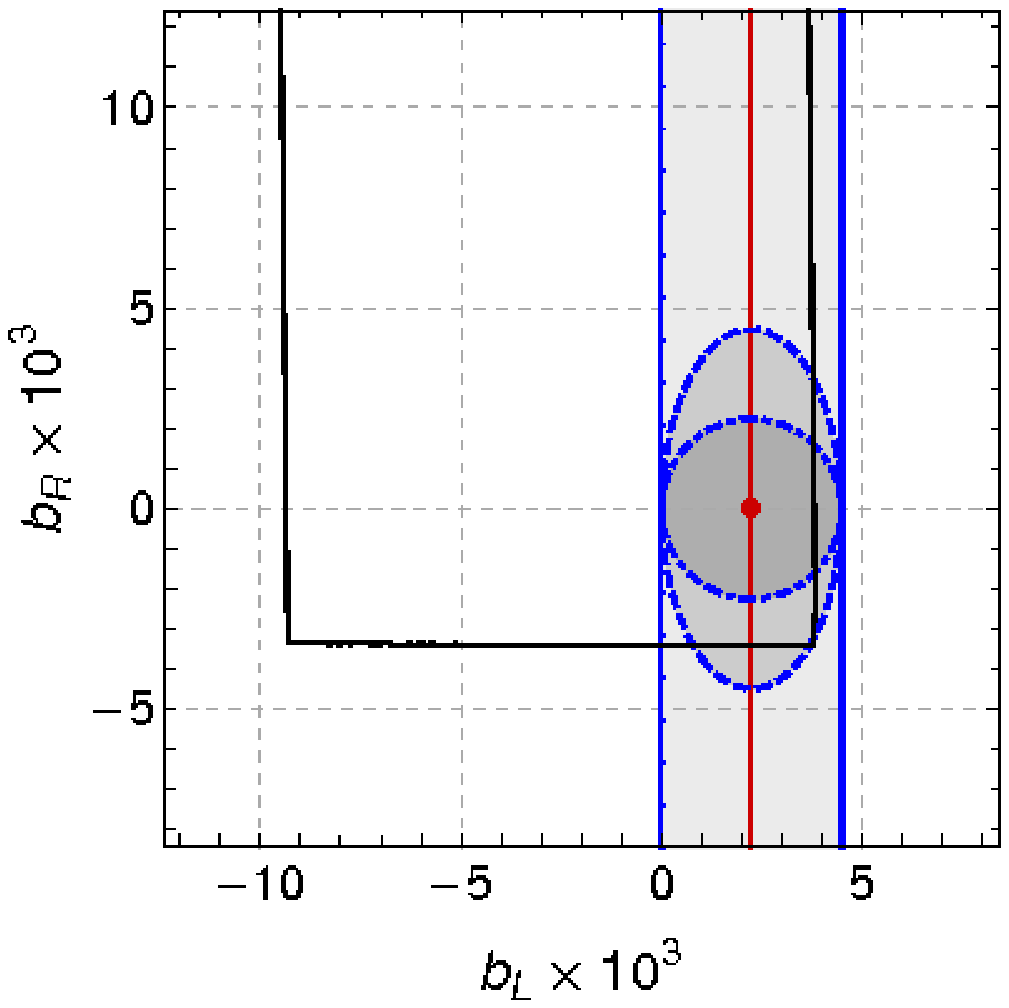}
\caption{\label{fig:DVtb}(color online)
The Death Valley regions
(shaded areas) of the $V^\pm\rightarrow t\bar{b}/\bar{t}b$ decay
for $M_{V^0}=1$~TeV and $g''=10$ (left) and $g''=20$ (right). The
dark gray area corresponds to the DV of $p=1$, the medium gray
area to $p=0.5$, and the light gray region to $p=0$. The red dot
($p=1$ and $p=0.5$) and the red line ($p=0$) indicate the
$(b_L,b_R)$ values for which the corresponding partial decay width
is equal to zero. The low-energy allowed region
for $p=0$, $\lambda_L=\lambda_R=0$, and $\Lambda=1$~TeV is
superimposed on the graphs.}
\end{figure}

The DV's for the quark decay of the charged resonance,
$V^\pm\rightarrow t\bar{b}/\bar{t}b$, are shown in
Fig.~\ref{fig:DVtb}.
As in the case of the $b\bar{b}$ channel the DV depends on $p$. If
$p=0$, its elliptical shape turns into the $b_R$-unbound strip.
The position of the zero width point does not depend on $p$, if
$p>0$. If $p=0$, the zero width point turns into a straight line
of a fixed $b_L$ value and any $b_R$ value. As in the previous
case, it is possible to hide the low-energy allowed regions inside
the corresponding DV; $\lambda_L=0.006$ would make the job as it
did in the case of $V^0\rightarrow b\bar{b}$.

\subsection{Scattering processes}
\label{subsec:ScatteringProcesses}

The main goal of the construction and study of the tBESS model is
to provide an effective tool for the description and analysis of
the possible experimental situation observed at the LHC and the
ILC. Even though the analysis of sensitivity of particular
scattering processes to the tBESS parameters is not within the
scope of this paper we would like to discuss two features of the
tBESS model which can prove important when such analysis will be
performed.

\subsubsection*{Hiding the peak}

The Death Valley effect can hide signals expected in scattering
processes. There might be new physics materialized through the
existence of the new vector resonances as well as nonzero values
of the $b$ parameters, yet it does not have to reveal itself in an
experiment. In particular, even if the tBESS resonances exist and
couple to the third quark generation we do not have to see a peak
in the scattering experiments for certain final states containing
top and/or bottom quarks. This would occur if the model parameters
happened to have their values inside the DV region. More
precisely, the region, in which the resonance peak in a scattering
process is lower than the peak due to the indirect couplings to
fermions, can slightly differ from the DV region. It is due to the
interference effects between signal and nonsignal amplitudes of
the process. Nevertheless, we will not elaborate on this in this
paper.

To illustrate the DV effect on the scattering amplitudes we have
plotted the cross sections for five processes: $e^-e^+\rightarrow
t\bar{t}/b\bar{b}/W^+W^-$ and $u\bar{d}\rightarrow t\bar{b}/W^+Z$.
The cross sections are evaluated for $M_{V^0}=1$~TeV and $g''=20$
at four different parameter space points (PSP) which are specified
in Table~\ref{tab:PSP}.
\begin{table}
\caption{Parameter space points (PSP) at which the cross sections
in Fig.~\ref{fig:XSforPSP} were calculated.} \label{tab:PSP}
\begin{ruledtabular}
  \begin{tabular}{cccccc}
     PSP & $p$ & $b_L$ & $b_R$ & $\lambda_L$ & $\lambda_R$ \\
     \hline
     1 & 0 & 0 & 0 & 0 & 0 \\
     2 & 0 & -0.01 & 0.03 & 0 & 0  \\
     3 & 0 & 0.009 & 0.03 & 0.006 & 0  \\
     4 & 0 & 0.0098 & 0.0034 & 0.006 & 0
  \end{tabular}
\end{ruledtabular}
\end{table}
The points were chosen to demonstrate how the tBESS resonance peak
behaves if PSP lies inside or outside the DV. Of course, the gauge-boson 
processes are sensitive to the choice of PSP only through
the resonance decay width. The nonzero values of $\lambda_L$ have
been chosen to shift the low-energy allowed region so that it
includes the given PSP. The cross section at the peak region is
not significantly affected by the $\lambda$-values, though.

\begin{figure*}
\includegraphics[scale=0.4]{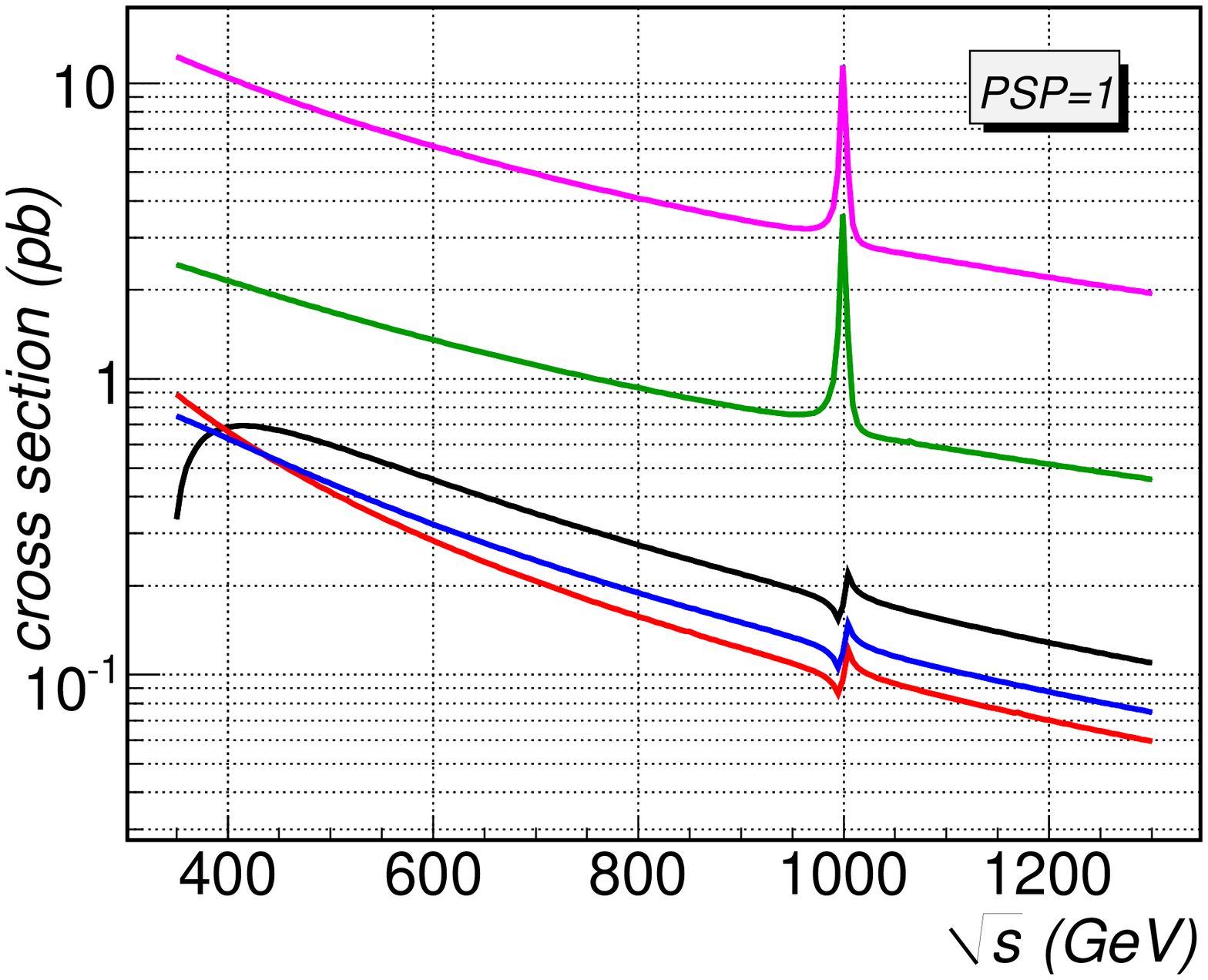}
\includegraphics[scale=0.4]{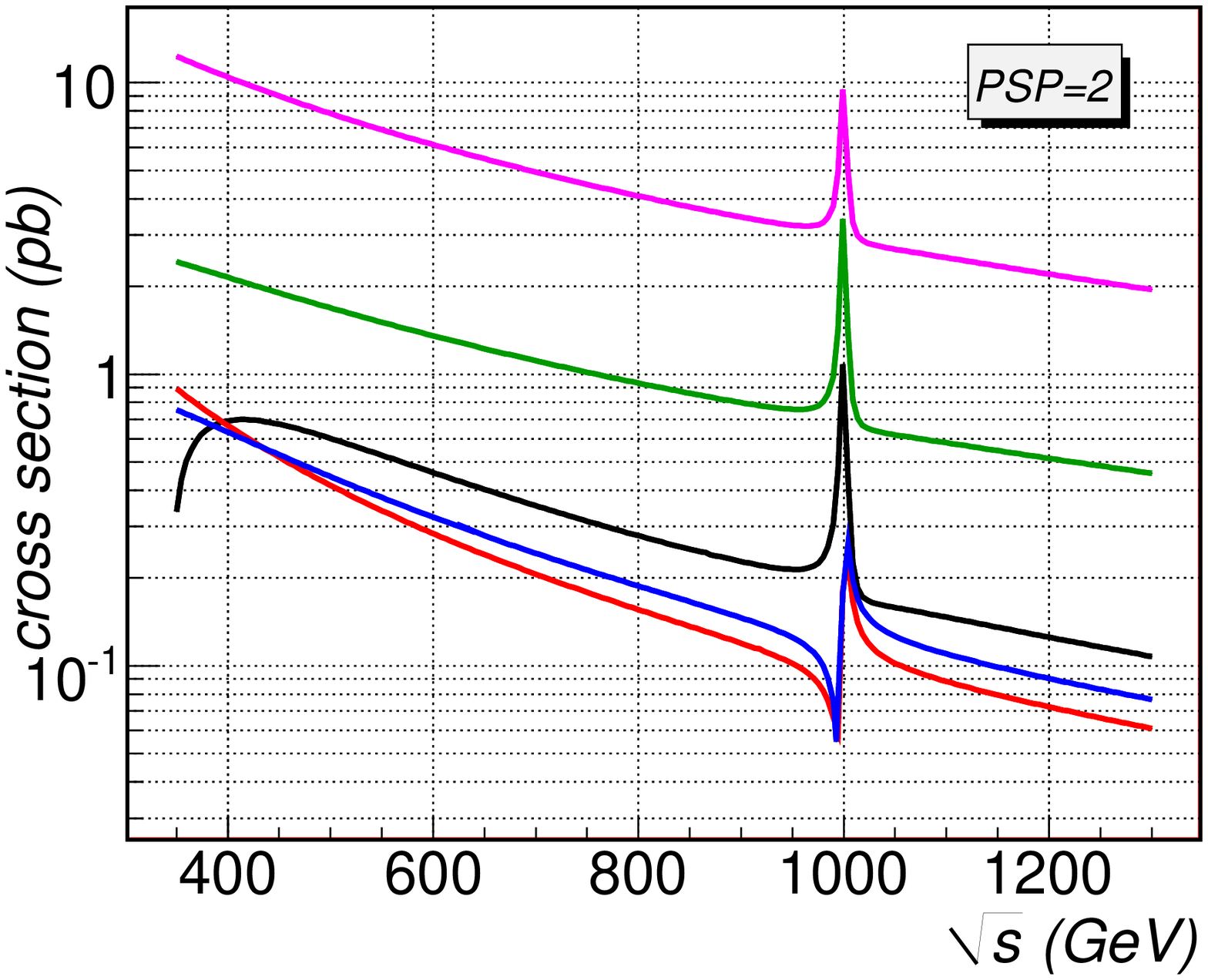}\\
\includegraphics[scale=0.4]{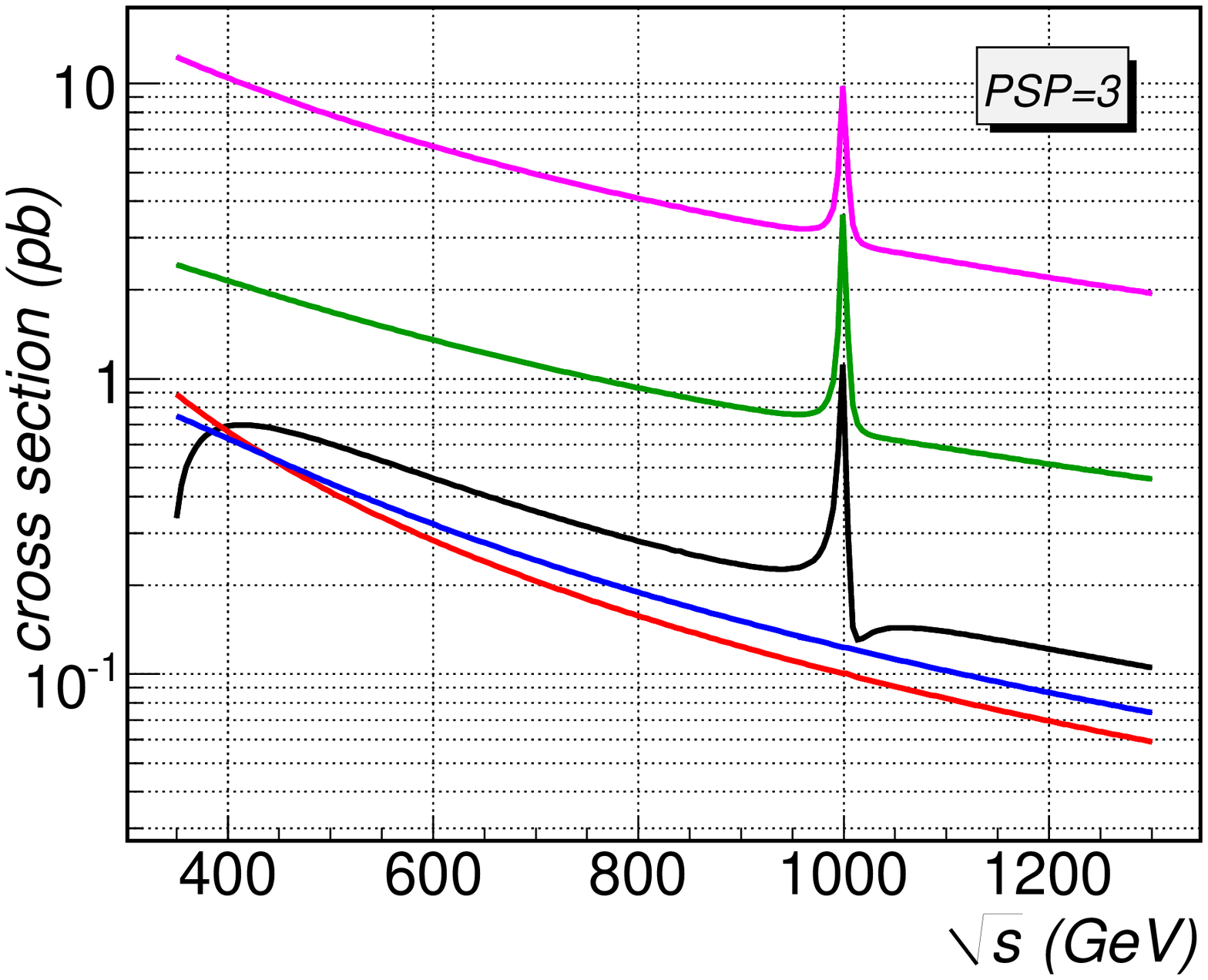}
\includegraphics[scale=0.4]{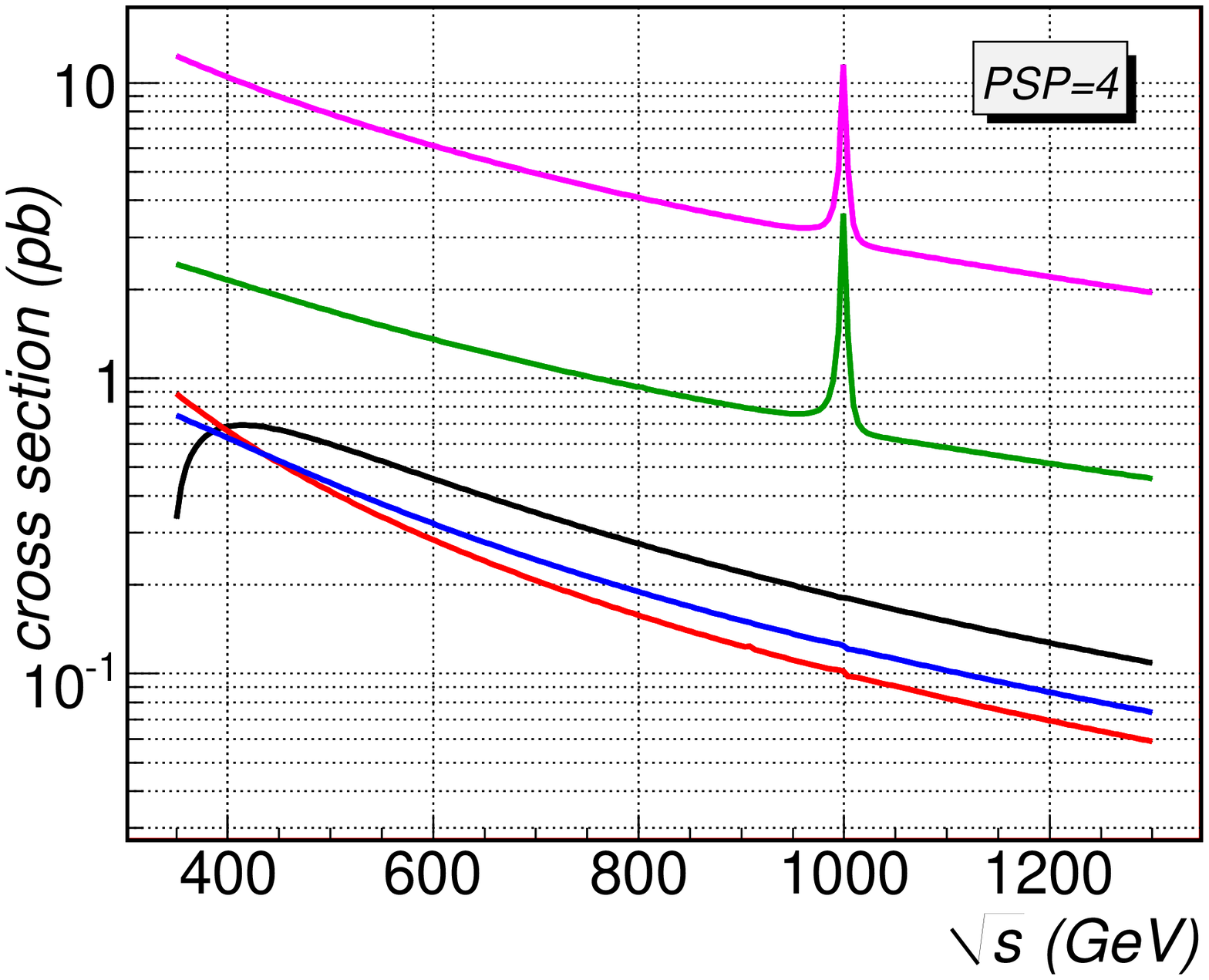}
\caption{\label{fig:XSforPSP}(color online) The cross sections of
the $e^-e^+\rightarrow t\bar{t}/b\bar{b}/W^+W^-$ and
$u\bar{d}\rightarrow t\bar{b}/W^+Z$ processes for $M_{V^0}=1$~TeV
and $g''=20$. The four graphs correspond to the four parameter
space points specified in Table~\ref{tab:PSP}. Each graph displays
plots of all five processes. From top to bottom:
$e^-e^+\rightarrow W^+W^-$ (magenta), $u\bar{d}\rightarrow W^+Z$
(green), $e^-e^+\rightarrow t\bar{t}$ (black),
$u\bar{d}\rightarrow t\bar{b}$ (blue), $e^-e^+\rightarrow
b\bar{b}$ (red).}
\end{figure*}

The PSP=1 graph in Fig.~\ref{fig:XSforPSP} shows the cross
sections of the five processes when there is \textit{no direct
coupling} of the vector resonance to fermions.
While there are clear $1$~TeV resonance peaks in the gauge-boson
channels the top/bottom channel processes exhibit only small
peaks.

PSP=2 was chosen far away from the DV's of all three top/bottom
channels. Thus we expect to see large $1$~TeV peaks in all five
cross sections. Indeed, the PSP=2 graph of Fig.~\ref{fig:XSforPSP}
shows exactly that behavior.

PSP=3 lies at the bottoms of the DV's for $b\bar{b}$ and
$t\bar{b}$ channels. On the other hand, for the $t\bar{t}$ channel
the PSP is far away from the channel's DV. In accordance with that
the PSP=3 graph in Fig.~\ref{fig:XSforPSP} shows the $1$~TeV
resonance peak only in the $e^-e^+\rightarrow t\bar{t}$ cross
section, other two top/bottom final state graphs being flat.

PSP=4 is localized at the bottom of the DV's of all three
top/bottom processes. Indeed, in their cross sections, no $1$~TeV
peak can be found in the PSP=4 graph of Fig.~\ref{fig:XSforPSP}.

Note that since $p=0$ for all PSP's $b_R$-related couplings are
effectively set to zero in the processes $e^-e^+\rightarrow
b\bar{b}$ and $u\bar{d}\rightarrow t\bar{b}$.

\subsubsection*{Drell-Yan processes}

The fundamental process for probing the mechanism of ESB is the
electroweak gauge boson (EWGB) scattering, $WW\rightarrow WW$,
where $W=W^\pm,Z$. No matter what is the theory behind ESB, it
must leave its footprints in all processes containing
$WW\rightarrow WW$ as a part of their Feynman diagrams. That is
why major attention in the literature has always been paid to the
processes which realize the EWGB scattering through $WW$ fusion,
either at the LHC or at the ILC~\cite{VLVLtoVLVL} 
(see also Ref.~\cite{snowmass} and references therein). Particularly,
if there is a vector resonance associated to the ESB sector, one
would expect that it strongly couples to the longitudinal
components of the massive electroweak gauge bosons and we should
detect its existence through these processes.

Beside the EWGB \textit{fusion processes}, the processes with the
\textit{associated production} of a resonance $R$,
$ee/qq\rightarrow RW$, where $R$ is radiated of the final EW
gauge boson, and decays subsequently into the pair of EW gauge
bosons, $R\rightarrow WW$, can also probe the ESB sector.

The answer to the question how the ESB vector resonance couples to
the SM fermions is very much model-dependent. There are many
strong ESB models where the EWPD, namely the limits on the
$\epsilon_3$ parameter, suppress the direct interactions of the
new vector resonances with fermions. For example, the BESS model
vector triplet is \textit{fermiophobic}. Also, the most common
Higgsless extra-dimensional theories, including the three-site
one~\cite{3siteExtraDimBESS}, are
fermiophobic~\cite{4siteExtraDimDBESS}. If this is the case, the
experimental search for the vector resonance is bound to the
fusion and associated production processes mentioned above.

In the case of \textit{nonfermiophobic} models, like the
degenerated BESS model~\cite{DBESS} and the four-site Higgsless
extra-dimensional model~\cite{4siteExtraDimDBESS}, the stronger
direct couplings to fermions bring up new candidate processes for
testing the ESB vector resonances. To discover the new resonances
and test their relationship to fermions the scope of candidate
processes can be widened to the EWGB fusion and the associated
resonance production where the EW gauge bosons at one or both ends
of resonance propagators are replaced with fermions. This also
includes the Drell-Yan processes at the LHC, as well as the
s-channel resonance production at the ILC. Indeed, these processes
were studied in the
literature~\cite{DominiciNuovoCim,4siteExtraDimDBESS} and found
promising.

Processes where the new resonance interacts with top quarks in
particular do not only provide supplemental opportunities to
discover the new vector resonances, they can also probe the
relationship between ESB physics and physics of the top
quark~\cite{Lee,snowmass}. Many papers focus on processes with
$ww\rightarrow t\bar{t}$ scattering involved~\cite{VLVLtoTT}.

Because of the nonuniversality of its interactions with the SM
fermions the tBESS vector triplet does not fit to either of the
two categories mentioned above. The tBESS model admits strong
direct couplings to top and bottom quarks and none to the light SM
fermions. Of course, there are the mixing-induced indirect
couplings to the light fermions which are suppressed. Thus, when
searching for candidate processes to probe the tBESS model one
would tend to avoid those where the vector triplet couples to
light fermions. However, the existing studies of the vector
resonances with nonuniversal couplings to
fermions~\cite{Han1,Han2,Han3} show that these naive expectations
are not always correct. At the LHC, the $WW\rightarrow tt$ fusion
process is overwhelmed by the QCD background. On the other hand,
the Drell-Yan processes and the resonance associated production
with top/bottom quark final states appear detectable. The
Drell-Yan processes can compete because the suppressed
interactions of the resonance with light quarks can be compensated
for by their higher luminosities in proton-proton collisions.

Our preliminary study~\cite{STM2008,InProgress} of sensitivity of
the LHC Drell-Yan processes to the tBESS resonances at
$M_{V^0}=1$~TeV suggests that while $pp\rightarrow
t\bar{t}X/b\bar{b}X$ is overwhelmed by the gluon-gluon background
and thus insensitive to the tBESS resonance, the $pp\rightarrow
t\bar{b}X/W^+W^-X/W^\pm ZX$ processes yield quite promising
signals.

\begin{figure}
\includegraphics[scale=0.45]{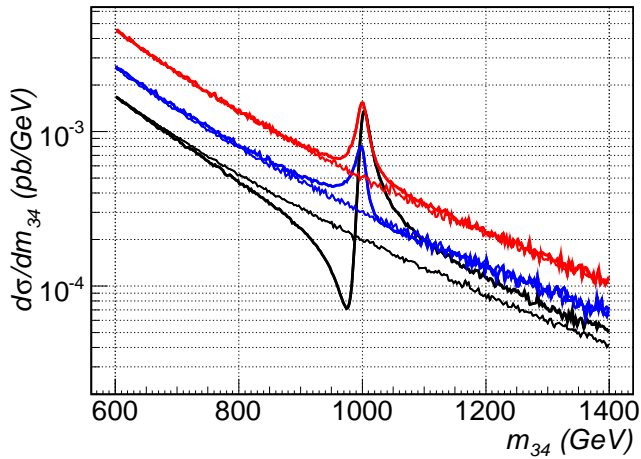}
\caption{\label{fig:XSforLHC}(color online) The invariant mass
distributions for the final state particles of the $pp\rightarrow
W^+W^-X$ (top red line), $pp\rightarrow (W^+Z+W^-Z)X$ (middle blue
line), and $pp\rightarrow (t\bar{b}+b\bar{t})X$ (bottom black
line) processes at the LHC for $\sqrt{s}=14$~TeV and
$M_{V^0}=1$~TeV, $g''=20$, $p=0.5$, $b_L=-0.072$, $b_R=0.074$,
$\lambda_L=\lambda_R=-0.03$. The thinner lines depict the SM
predictions assuming $M_{Higgs}=115$~GeV.}
\end{figure}

In this paper we do not aim to perform systematic study of the
tBESS model testing at either of the existing or future colliders.
Nevertheless, as an illustration, in Fig.~\ref{fig:XSforLHC} we
show the invariant mass distributions for the final state
particles of the $pp\rightarrow t\bar{b}X/W^+W^-X/W^\pm ZX$
processes at the LHC.
The collision energy is $\sqrt{s}=14$~TeV and $M_{V^0}=1$~TeV.
Other tBESS parameters read $g''=20$, $p=0.5$, $b_L=-0.072$,
$b_R=0.074$. If $\lambda_L=\lambda_R=-0.03$ this PSP finds itself
in the low-energy allowed region of the tBESS parametric space,
away from the DV's of all decay channels. The mass of the charged
resonance is $M_{V^\pm}=999.84$~GeV. The only cuts applied to all
processes exclude the forward and backward scattering angles for
which their cosines are either below $-0.99$ or above $0.99$. The
Cabibbo-Kobayashi-Maskawa matrix mixing is ignored in the calculations. 
The total cross
sections obtained under these conditions read
\begin{eqnarray}
  \sigma(t\bar{b}+b\bar{t}) &=& \phantom{3}4.18\;(\phantom{3}4.09)\;\;\mbox{pb},
  \nonumber\\
  \sigma(W^+Z+W^-Z) &=& 10.75\;(10.52)\;\;\mbox{pb},
  \nonumber\\
  \sigma(W^+W^-) &=& 31.85\;(31.29)\;\;\mbox{pb}.
  \nonumber
\end{eqnarray}
The values in the round brackets correspond to the SM with
$M_{Higgs}=115$~GeV. The cross sections of individual subprocesses
are shown in Table~\ref{tab:XSsubprocesses}.
\begin{table}[h]
\caption{The tree-level cross sections of individual subprocesses
contributing to the processes calculated in
Fig.~\ref{fig:XSforLHC}. $M_{Higgs}=115$~GeV is assumed for the
SM.} \label{tab:XSsubprocesses}
\begin{ruledtabular}
  \begin{tabular}{cccccc}
  \multicolumn{6}{c}{$pp\rightarrow t\bar{b}X/b\bar{t}X$}\\
  \hline
  $\sigma$ (pb) & $u\bar{d}$ & $c\bar{s}$ & $d\bar{u}$ & $s\bar{c}$ & \\
  tBESS & 1.05 & 0.16 & 0.73 & 0.16 & \\
  SM & 1.02 & 0.15 & 0.72 & 0.15 & \\
  \hline
  \multicolumn{6}{c}{$pp\rightarrow W^+ZX/W^-ZX$}\\
  \hline
  $\sigma$ (pb) & $u\bar{d}$ & $c\bar{s}$ & $d\bar{u}$ & $s\bar{c}$ & \\
  tBESS & 2.63 & 0.45 & 1.84 & 0.45 & \\
  SM & 2.57 & 0.44 & 1.80 & 0.44 & \\
  \hline
  \multicolumn{6}{c}{$pp\rightarrow W^+W^-X$}\\
  \hline
  $\sigma$ (pb) & $u\bar{u}$ & $d\bar{d}$ & $s\bar{s}$ & $c\bar{c}$ & $b\bar{b}$ \\
  tBESS & 7.00 & 5.62 & 1.85 & 1.08 & 0.38 \\
  SM & 6.88 & 5.52 & 1.81 & 1.06 & 0.37
  \end{tabular}
\end{ruledtabular}
\end{table}
The \verb+CTEQ6L1+ parton distribution functions were used to
obtain these results.

The vector resonance decay widths for the PSP at which the $pp$
processes were calculated are shown in
Table~\ref{tab:DecayWidths}.
\begin{table}[h]
\caption{The partial decay widths of the vector resonance triplet
at $M_{V^0}=1$~TeV, $g''=20$, $p=0.5$, $b_L=-0.072$, $b_R=0.074$,
$\lambda_L=\lambda_R=-0.03$.} \label{tab:DecayWidths}
\begin{ruledtabular}
  \begin{tabular}{c|ccccc|c}
     $V^0\rightarrow$ & $W^+W^-$ & $t\bar{t}$ & $b\bar{b}$ & $u\bar{u}$ & $d\bar{d}$ & total\\
     width (GeV) & 5.29 & 8.98 & 5.79 & 0.007 & 0.004 & 20.09 \\
     \hline
     $V^+\rightarrow$ & $W^+Z$ & \multicolumn{2}{c}{$t\bar{b}$} & \multicolumn{2}{c|}{$u\bar{d}$} & total\\
     width (GeV) & 5.40 & \multicolumn{2}{c}{13.10} & \multicolumn{2}{c|}{0.010} & 18.53 \\
  \end{tabular}
\end{ruledtabular}
\end{table}

Let us note that the $pp\rightarrow (t\bar{b}+b\bar{t})X$ process
is sensitive to the direct fermion couplings through the $V^\pm
tb$ vertex, the $pp\rightarrow (W^+Z+W^-Z)X$ process is only
sensitive to $g''$ through the triple gauge vertex of $V^\pm W^\pm
Z$, and the $pp\rightarrow W^+W^-X$ process is sensitive to the
fermion couplings through the $V^0bb$ vertex and to $g''$ through
the triple gauge vertex of $V^0W^+W^-$. In the $pp\rightarrow
W^+W^-X$ case the sensitivity to fermion couplings is only through
the $b\bar{b}\rightarrow W^+W^-$ component which contributes just
a small fraction of the cross section. Nevertheless, this
subprocess contributes significantly to the resonance peak.

In the ILC s-channel production the resonance must be produced
through the annihilation of the light fermions which is a
disadvantage for this kind of process. Nevertheless,
Fig.~\ref{fig:XSforPSP} suggests that it might be worthwhile to
study sensitivity of the ILC $e^-e^+\rightarrow V^0\rightarrow 2$
processes with the electroweak gauge bosons or top/bottom quarks
in the final states. Our preliminary work~\cite{InProgress,18KSF}
on this issue further supports this hope.

All cross section calculations in this section were performed
at the tree-level using the CompHEP software~\cite{CompHEP}. For
that sake, we have implemented the tBESS Lagrangian into the
COMPHEP as one of its models.

\section{Conclusions}
\label{sec:End}

The effective Lagrangian, the so-called top-BESS model, of an
alternative scenario of ESB has been formulated and investigated.
It is the effective description of beyond the SM hypotheses where
new strong interactions are responsible for ESB. The tBESS model
singles out the direct coupling of the vector triplet to the third
quark generation only. Therefore, it is a suitable effective
Lagrangian for theories where the top (and perhaps also bottom)
quark play an outstanding role in new physics beyond the SM.

There is no direct coupling of the tBESS vector triplet to the
light SM fermions. Thus, the vector triplet can couple to the
light fermions only through indirect interactions induced by the
mixing of the vector triplet with the electroweak gauge bosons.

The study of the electroweak gauge-boson scattering implies that
the no-Higgs SM unitarity restriction of 1.7~TeV can be somewhat
raised by the introduction of the tBESS vector triplet for the
limited choice of the tBESS free parameters only. For example,
when $M_{V^0}=1$~TeV, the tBESS unitarity up to $1.7$~TeV, at
least, is guaranteed for $g''\stackrel{>}{\sim} 3$ and for any
values of the fermion sector parameters $b_L$, $b_R$, $\lambda_L$,
$\lambda_R$ and $p$. The unitarity allowed parameter region
quickly shrinks if we require the unitarity limit higher then
$2$~TeV. It seems that the strong couplings of the vector triplet
with fermions might influence these conclusions. However, the
analysis of the top/bottom quark scattering amplitudes would be
required to settle this question. In addition, large values of the
$b$ and $\lambda$ parameters are admissible only in the
fine-tuning regime.

Confrontation of the tBESS model with the EWPD results in the
low-energy limits on the model's parameters. The $\epsilon_3$
parameter can accommodate any value of $g''$ with only quite low
probability, e.g.\ $10\%$ when $M_H$ is set to 1~TeV in
approximating the loop contributions to $\epsilon_3$.
Nevertheless, there are good reasons not to take these numbers too
seriously. For example, they can be altered by adding a new direct
interaction of the vector triplet with light fermions to the tBESS
model. There is also an independent lower limit on $g''$ set by
the D0 measurements of the $WWZ$, $g''\geq 3.4$ ($95\%$~C.L.),
which plays no significant role under the given circumstances.

The epsilon analysis combined with the $B\rightarrow X_s\gamma$
measurement restricts the expressions $b_L-2\lambda_L$ and
$b_R+2\lambda_R$. The situation is complicated by the fact that
due to its definition $\epsilon_b$ can be used to extract
low-energy limits only if $p=0$ or $b_R=-2\lambda_R$. To obtain
the restrictions for more general case of the tBESS model, we have
used the measurements of the branching ratio for $Z\rightarrow
b\bar{b}$ and the total decay width of the $Z$ boson. There are no
low-energy limits on the individual $b$ and $\lambda$ parameters.
However, if the fine-tuning of the $b$ and $\lambda$ parameters
should not go below $10\%$ then $b_L$ and $\lambda_L$ might be as
large as about $\pm 0.1$. Analogically, $b_R$ and $\lambda_R$ can
be as large as about $\pm 0.7$ when $p=0$ or about $\pm 0.08$ when
$p=1$. These numbers are $\Lambda$ dependent and, to a lesser
extent, $g''$ dependent, though.

If the values of the $b_L$ and $b_R$ parameters lie in the Death
Valley region, the top/bottom partial decay widths of the vector
resonances diminish below the no-direct-coupling value. It is a
consequence of the interplay of the direct and indirect fermion
couplings of the vector triplet. If this occurred the resonance
peak in a process where $V$ decays to top and/or bottom quarks
could disappear even though the resonance exists and couples
directly to the third quark generation.

Our calculations suggest that there are acceptable values of the
tBESS parameters which can result in detectable signals at the LHC
and/or the ILC. In particular, despite what would be the one's
first guess, it seems to be worthwhile to study the LHC Drell-Yan
processes and the ILC $e^+e^-\rightarrow R$ processes with
top/bottom quarks and EW gauge bosons in their final states in
order to probe the top-BESS model. However, this is far from
conclusive. It would need to perform a more systematic study
focused on the process analysis. The next step in this direction
might include more systematic scan of the parametric space, more
realistic final states and cuts, and the inclusion of the
backgrounds, at least.

\begin{acknowledgments}
The work of M.G.\ and J.J.\ was supported by the Research Program
MSM6840770029 and by the project International Cooperation
ATLAS-CERN of the Ministry of Education, Youth and Sports of the
Czech Republic. J. J.\ was also supported by the NSP grant of the Slovak
Republic. M.G.\ and I.M.\ were
supported by the Slovak CERN Fund. We would also like to thank the
Slovak Institute for Basic Research for their support.
\end{acknowledgments}

\appendix

\section{Transformation relations}
\label{app:Trafos}

The transformation relations of the basic mathematical objects
used to build the tBESS effective Lagrangian are summarized in
Table~\ref{tab:trafos}. Recall that the weak hypercharge $Y =
T_R^3+(B-L)/2$, where $T_R^3$ is the third $SU(2)_R$ generator.
Thus when $B-L=0$ then $Y=T_R^3\equiv Y_R$.
\begin{table}
\caption{\label{tab:trafos} The transformation relations of the
basic mathematical objects used to build the tBESS effective
Lagrangian.}
\begin{ruledtabular}
\begin{tabular}{ccc}
 Object & Global\footnotemark[1]\footnotemark[3]
        & Local\footnotemark[2]\footnotemark[3] \\
\hline
    $\BW_\mu = i g W_\mu^a\tau^a$ & $g_L \BW_\mu g_L^\dagger$ &
    $g_L \BW_\mu g_L^\dagger +g_L\pard_\mu g_L^\dagger$ \\
    $\BB_\mu = i g' B_\mu Y$ & $g_{R} \BB_\mu g_{R}^\dagger$ &
    $\BB_\mu +g_{Y}\pard_\mu g_{Y}^\dagger$ \\
    $\BV_\mu=ig''V_\mu^a\tau^a/2$ & $h^\dagger\BV_\mu h$ &
    $h^\dagger\BV_\mu h + h^\dagger\pard_\mu h$ \\
    $\xi_L$ & $g_L\; \xi_L \;h$ & $g_L(x)\; \xi_L \;h(x)$ \\
    $\xi_R$ & $g_R\; \xi_R \;h$ & $g_{Y_R}(x)\; \xi_R \;h(x)$ \\
    $U = \xi_L\xi_R^\dagger$ & $g_L\; U \;g_R^\dagger$ & $g_L(x)\; U \;g_{Y_R}^\dagger(x)$ \\
    $\bar{\omega}_\mu^{\parallel,\perp}$ &
    $h^\dagger\; \bar{\omega}_\mu^{\parallel,\perp}\; h$ &
    $h^\dagger(x)\; \bar{\omega}_\mu^{\parallel,\perp}\; h(x)$ \\
    $\psi_L$ & $g_{BL}\;g_L\;\psi_L$ & $g_{Y}(x)\;g_L(x)\;\psi_L$ \\
    $\psi_R$ & $g_{BL}\;g_R\;\psi_R$ & $g_{Y}(x)\;\psi_R$ \\
    $\chi_L=\xi_L^\dagger\psi_L$ & $g_{BL}\;h^\dagger\;\chi_L$ & $g_{Y}(x)\;h^\dagger(x)\;\chi_L$ \\
    $\chi_R=\xi_R^\dagger\psi_R$ & $g_{BL}\;h^\dagger\;\chi_R$ & $g_{BL}(x)\;h^\dagger(x)\;\chi_R$ \\
\end{tabular}
\end{ruledtabular}
 \footnotetext[1]{$SU(2)_L\times SU(2)_R\times U(1)_{B-L}\times SU(2)_{HLS}$}
 \footnotetext[2]{$SU(2)_L\times U(1)_{Y}\times SU(2)_{HLS}$}
 \footnotetext[3]{$g_L\in SU(2)_L$, $g_R\in SU(2)_R$, $h\in
SU(2)_{HLS}$, $g_Y\in U(1)_Y$, $g_{BL}\in U(1)_{B-L}$, and
$g_{Y_R}\in U(1)_Y$ if $B-L=0$}
\end{table}

\section{Low-energy limits from the epsilon analysis}
\label{app:LElimits}

In deriving the low-energy limits from the epsilon analysis we
follow the approach of \cite{LowEBESS}.

The fermion Lagrangian describing the anomalous interactions of
the electroweak gauge bosons with the top and bottom quarks reads
\begin{eqnarray}
\label{anomLagrangian}
  \!\!\!\!\!\!\!{\cal L}_{(t,b)}^{anom} &=& \phantom{-}{\cal L}_{(t,b)}^{SM}
  - \sum_{h=L,R}\left\{
    \frac{e}{\sqrt{2}s_\theta}\kappa_h^{Wtb}(\bar{t}_h\Wslash^+ b_h +\mbox{h.c.})\right.
  \nonumber\\
  && \left. + \frac{e}{2 s_\theta c_\theta}
       \left[\kappa_h^{Ztt}(\bar{t}_h\Zslash t_h) +
             \kappa_h^{Zbb}(\bar{b}_h\Zslash b_h) \right]\right\},
\end{eqnarray}
where $\kappa$`s parameterize the deviations from the SM
\begin{eqnarray}
  {\cal L}_{(t,b)}^{SM}
  & = & - \frac{2}{3} e \;\bar{t} \Aslash t
        + \frac{1}{3} e \;\bar{b} \Aslash b
        - \frac{e}{\sqrt{2}s_\theta}
        (\bar{t}_L \Wslash^+ b_L + \mbox{h.c.})
  \nonumber\\
   && - \frac{e}{2 s_\theta c_\theta} \left[\left(1-\frac{4}{3} s_\theta^2 \right)
        (\bar{t}_L \Zslash t_L)
      - \frac{4}{3} s_\theta^2 ({\bar t}_R \Zslash t_R)\right]
   \nonumber \\
   & & +\frac{e}{2 s_\theta c_\theta} \left[\left(1-\frac{2}{3} s_\theta^2 \right)
       (\bar{b}_L \Zslash b_L) -
       \frac{2}{3} s_\theta^2 (\bar{b}_R \Zslash b_R)\right],
   \nonumber
\end{eqnarray}
where electric charge $e$ and sinus theta $s_\theta$ are defined
as
\begin{eqnarray}
 e = \sqrt{4\pi \alpha(M_Z)}, \\
 s_\theta^2 c_\theta^2 = \frac{\sqrt{2}e^2}{8G_F M_Z^2}.
\end{eqnarray}
In the case of the tBESS model the $\kappa$ parameters read
\begin{eqnarray}
\label{kappas} \kappa_L^{Wtb} & = &
-\left(\frac{b_L}{2}-\lambda_L\right)
    \left(1-h\right)-h, \nonumber \\
\kappa_R^{Wtb} & = & p\left(\frac{b_R}{2}+\lambda_R\right)
    \left(1-h\right), \nonumber \\
\kappa_L^{Ztt} & = & -\left(\frac{b_L}{2}-\lambda_L\right)
    -\frac{4}{3}h, \nonumber \\
\kappa_R^{Ztt} & = & \left(\frac{b_R}{2}+\lambda_R\right)
   -\frac{4}{3}h, \nonumber \\
\kappa_L^{Zbb} & = & \left(\frac{b_L}{2}-\lambda_L\right)
    +\frac{2}{3}h, \nonumber \\
\kappa_R^{Zbb} & = & -p^2\left(\frac{b_R}{2}+\lambda_R\right)
    +\frac{2}{3}h,
\end{eqnarray}
and
\begin{equation}
\label{renormalization}
h=\frac{s_\theta^2}{c_{2\theta}}\left(\frac{g}{g''}\right)^2,
\end{equation}
where $c_{2\theta}=c_\theta^2-s_\theta^2$.

The Lagrangian (\ref{anomLagrangian}) can be confronted with the
model independent epsilon analysis of the electroweak precision
data \cite{EpsilonMethod1,EpsilonMethod2}. The new physics tree-level
contributions to the epsilon parameters are functions of the
$\kappa$`s. Beside that, there are loop-level contributions
$\delta\epsilon^{loops}$. Thus,
\begin{equation}
  \epsilon=\epsilon^{tree}+\delta\epsilon^{loops},
\end{equation}
where the loop contributions are of two kinds:
the SM ones~\cite{4siteBESS}
and the new physics ones,
$\delta\epsilon^{loops}\approx\delta\epsilon^{SM}+\delta\epsilon^{NP}$.
We derive limits from the values of three epsilons ---
$\epsilon_1, \epsilon_3$, and $\epsilon_b$ --- including some
radiative corrections as indicated in Table~\ref{tab:epsilons}.
\begin{table}
\caption{\label{tab:epsilons} The contributions to the individual
epsilons. The included contributions are denoted by the checkmark
(\Checkmark\ ), the left-out (not calculated) contributions are
denoted by the cross mark (\bm{$\times$}).}
\begin{ruledtabular}
\begin{tabular}{cccccc}
 & Tree & SM loops & \multicolumn{3}{c}{NP loops} \\
 & & & $Wtb$ & $Ztt$ & $Zbb$ \\
\hline
 $\epsilon_1$ & \mbox{\Checkmark} & \mbox{\Checkmark} & \mbox{\Checkmark} & \mbox{\Checkmark} & $\bm{\times}$ \\
 $\epsilon_3$ & \mbox{\Checkmark} & \mbox{\Checkmark} & $\bm{\times}$ & $\bm{\times}$ & $\bm{\times}$ \\
 $\epsilon_b$ & \mbox{\Checkmark} & \mbox{\Checkmark} & \mbox{\Checkmark} & \mbox{\Checkmark} & $\bm{\times}$ \\
\end{tabular}
\end{ruledtabular}
\end{table}

The restriction on $g''$ can be obtained from $\epsilon_3$. In the
case of tBESS
\begin{equation}
\label{eps3}
    \epsilon_3 = \left(\frac{g}{g''}\right)^2 + \delta\epsilon_3^{SM},
\end{equation}
where the radiative correction beyond the SM have not been
included.
When $\delta\epsilon_3^{SM}$ is calculated,
$M_H\sim$~TeV should be considered in order to imitate a strong
ESB physics. In our calculations, we consider $M_H=1$~TeV, $2$~TeV,
as well as $300$~GeV, for the sake of comparison. All these
values result in $\delta\epsilon_3^{SM}$ larger than the experimental
$\epsilon_3$. The consequences are discussed in Sec.~\ref{subsec:Limits}.

An expression, analogical to~(\ref{eps3}), was obtained for BESS
in \cite{LowEBESS}. In BESS though, $\epsilon_3$ is also a
function of parameter $b$ which parameterizes the direct universal
coupling of the vector triplet to the left fermions
\[ \epsilon_3 = -\frac{b}{2}+\left(\frac{g}{g''}\right)^2 + \delta\epsilon_3^{SM}. \]
Thus, due to the universality, the limit for $g''$ depends on $b$
and \textit{vice versa} (see Fig.~1 in~\cite{LowEBESS}). When we
change $g''$ from $10$ to $\infty$ the interval of the allowed $b$
values shifts by about $146\%$ of its length. In contrast,
the tBESS allowed interval for $b_L-2\lambda_L$, when
$b_R+2\lambda_R=0$, will shift by about $1\%$ only (see
Fig.~\ref{fig:blambdaconts1}). Here, the sensitivity to $g''$
enters only through the $h$, as can be seen in
(\ref{renormalization}).

Each of the parameters $\epsilon_1$ and $\epsilon_b$ restricts the
combinations $b_L-2\lambda_L$ and $b_R+2\lambda_R$. The
$\epsilon_1$ has no tree contribution, so we have
\begin{equation}\label{eq:eps1}
  \epsilon_1 = \delta\epsilon_1^{SM} + \delta\epsilon_1^{NP}.
\end{equation}
On the other hand,
\begin{equation}\label{eq:epsb}
 \epsilon_b = \epsilon_b^{tree} + \delta\epsilon_b^{SM} + \delta\epsilon_b^{NP},
\end{equation}
where
\begin{equation}\label{eq:epsbtree}
 \epsilon_b^{tree} = -\kappa_L^{Zbb}+\kappa_R^{Zbb}.
\end{equation}
For the NP loop contributions to the $\epsilon_{1}$ and
$\epsilon_{b}$ the relations from \cite{LariosKappaAnalysis} have
been adopted to obtain
\begin{eqnarray}
  \delta\epsilon_1^{NP} &=& \frac{3 m_t^2 G_F}{2 \sqrt{2} \pi^2}\ln\frac{\Lambda^2}{m_t^2}
                       \left[ \kappa_L^{Wtb}\left( 1+{\kappa_L^{Wtb}}\right)\right.
  \nonumber\\
                   & & +
  \left. \left(\kappa_R^{Ztt}-\kappa_L^{Ztt}\right)\left(1-\kappa_R^{Ztt}+\kappa_L^{Ztt}\right) \right],
  \\
  \delta\epsilon_b^{NP} &=& \frac{m_t^2 G_F}{2 \sqrt{2} \pi^2} \ln\frac{\Lambda^2}{m_t^2}
  \nonumber\\
  && \times
  \left[\left(\kappa_L^{Ztt}-\frac{1}{4}\kappa_R^{Ztt}\right)\left(1+2\kappa_L^{Wtb}\right)\right],
\end{eqnarray}
where $\Lambda$ is the cut-off of the low-energy
top-BESS model and it is set to 1 and 2 TeV. When calculating the SM
radiative corrections we set $\Lambda=M_H$. The loop
contributions coming from the anomalous $Zb\bar{b}$ vertex are not
considered here. The mass of the top quark considered in the
calculations is $m_t=172.7$~GeV.

The allowed values of the expressions $b_L-2\lambda_L$ and
$b_R+2\lambda_R$ are given by the intersections of the
$\epsilon_1$ and $\epsilon_b$ restrictions.
However, recall that the $\epsilon_b$
restrictions can be applied only if $p=0$ or $b_R=-2\lambda_R$.

\begin{figure}
\includegraphics[scale=0.48]{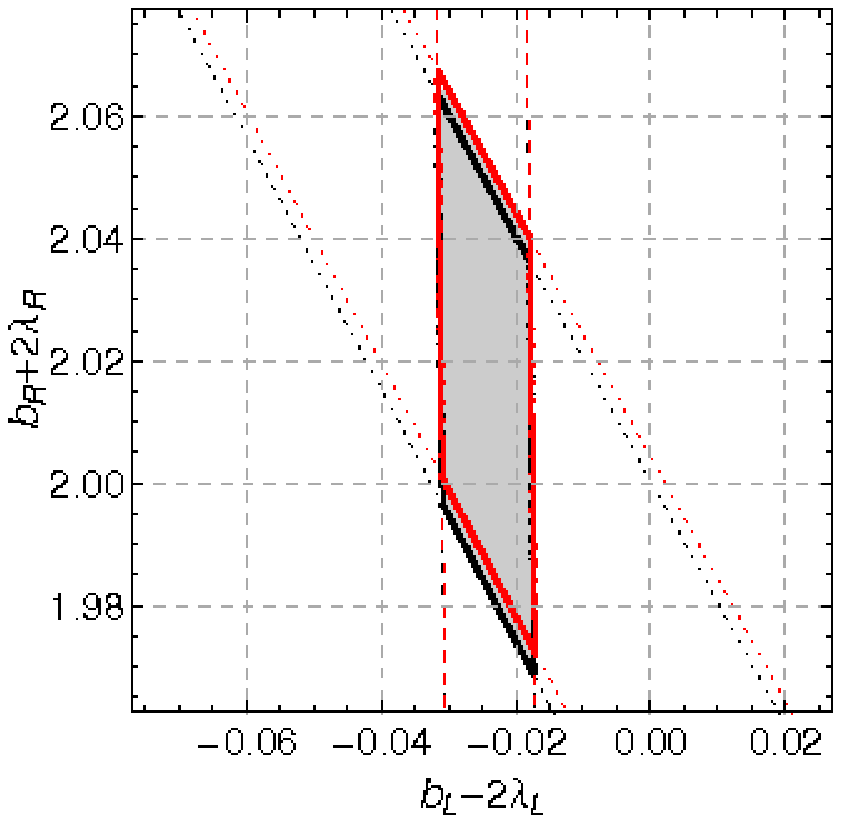}\hfill
\includegraphics[scale=0.48]{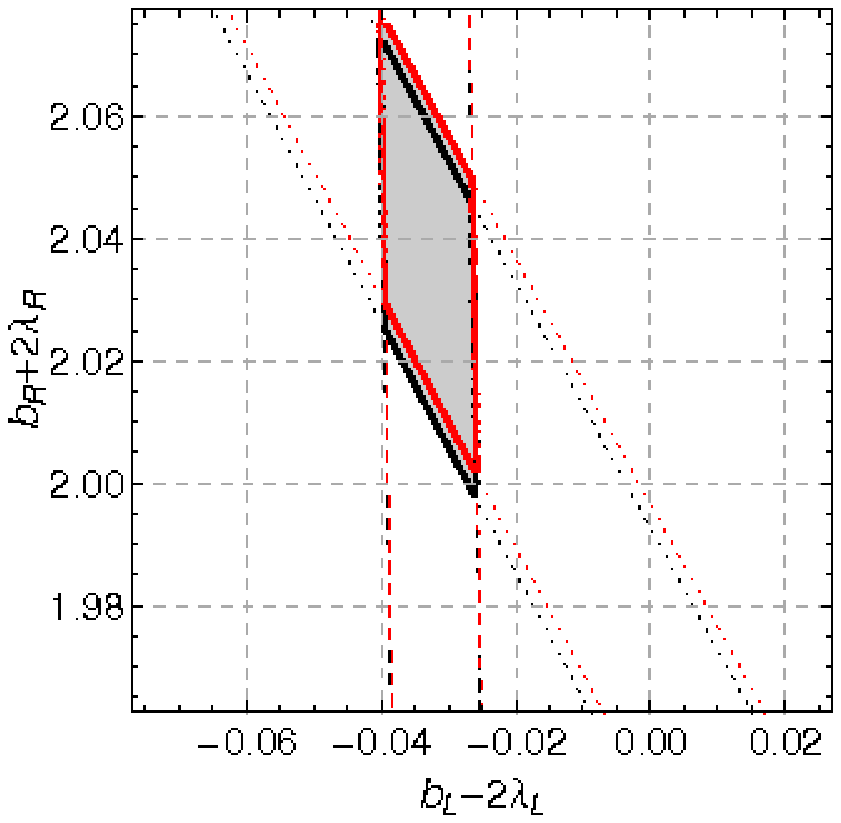}
\caption{\label{fig:LElimits2}(color online) One of the four
intersections of the $90\%$~C.L. $\epsilon_1$-allowed region
(tilted strip) with the $90\%$~C.L. $\epsilon_b$-allowed region
(vertical strip). Here, $p=0$ is assumed. Other two intersections
are excluded by the $B\rightarrow X_s\gamma$ measurement and the
fourth intersection is depicted in Fig.~\ref{fig:blambdaconts1}.
The black contours correspond to $g''=10$ and the red ones to
$g''\rightarrow\infty$. The cut-off scales considered are
$\Lambda=1$~TeV (left) and $\Lambda=2$~TeV (right).}
\end{figure}

If $p=0$, the intersections form four disconnected
regions. The intersection, which contains the origin, is depicted
in Fig.~\ref{fig:blambdaconts1}. Two other intersections are ruled
out by the $B\rightarrow X_s\gamma$ measurement. The fourth
intersection and its dependence on $g''$ and $\Lambda$ is depicted
in Fig.~\ref{fig:LElimits2}.
All the shown intersections of Fig.~\ref{fig:LElimits2} lie
completely inside the $B\rightarrow X_s\gamma$ allowed area.
Despite that, we have not considered the fourth intersection
values for the tBESS parameters in our analysis. The main reason
is that the allowed interval of $b_R+2\lambda_R$ is too narrow.
Using the values of $b_R$ and $\lambda_R$ of this region would
correspond to fine-tuning below $10\%$, at least.

\section{Low-energy limits from the \bm{$\Gamma(Z\rightarrow b\bar{b})$} decay}
\label{app:LElimitsGammab}

To derive the low-energy limits from partial decay width
$\Gamma(Z\rightarrow b\bar{b})$ the Eq.~(11) of~\cite{EpsilonMethod2}
has been used
\begin{eqnarray}\label{eq:Gammab}
\Gamma_b & = & \frac{G_F {M_Z}^3}{6 \pi \sqrt{2}} \beta
\left(\frac{3-\beta^2}{2}{g_{bV}}^2 + \beta^2 {g_{bA}}^2
\right) \nonumber \\
        & & \times N_C R_{QCD} \left(1+ \frac{\alpha_e}{12\pi} \right),
\end{eqnarray}
where $\beta = \sqrt{1-4m_b^2/M_Z^2}$, $R_{QCD} = 1+1.2 a - 1.1a^2 -13a^3$
is the QCD correction factor, $a = \alpha_s(M_Z)/\pi$,
and $g_{bV}$ and $g_{bA}$ are vector and axial-vector couplings of
the $b$ quark. We approximate the couplings by
\begin{eqnarray}
 g_{bV} & = & g_{bV}^{tBESS,tree} + \delta g_{bV}^{SM,loop}, \label{eq:gbV}\\
 g_{bA} & = & g_{bA}^{tBESS,tree} + \delta g_{bA}^{SM,loop}, \label{eq:gbA}
\end{eqnarray}
where the first terms are the top-BESS tree-level couplings
\begin{eqnarray}
 g_{bV}^{tBESS,tree} & = & \left(1 - 4/3 \;s_\theta^2
                             -\kappa_L^{Zbb} - \kappa_R^{Zbb} \right)/2,
 \\
 g_{bA}^{tBESS,tree} & = & \left(1 - \kappa_L^{Zbb} + \kappa_R^{Zbb} \right)/2,
\end{eqnarray}
and the second terms are the SM loop contributions
which can be expressed in terms of the epsilon analysis
\begin{eqnarray}
 \delta g_{bV}^{SM,loop} & = & \frac{1}{2} \left(
 1+\frac{\epsilon_1^{SM}}{2}\right) \left(1-\frac{4}{3}(1+\Delta k)s_\theta^2
 +\epsilon_b^{SM}\right)
 \nonumber \\
 & & - g_{bV}^{SM,tree},
 \\
 \delta g_{bA}^{SM,loop} & = & \frac{1}{2} \left(
 1+\frac{\epsilon_1^{SM}}{2}\right) (1+\epsilon_b^{SM})
 - g_{bA}^{SM,tree},
\end{eqnarray}
where $g_{bV}^{SM,tree} = (1-4/3\;s_\theta^2)/2$, $g_{bA}^{SM,tree}=1/2$, and
\begin{equation}
 \Delta k = \frac{\epsilon_3^{SM}-c_\theta^2 \epsilon_1^{SM}}{c_{2\theta}}.
\end{equation}
The SM epsilons are equal to $\delta\epsilon^{SM}$'s
of~\cite{4siteBESS}. If the loop corrections were not considered
in the Eqs.~(\ref{eq:gbV}) and (\ref{eq:gbA}), each of the
resulting stripes in Fig.~\ref{fig:blambdaconts1} would shift to
the right by its width.

The low-energy limits for the fermion parameters are based on the
experimental values~\cite{PDG2010}
\begin{eqnarray}
 \mbox{B.R.}(Z\rightarrow b\bar{b}) &=& (0.1512\pm 0.0005),
 \\
 \Gamma_{tot}(Z)\;\;\;\;\; &=& (2.4952 \pm 0.0023)~\mbox{GeV}.
\end{eqnarray}


\end{document}